\documentclass[letterpaper,journal]{IEEEtran}
\usepackage{amsmath,amssymb,amsfonts,amsthm, bm}
\usepackage{algorithmic}
\usepackage{algorithm}
\usepackage{array}
\usepackage[caption=false,font=footnotesize,labelfont=rm,textfont=rm]{subfig}
\usepackage{textcomp}
\usepackage{longtable}
\usepackage{tablefootnote}
\usepackage{stfloats}
\usepackage{url}
\usepackage{verbatim}
\usepackage{graphicx}
\usepackage{cite}
\usepackage{diagbox}
\usepackage[dvipsnames]{xcolor}
\usepackage{threeparttable}
\usepackage{multirow} 
\usepackage[colorlinks = true,
            linkcolor = MidnightBlue,
            urlcolor  = teal,
            citecolor = NavyBlue,
            anchorcolor = black]{hyperref}
\usepackage{booktabs}         

\usepackage{enumitem} 
\def\BibTeX{{\rm B\kern-.05em{\sc i\kern-.025em b}\kern-.08em
    T\kern-.1667em\lower.7ex\hbox{E}\kern-.125emX}}
\usepackage{balance}
\usepackage{etoolbox}

\definecolor{b}{rgb}{0,0,0} 
\makeatletter
\newcommand*{\new}{\@ifnextchar\bgroup{\new@}{\color{b}}}
\newcommand*{\new@}[1]{{\textcolor{b}{#1}}}
\makeatother 

\definecolor{c}{rgb}{0,0,0} 
\makeatletter
\newcommand*{\newer}{\@ifnextchar\bgroup{\newer@}{\color{c}}}
\newcommand*{\newer@}[1]{{\textcolor{c}{#1}}}
\makeatother 

\begin{document}
\bstctlcite{setting}
\title{A Unified Multicarrier Waveform Framework for Next-generation Wireless Networks: Principles, Performance, and Challenges}

\author{Xingyao Zhang, Haoran Yin, Yanqun Tang, Yao Ge,~\IEEEmembership{Member,~IEEE,}\\ Yong Zeng,~\IEEEmembership{Fellow,~IEEE,}  Miaowen Wen,~\IEEEmembership{Senior Member,~IEEE,} Zilong Liu,~\IEEEmembership{Senior Member,~IEEE,} \\Yong Liang Guan,~\IEEEmembership{Senior Member,~IEEE,} H\"useyin Arslan,~\IEEEmembership{Fellow,~IEEE,} and Giuseppe Caire,~\IEEEmembership{Fellow,~IEEE}

\thanks{This work was supported in part by the Shenzhen Science and Technology Major Project under Grant KJZD20240903102000001 and in part by the Science and Technology Planning Project of Key Laboratory of Advanced IntelliSense Technology, Guangdong Science and Technology Department under Grant 2023B1212060024. (Xingyao Zhang and Haoran Yin are Co-first authors with
equal contribution.) (\textit{Corresponding author: Yanqun Tang.})

Xingyao Zhang, Haoran Yin, and Yanqun Tang are with the School of Electronics and Communication Engineering, Sun Yat-sen University, Shenzhen 518107, China, and also with the Guangdong Provincial Key Laboratory of Sea-Air-Space Communication, China (e-mail: zhangxy956@mail2.sysu.edu.cn; yinhr6@mail2.sysu.edu.cn; tangyq8@mail.sysu.edu.cn);

Yao Ge is with the AUMOVIO-NTU Corporate Lab, Nanyang Technological University, Singapore 639798 (e-mail: yao.ge@ntu.edu.sg);

Yong Zeng is with the National Mobile Communications Research Laboratory and Frontiers Science Center for Mobile Information Communication and Security, Southeast University, Nanjing 210096, China, and also with the Purple Mountain Laboratories, Nanjing 211111, China (email: yong$\_$zeng@seu.edu.cn);

Miaowen Wen is with the School of Electronic and Information Engineering, South China University of Technology, Guangzhou 510641, China (e-mail: eemwwen@scut.edu.cn);

Zilong Liu is with the School of Computer Science and Electronics Engineering, University of Essex, Colchester CO4 3SQ, U.K. (e-mail: zilong.liu@essex.ac.uk);

Yong Liang Guan is with the School of Electrical and Electronics Engineering, Nanyang Technological University, Singapore 639798 (e-mail: eylguan@ntu.edu.sg);

Hüseyin Arslan is with the Department of Electrical and Electronics Engineering, Istanbul Medipol University, Istanbul 34810, Turkey (email: huseyinarslan@medipol.edu.tr);

Giuseppe Caire is with the Faculty of Electrical
Engineering and Computer Science, Technical University of Berlin,  Berlin 10587, Germany (e-mail:caire@tu-berlin.de).

} 
}

\maketitle

\begin{abstract}
    Next-generation wireless networks require enhanced flexibility, efficiency, and reliability in physical layer waveform design to address the challenges posed by heterogeneous channel conditions and stringent quality-of-service demands. To this end, this paper proposes a unified multicarrier waveform framework that provides a systematic characterization and practical implementation guidelines to facilitate waveform selection for the sixth-generation (6G) mobile networks and beyond. We commence by examining the design principles of the state-of-the-art waveforms, which are categorized into one-dimensional modulation waveforms (e.g., orthogonal frequency division multiplexing (OFDM) and affine frequency division multiplexing (AFDM)) and two-dimensional modulation waveforms (e.g., orthogonal time frequency space (OTFS)). Their inherent resilience against various channel-induced interference is further studied, revealing their distinct suitability in diverse channel conditions. Furthermore, an in-depth performance analysis is presented by comparing their key performance indicators (KPIs), followed by an extensive exploration of these advanced waveforms in various applications. Consequently, this work aims to serve as a pivotal reference for waveform adoption in future 6G standardization and network deployment.
\end{abstract}


\vspace{-4pt}
\section{Introduction}
\label{sec1}
\subsection{Background}
\label{sec1-1}
The fifth-generation (5G) mobile communication network, commercially launched in 2019, has enabled three cornerstone service categories: enhanced mobile broadband (eMBB), massive machine-type communications (mMTC), and ultra-reliable low-latency communications (URLLC)\cite{giordani20206g}. Building upon this foundation, the sixth-generation (6G) mobile communication network is expected to support a broader set of application scenarios, including native integration of artificial intelligence (AI) and communications, integrated sensing and communication (ISAC), and truly ubiquitous connectivity\cite{ITU2023,wang2023road}. To meet these emerging demands, 6G envisions a tightly integrated terrestrial-non-terrestrial network (NTN) architecture characterized by ultra-high performance, pervasive AI and machine learning (ML), and an Internet of Everything (IoE) that extends system capabilities beyond conventional communications\cite{pennanen20256g,zeng2024tutorial}. This seamless, ubiquitous connectivity will further enhance human-centric services by converging perceptual and communicative functions within a unified network fabric.

Achieving these ambitious use cases requires meeting a comprehensive set of key performance indicators (KPIs) defined by IMT-2030. For example, 6G targets peak data rates beyond 1 Tbps, end-to-end latencies on the order of 10 microseconds for time-critical services, connection densities spanning
$10^6$ to $10^8$ devices per square kilometer to support a truly ubiquitous IoE, and energy-efficiency improvements of 10 to 100 times over 5G to sustain large-scale, pervasive operation\cite{chafii2023twelve}. Since these requirements exceed the boundaries of current wireless capabilities, they demand a fundamental rethinking and redesign of the physical layer (PHY).

Multicarrier waveforms form the core of modern wireless PHY design. Among many others, orthogonal frequency division multiplexing (OFDM)\cite{solaija2024orthogonal} has been the foundational waveform since the 4G time. Although it has facilitated high data rates and robustness against multipath fading, its inherent limitations hinder its ability to satisfy the stringent demands of 6G\cite{lu2023delay}. Specifically, OFDM faces four major drawbacks that limit its effectiveness for next-generation networks. First, its high out-of-band emission (OOBE), resulting from the sinc-shaped subcarrier spectra, leads to significant spectrum leakage. Second, the spectral efficiency is further compromised by the cyclic prefix (CP), which is inserted to combat inter-symbol interference (ISI), resulting in redundant overhead and diminished resource utilization. Third, OFDM exhibits a high peak-to-average power ratio (PAPR), which can distort transmitted signals and reduce the effective communication range. Fourth, OFDM is sensitive to large Doppler shifts, as the orthogonality among different subcarriers may be destroyed in high-mobility channels, leading to a significant increase in inter-carrier interference (ICI). Considering that 6G will move towards millimeter wave (mmWave) and terahertz (THz) bands\cite{akdeniz2014millimeter,THz,lu2024tutorial}, it is necessary to explore alternative multicarrier waveform schemes under time-varying channels to support high mobility scenarios with large Doppler spread. 

To this end, many alternative new waveforms emerged. As a matter of fact, it is difficult to satisfy diverse use cases like massive connectivity and perception with just a single waveform in next-generation wireless networks. \textbf{\textit{Subsequently, how to make proper waveform selection becomes an important issue that needs to be carefully considered.}} Therefore, it is essential to systematically review existing waveform schemes, summarize multicarrier waveforms in terms of their principles, performance, and applications, and develop a unified framework to address the challenges outlined by IMT-2030.

\vspace{-7pt}
\subsection{Related Works}
\label{sec1-2}

In recent years, a variety of novel multicarrier waveform schemes have emerged, which rely on optimizing OFDM at various levels to mitigate interference arising from channel delay and/or Doppler spreads. In 5G uplink transmissions, discrete Fourier transform spread OFDM (DFT-s-OFDM)\cite{sahin2016flexible} incorporates a DFT precoding step before conventional OFDM modulation, which spreads the input data symbols across multiple subcarriers in the frequency domain and reduces the PAPR substantially. Another representative class of multicarrier technology is filter bank multicarrier (FBMC)\cite{nissel2017filter}, including filtered orthogonal frequency division multiplexing (F-OFDM)\cite{zhang2015filteredofdm}, generalized frequency division multiplexing (GFDM)\cite{michailow2014generalized}, and universal filtered multicarrier (UFMC)\cite{vakilian2013universalfiltered}. Among them, FBMC utilizes per-subcarrier filtering to minimize OOBE and enhance spectral efficiency by replacing the rectangular window used in OFDM. However, these schemes are not well adapted to the delay-Doppler spreading in doubly-dispersive channels (DDC). Recently, multicarrier modulation schemes in the chirp domain or delay-Doppler (DD) domain have emerged as promising alternatives.

For chirp-based schemes\cite{sui2025multifunctional}, the theoretical foundations of multi-chirp basis modulation have been extensively explored, including the fractional Fourier transform (FrFT)\cite{candan2000discrete}, discrete Fresnel transform (DFnT)\cite{james1996generalized}, and discrete affine Fourier transform (DAFT)\cite{soo-changpei2000closedform}. Building on these transforms, a FrFT-based multicarrier system, termed FrFT-OFDM hereafter\cite{martone2001multicarrier}, employs time-varying subchannel carriers to decompose frequency distortions in time-varying channels. For enhanced compatibility with OFDM, orthogonal chirp division multiplexing (OCDM)\cite{ouyang2016orthogonal} utilizes DFnT for high-speed communications, whereas affine frequency division multiplexing (AFDM)\cite{bemani2021afdm, bemani2021affine, bemani2023affinea, yin2025affineb} exploits parameterizable DAFT to address both channel delay and Doppler spreads in the chirp domain. Notably, among chirp-based schemes, AFDM offers high flexibility, low pilot overhead, and full diversity in linear time-varying (LTV) channels.

For the DD-based schemes, orthogonal time frequency space (OTFS)\cite{hadani2017orthogonal} stands out as a representative waveform, which leverages the inverse symplectic finite Fourier transform (ISFFT) to modulate a two-dimensional (2D) grid of information symbols directly in the DD domain. Another example is orthogonal delay-Doppler modulation (ODDM)\cite{lin2022orthogonal}, which designs orthogonal pulses tailored to fine DD resolutions. There are several waveforms that perform modulation in other 2D domains beyond the traditional time-frequency (TF) plane. For instance, orthogonal time-sequency multiplexing (OTSM)\cite{thaj2021orthogonal} maps information symbols to the delay-sequency domain via cascaded time-division multiplexing (TDM) and Walsh-Hadamard sequence multiplexing. Similarly, orthogonal delay-scale space (ODSS) modulation\cite{k.p.2022orthogonal} is tailored for wideband time-varying channels in the delay-scale domain. To further decouple multipath components, delay-Doppler alignment modulation (DDAM)\cite{xiao2025rethinking} manipulates channel delay and Doppler spreads through delay-Doppler compensation and path-based beamforming, thereby converting time-varying channels into time-invariant, ISI-free equivalents. Unlike DDAM, interleave frequency division multiplexing (IFDM)\cite{chi2024interleave} provides an alternative waveform design approach, wherein signals experience sufficient statistical fading, enhanced by advanced detectors such as memory approximate message passing (MAMP)\cite{liu2022memory} for improved performance.

\begin{table*}[tbhp]
\caption{Existing Studies on Multicarrier Waveforms}
\vspace{-4pt}
\renewcommand\arraystretch{1.1}
\centering
\label{tab1-1}
\begin{threeparttable}
\begin{tabular}{|c|c|m{2cm}|m{8.2cm}|m{4cm}|}
\hline
\textbf{Refs} & \textbf{Year} & \textbf{Scenarios /Background}                       & \textbf{Main contributions}                                                                                                                                                                                      & \textbf{Covered waveforms}                                                                                                                            \\ \hline
\cite{banelli2014modulation}     & 2014 & DDC                          & Presented an overview of 5G alternative modulation schemes, compared the achievable spectral efficiency (ASE) of these modulation schemes in a cellular environment.                                        & OFDM, FBMC, FTN/TFS\tnote{1}, SCM\tnote{2}.                                                                                                                                \\ \hline
\cite{sahin2014survey}     & 2014 & DDC                 & Provided a unified framework for multicarrier schemes with Gabor systems by emphasizing symbols, lattice, and filters.                                                                                  & Orthogonal, biorthogonal, and non-orthogonal schemes.                                                                                                \\ \hline
\cite{farhang-boroujeny2016ofdm}     & 2016 & Multiple-Input Multiple-Output (MIMO)                                       & Established a common framework based on the OFDM principle to derive and analyze GFDM and circular FBMC (C-FBMC), facilitating efficient transceiver implementation, MIMO extension, and OOBE analysis. & OFDM, GFDM, C-FBMC, UFMC.                                                                                                                         \\ \hline
\cite{dealmeida20195g}     & 2019 & IoT scenarios                              & Presented an overview of 5G PHY waveforms to address the IoT requirements.                                                                                                                              & OFDM, FBMC, GFDM, UFMC, F-OFDM.                                                                                                                      \\ \hline
\cite{gaudio2022otfs}     & \new{2022} & \new{DDC}                & \new{Presented a fair comparison between OFDM and OTFS in terms of achievable communication rate.}                                                                                                        & \new{OFDM, OTFS.}                                                                                                                                        \\ \hline
\cite{kebede2022multicarrier}     & \new{2022} & \new{Time-dispersive channels}                   & \new{Introduced the integrated research on 5G multicarrier and multi-access technologies, followed by a comprehensive overview of waveform KPIs and applications.}                                              & \new{OFDM, FBMC, GFDM, UFMC, F-OFDM. }                                                                                                                     \\ \hline
\cite{yuan2023new}     & 2023 & DDC                 & Presented an in-depth review of OTFS in the 6G era, encompassing fundamentals, recent advancements, and future directions.                                                                                  & OFDM, OTFS.                                                                                                                                          \\ \hline
\cite{lin2023multicarrier} &\new{2023}  & \new{DDC} & \new{Provided a comprehensive overview on ODDM, and analyzed the equivalent sampled DD domain (EDSS) channels, pulse (bi)orthogonality, and transmission strategies.}  & \new{OTFS, ODDM.}  \\ \hline
\cite{dube2024overview}     & \new{2024} & \new{Underwater acoustic communication} & \new{Surveyed multicarrier waveform schemes in underwater acoustic communication and compared their performance.}                                                                                              & \new{OFDM, ODSS, ODDM, OTSM, OCDM.}     \\ \hline
\cite{li2024integrated}     & 2024 & DDC                 & Proposed a waveform design framework based on unified matrix (UM) by utilizing discrete fractional Fourier transform (DFrFT), and introduced UM-OTFS and effective detection schemes.                                                           & Candidate integrated sets: OTFS, ODDM, OFDM, GFDM, OCDM, AFDM, ODSS, OTSM. \\ \hline
\cite{rou2024orthogonal}     & 2024 & DDC                 & Provided a comparative study of AFDM and OTFS by analyzing KPIs of waveforms in the context of ISAC.                                                                                                      & OFDM, OTFS, AFDM, GFDM.                                                                                                                              \\ \hline
\cite{singh2024generalized}     & 2024 & ISAC in vehicular environments             & Proposed a waveform design framework called generalized adaptive spreading modulation (GASM) for vehicular applications.                                                                                & OFDM, FrFT-OFDM, DFT-s-OFDM, GFDM, OTFS, OCDM, GASM.                                                                                                 \\ \hline
\cite{solaija2024orthogonal}     & 2024 & DDC                 & Proposed an extensible OFDM-based waveform design framework for 6G, indicating that the different waveforms being proposed share various slices/blocks.                                                 & OFDM, OCDM, OTFS, OTSM, DFT-s-OFDM, MC-CDMA\tnote{3}, LoRa-Chirp\tnote{4}, NOMA\tnote{5}.                                                                                       \\ \hline
\cite{xiao2025rethinking}     & 2024 & DDC                 & Proposed a new waveform design scheme based on DDAM, which reduces PAPR and complexity by applying DDAM to other modulations.                                                                           & OFDM, OTFS, DDAM.                                                                                                                                    \\ \hline
\cite{yuan2024otfs}     & 2024 & DDC                 & Proposed the concept of DD domain ISAC, integrating various new waveform technologies into DD domain waveforms.                                                                                         & OTFS, ODDM, DDAM.                                                                                                                                     \\ \hline
\cite{zhou2024overview}  &\new{2024} &\new{DDC} & \new{Introduced the relationship between different modulation schemes and compared BER performance.} & \new{OFDM, OCDM, AFDM, OTFS, OTSM.} \\ \hline
\cite{ammarboudjelal2025common} & \new{2025} &\new{DDC} & \new{Proposed a unified transceiver framework based on SC-IFDM\tnote{6}, which ensures seamless waveform adaptation and orthogonal coexistence in the DFT domain.} & \new{OFDM, OTFS, OCDM, AFDM, FMCW\tnote{7}, SC-IFDM.} \\ \hline
\new{\textbf{This Work}} & \new{2025} &\new{DDC} & \new{Proposed a unified waveform framework, with existing waveforms categorised by dimension; clarified the underlying principles of each waveform from the perspective of modulation-domain ISI; discussed and compared the KPIs of various waveforms, paving the way for addressing diverse challenges.} &\new{OFDM, DFT-s-OFDM, FrFT-OFDM, OCDM, IFDM, AFDM, FBMC, OTFS, ODDM, DDAM, OTSM, ODSS.} \\ \hline
\end{tabular}
\begin{tablenotes}
        \footnotesize
        \renewcommand\arraystretch{0.4}
        \begin{tabular}{p{0.45\linewidth} p{0.5\linewidth}}
        \item[1] FTN: Faster-than-Nyquist; TFS: Time-frequency-packed signaling.&
        \item[5] NOMA: Non-orthogonal multiple access.\\
        \item[2] SCM: Single-carrier modulation.&
        \item[6] SC-IFDM: Single-carrier interleaved frequency division multiplexing.\\
        \item[3] MC-CDMA: Multicarrier code division multiple access.&
        \item[7] FMCW: Frequency modulated continuous wave.\\
        \item[4] LoRa-Chirp: Long range chirp modulation.&
        \item[]\\
        \end{tabular}
\end{tablenotes}
\vspace{-12pt}
\end{threeparttable}
\end{table*}

Consequently, numerous researchers have examined waveform frameworks, performance comparisons, and application adaptations for these waveforms, as illustrated in Table \ref{tab1-1}. In \cite{sahin2014survey}, a unified multicarrier framework based on the Gabor system is introduced, normalizing multicarrier modulation through symbols, lattices, and filters. On this basis, some researches are conducted with emphasis on OFDM and a series of multicarrier modulations based on filter banks, in which waveforms such as FBMC, GFDM, and UFMC are thoroughly investigated and summarized\cite{tao2015survey, medjahdi2017road, zhang2017study, hammoodi2019green, dealmeida20195g}. 

Regarding recent developments, two contrasting perspectives emerge: one perspective maintains that OFDM will continue to serve as a cornerstone for next-generation systems\cite{solaija2024orthogonal, ammarboudjelal2025common,dai2026tutorial}, whereas the other advocates the universality of novel waveforms, such as viewing various 2D modulation schemes as OTFS variants\cite{deng2025unifyingb}. \new{It can also be observed that shifting waveform design from the TF domain to the DD domain is an effective approach to counter ISI and ICI, as indicated in \cite{lin2023multicarrier}, which unveils that ODDM can achieve effective transmission by leveraging the orthogonality of pulses.} Particularly noteworthy is the comparative analysis of AFDM and OTFS in ISAC systems\cite{rou2024orthogonal}, exemplifying chirp-domain- and DD-domain-based waveform designs. In light of this, this paper advocates that the efficacy of specific waveforms must be contextualized within their intended application scenarios. Regardless of whether one-dimensional (1D) or 2D modulation is involved, leveraging waveform advantages and comprehending their interference-mitigation principles are crucial to empowering future applications.

Against the aforementioned background, this article proposes a comprehensive unified framework for multicarrier waveforms, systematically reviewing and summarizing state-of-the-art schemes. We focus on analyzing the effects of interference in PHY multicarrier communications and the measures of waveforms to counteract the interference, and provide an overview of the research points of waveforms from multiple application dimensions, so as to provide an open horizon for future research.

\vspace{-5pt}
\subsection{Contributions}
\label{sec1-3}
The contributions of this paper are summarized as follows:

\begin{itemize}
    \item{First, we review various state-of-the-art multicarrier waveforms applicable to next-generation wireless networks and re-examine these waveforms from a modulation-dimension perspective. In order to better illustrate the mathematical representation of the unified framework, we provide a unified representation of the channel model based on wideband doubly-dispersive channels, strengthened by a unified time-domain input-output relationship.}
    \item{Next, we introduce a unified waveform design framework by classifying them as 1D and 2D waveforms, and clarify the general principles of multicarrier waveform design. Specifically, 1D waveforms include OFDM, DFT-s-OFDM, FrFT-OFDM, OCDM, IFDM, and AFDM; while 2D waveforms include FBMC, OTFS, ODDM, DDAM, OTSM, and ODSS.}
    \item{Then, the channel-induced interference is analyzed and generalized as modulation-domain ISI (MD-ISI). We reveal that the goal of the key technologies of PHY waveform design, such as synchronization, channel estimation, and modulation-domain pulse shaping, is essentially to suppress the MD-ISI.}
    \item{\new{In addition, the communication and sensing KPIs for waveforms are discussed, including CP overhead, bit error rate (BER), PAPR, spectral efficiency, modulation complexity, ambiguity functions, and hardware and radio frequency (RF) impairments.} By comparing the KPIs of waveforms, we analyze the performance of waveforms in different aspects, which provides effective guidance for waveform selection.}
    \item{Finally, the key applications and challenges are discussed in terms of waveform KPIs, which provide a detailed guideline for the advancement of communication, sensing, and their integration. In particular, we summarize the potential of waveforms in each of these applications by reviewing their relevant influential works.}
\end{itemize}

\new{Compared to prior unified frameworks such as those based on Gabor systems, matrix forms, or DD domain modeling, the main innovations of this paper include:}

\begin{itemize}[label=\checkmark] \new
    \item \textbf{New Waveforms and New Framework:} This paper introduces a novel framework for existing 12 modulation schemes, including the new waveforms like IFDM and DDAM. 
    \item \textbf{MD-ISI Abstraction:} This paper introduces a novel abstraction of MD-ISI, which simplifies the analysis of interference across different waveforms.
    \item \textbf{New KPI Set for Performance Analysis:} A new KPI set is proposed, which includes ambiguity functions in addition to conventional metrics such as PAPR and BER.
\end{itemize}

\new
{These innovations distinguish our work from existing literature and provide a deeper, more applicable understanding of waveform classification, interference mitigation, and performance optimization in future communication systems.
}

\vspace{-12pt}
\subsection{Organization and Notation}
\label{sec1-4}
The remainder of this article is organized as follows, as shown in the Table \ref{tab1-2}. Section \ref{sec2} presents a unified framework for multicarrier modulations. Section \ref{sec3} clarifies the basics and principles of existing multicarrier schemes. Section \ref{sec4} presents the generalized modulation-domain interference analysis of the waveforms, followed by Section \ref{sec5} presents the KPI analysis of the waveforms. In Section \ref{sec6}, advanced applications are provided. Finally, Section \ref{sec7} provides the open challenges and future directions, and Section \ref{sec8} concludes the paper.

$Notations$: The following notations will be followed in this paper: $a,\mathbf{a}, \mathbf{A}$ represent a scalar, vector, and matrix, respectively. $(\cdot)^T$ is the transpose operator, $(\cdot)^H$ is the Hermitian transpose operation, $\otimes$ is the Kronecker product, $\delta$ is the Kronecker delta function,  $\operatorname{vec}(\mathbf{A})$ is the column-wise vectorization of the matrix $\mathbf{A}, \operatorname{vec}_{M, N}^{-1}(\mathbf{r})$ is the matrix formed by folding a vector $\mathbf{r}$ into an $M \times N$ matrix by filling it column wise, and $\mathbf{I}$ denotes the identity matrix.

\begin{table}[tbph]
\caption{Outline of the paper}
\renewcommand\arraystretch{1}
\centering
\label{tab1-2}
\begin{tabular}{|c|}
\hline
\textbf{Section I. Introduction}                                                                                                                                                                                                         \\ \hline
\begin{tabular}[c]{@{}ll@{}}A. Background  &B. Related Works\\C. Contributions  &D. Organization and Notation\end{tabular}                                                                                                                 \\ \hline \hline
\textbf{Section II. A Unified Framework for Multicarrier Modulation}                                                                                                                                                                     \\ \hline
\begin{tabular}[c]{@{}l@{}}A. Channel Model\quad\quad B. Unified Signal Representation\\ C. Unified Time-Domain Input-Output Relationship\end{tabular}                                                                                             \\ \hline \hline
\textbf{Section III. Basics and Principles of Existing Multicarrier Waveforms}                                                                                                                                                                    \\ \hline
\begin{tabular}[c]{@{}l@{}}A. 1D Modulation Waveforms\\ B. 2D Modulation Waveforms\end{tabular}                                                                                                                                          \\ \hline \hline
\textbf{Section IV. Generalized Modulation-domain Interference Analysis}                                                                                                                                                                 \\ \hline
\begin{tabular}[c]{@{}l@{}}A. Modulation-domain ISI\\ B. Key Solutions against Modulation-domain ISI\end{tabular}                                                                                                                        \\ \hline \hline
\textbf{Section V. Comparative KPI-based Analysis}                                                                                                                                                                                       \\ \hline
\begin{tabular}[c]{@{}ll@{}}A. CP Overhead &B. Bit Error Rate\\ C. Peak-to-Average Power Ratio &D. Spectral Efficiency\\ E. Modulation Complexity &F. Ambiguity Functions\\\new{G. Hardware and RF Impairments }& \new{H. Summary and Practical Insights}\end{tabular}                                                                                           \\ \hline \hline
\textbf{Section VI. Advanced Applications of Multicarrier Waveforms}                                                                                                                                                                     \\ \hline
\begin{tabular}[c]{@{}l@{}}A. Multiple Access\quad\quad B. MIMO\quad\quad C. Full Duplex\\D. Security and Privacy\quad\quad E. Index Modulation\\ F. Integrated Sensing and Communications \\ G. Integrated Localization and Communications \end{tabular} \\ \hline \hline
\textbf{Section VII. Open Challenges and Future Directions}                                                                                                                                                                              \\ \hline
\begin{tabular}[c]{@{}l@{}}A. Theoretical Limits\quad\quad B. Application Challenges\\ C. Implementation Challenges\end{tabular}                                                                                                                   \\ \hline \hline
\textbf{Section VIII. Conclusions}                                                                                                                                                                                                       \\ \hline
\end{tabular}
\vspace{-14pt}
\end{table}

\section{A Unified Framework for Multicarrier Modulation}
\label{sec2}

\subsection{Channel Model}
\label{sec2-1}
We consider a communication system with
bandwidth $B$, carrier frequency $f_{c}$ and total duration $T$, i.e., signal time $t \in [0, T]$ and frequency $f\in [0, B]$. Next, we discuss several commonly used channel models\cite{hlawatsch2011wireless,bello1963characterization}.

\subsubsection{Wideband doubly-dispersive channels}
Wireless channels typically exhibit dispersion in both time and frequency domains in general. Specifically, high data rates and multipath propagation give rise to time dispersion (i.e., frequency-selectivity) of wireless channels, where the transmitted signals arrive at the receiver via different paths with varying delays, resulting in ISI. Meanwhile, carrier frequency offset (CFO) and Doppler spread caused by relative motion between the transmitter, receiver, or objects in the environment introduce frequency dispersion (i.e., time-selectivity) of wireless channels, leading to time-varying fading and ICI, especially problematic in multicarrier systems.

In wideband communication systems, where the transmission bandwidth $B$ is non-negligible compared to the carrier frequency $f_{c}$. Different frequency components within the wideband signal experience different Doppler shifts, leading to a non-uniform Doppler spread with time-scaling effects. Specifically, the wideband time-delay channel representation $h(t, \tau)$ can be given by 
\begin{equation}\new
h(t, \tau)=\sum_{i=1}^{P} h_{i}(t) \delta(\tau-\underbrace{\left(\tau_{i}-\alpha_{i} t\right)}_{\tau_{i}(t)}) ,
\label{eqyhr0413.1}
\end{equation}
where $P$ denotes the number of propagation paths. The parameters \new{$h_{i}(t)$}, $\tau_{i}\in [0, \tau_{\max}]$ and $\alpha_{i}\in [-\alpha_{\max},\alpha_{\max}]$ represent the complex channel gain, delay and Doppler scale factor of the $i$-th path, respectively. $\tau_{\max}$ and $\alpha_{\max}$ are the corresponding maximum channel delay and maximum Doppler scale factor. \new{Here, we assume that the time variation of $h_{i}(t)$ is mainly caused by the Doppler shift as $\nu_i = \alpha_{i} f_{c}$ and the channel tap coefficient $h_i$ remains constant during the stationary time of the channel, i.e., $h_{i}(t) = h_{i} e^{j 2 \pi \alpha_{i} f_{c} t}$.} For simplicity, assume that $h_{i}$ is time-independent\cite{xiao2025integrated}. Note that the equivalent delay $\tau_{i}(t)=\tau_{i}-\alpha_{i}t$ is time-dependent, resulting in a significant time-scaling effect. The Doppler scale factor ${\alpha _i} = {\upsilon _i}/ c$, where ${{\upsilon _i}}$ is the relative velocity of the $i$-th path between the transmitter and receiver, and $c$ is the propagation medium speed. 

To further investigate the time-scaling characteristics in wideband systems, we convert the time-delay domain channel $h(t, \tau)$ into its TF domain representation $H(t,f)$ as follows:
\begin{align}
    H(t,f) &= \int_\tau  {h(t,\tau ){e^{ - j2\pi f\tau }}\text{d}\tau } \nonumber\\
    &\new{ = \sum\limits_{i = 1}^P {{h_i(t)}\int_\tau  {\delta [\tau  - ({\tau _i} - {\alpha _i}t)]{e^{ - j2\pi f\tau }}\text{d}\tau } } \nonumber}\\
    & = \sum\limits_{i = 1}^P {{h_i}\int_\tau  {\delta [\tau  - ({\tau _i} - {\alpha _i}t)]{e^{j2\pi {\nu _i}t}}{e^{ - j2\pi f\tau }}\text{d}\tau } } \nonumber\\
    & = \sum\limits_{i = 1}^P {{h_i}{e^{ - j2\pi f{\tau _i}}}{e^{j2\pi ({\nu _i} + {\alpha _i}f)t}}}. \label{Channel_TF}
\end{align}
It is important to note that in wideband systems, the equivalent Doppler frequency shift, given by ${\nu _i}(f) = {\nu _i} + {\alpha _i}f$, is both non-uniform and frequency-dependent. This phenomenon is also known as the Doppler squint effect (DSE) in wideband transmissions \cite{duan2025channela,wang2023doppler,li2026affine}. Wideband DDC commonly arises in underwater acoustic (UWA), low earth orbit (LEO) satellite, and ultra-wideband (UWB) radio communications. Most of the existing multicarrier techniques tend to perform poorly in doubly-dispersive wideband channels, where the Doppler effect manifests as a time-scaling distortion in the received waveform. Therefore, advanced modulation and signal processing techniques must be explored to effectively suppress the impact of time-scaled wideband DDC.

{\footnotesize
\renewcommand{\arraystretch}{1.1}
\setlength{\tabcolsep}{4.5pt}
\begin{table*}[t]\new
\caption{Representative 5G/6G Use Cases and Their Channel-model Implications}
\label{tab:usecases}
\centering
\begin{tabular}{m{4cm} m{2.2cm} m{3.8cm} m{4cm} m{2.2cm}}
\toprule
\textbf{Use Case} & \textbf{Typical Band} & \textbf{Mobility} & \textbf{Channel Characteristics} & \textbf{Model} \\
\midrule
Urban eMBB (FR1, sub-6) & 3.5\,GHz, 100\,MHz & 0–60\,km/h; low–mid Doppler & Frequency-selective, rich scattering & TDC \\
mmWave eMBB (FR2) & 26–39\,GHz, 400\,MHz & 0–60\,km/h; beam tracking & Sparse clusters, short delay spread; blockage prone & FDC \\
URLLC factory (industrial IoT) & 3.5–6\,GHz & Pedestrian; very low Doppler & Metallic multipath; frequency-selective & TDC \\
High-speed rail & 3.5/4.9\,GHz & 200–350\,km/h; high Doppler & Doubly selective; time variation dominates & Narrowband DDC \\
V2X highway (mmWave sidelink) & 28/60\,GHz & Relative speeds 200+\,km/h; very high Doppler & Sparse line of sight (LoS) + specular; blockage; fast time selectivity & Narrowband DDC\\
LEO NTN downlink & S/Ka bands & High Doppler, drift; long delays & Strong DD selectivity; large delay/Doppler spreads & Wideband DDC \\
UAV U2X (air-to-ground) & 2–7\,GHz/mmWave & 10–120\,km/h; Rician LoS & Mix of LoS and specular; geometry-driven & Narrowband DDC \\
ISAC automotive (77\,GHz) & 76–81\,GHz & Relative speed up to 140\,km/h & Short delay spread; sensing metrics matter & Narrowband DDC \\
Sub-THz backhaul & 100–140\,GHz & Static/slow & LoS/non-line of sight (NLoS) sparse; phase noise/CFO critical & FDC \\
UWA & 10–100\,kHz & 0–5\,m/s; low speed but strong Doppler spread (sound propagation) & Very long delay spread; temperature/salinity dependent & Wideband DDC \\
UWB  & 3.1–10.6\,GHz & Static/low mobility & Extremely short pulses; dense multipath but small Doppler & Wideband DDC \\
Maritime links & 2–6\,GHz & Low–mid; sea motion & Two-ray effects; long delays & TDC \\
\bottomrule
\end{tabular}
\end{table*}
}

\subsubsection{Narrowband doubly-dispersive channels}
Since most terrestrial mobile communications experience time-varying narrowband channels, the time contractions or dilations caused by the Doppler effect can be well approximated by uniform Doppler frequency shifts.
Actually, the wideband DDC can be simplified to narrowband if the communication systems satisfy the following two conditions: (1) The transmission bandwidth $B$ is small relative to the carrier frequency $f_{c}$, i.e., $\frac{B}{f_{c}} \ll  1$; (2) The relative position of the transmitter and receiver does
not change significantly relative to the positional resolution of the transmitted signal, i.e., ${\alpha _i} = {{{\upsilon _i}} \mathord{\left/
 {\vphantom {{{\upsilon _i}} c}} \right.
 \kern-\nulldelimiterspace} c} \ll {1 \mathord{\left/
 {\vphantom {1 {BT}}} \right.
 \kern-\nulldelimiterspace} {BT}},i = 1,2, \cdots ,P$. \new{Under the above two conditions, the time-variation of $\tau_i(t)$ can be neglected during the channel stationary time. Similar to the wideband DDC case, the time variation of $h_{i}(t)$ is mainly caused by the Doppler shift $\nu_i$, i.e., $h_{i}(t) = h_{i} e^{j 2 \pi \nu_i t}$. Here, for the equivalent Doppler frequency shift $\nu_i(f) = {\nu _i} + {\alpha _i}f$, the narrowband condition ensures the frequency-dependent term ${\alpha _i}f$ is much smaller than the constant Doppler shift $\nu_i$ (i.e., ${\alpha _i}f\ll\nu_i$), and the extra term ${\alpha _i}f$ in the exponential case be neglected.} Therefore, the wideband TF response $H(t,f)$ in (\ref{Channel_TF}) can be simplified as
\begin{align}
    H(t,f) \new{\approx \sum\limits_{i = 1}^P {{h_i(t)}{e^{ - j2\pi f{\tau _i}}}} }\approx \sum\limits_{i = 1}^P {{h_i}{e^{ - j2\pi f{\tau _i}}}{e^{j2\pi {\nu _i}t}}},
\end{align}
 and its corresponding time-delay domain representation ${h(t,\tau )}$ in (\ref{eqyhr0413.1}) can be reduced to
\begin{equation}
h(t, \tau) \new{\approx\sum_{i=1}^{P} h_{i}(t) \delta(\tau-\tau_{i}) } \approx\sum_{i=1}^{P} h_{i} \delta(\tau-\tau_{i}) e^{j 2 \pi \nu_{i} t},
\label{eqyhr0414.1}
\end{equation}
\new{by replacing $\tau_i(t)$ with $\tau_i$.}

Although 6G systems usually involve broadband transmission, narrowband DDC modeling is still applicable in specific scenarios (e.g., low to medium speed mobility or limited bandwidth allocation)\cite{zhang2022general}. This is compatible with two representative classes of recently proposed novel waveforms, namely AFDM and OTFS, which are modulated in the chirp domain and DD domain, respectively. As strong candidates for next-generation communication networks\cite{rou2024orthogonal}, these two waveforms are expected to address the Doppler effect in narrowband DDC and achieve high-speed and reliable transmission.

\subsubsection{Time-dispersive channels}
Some low-mobility or stationary wireless communication systems exhibit only multipath effect without Doppler shift, such as indoor wireless local area networks (WLANs) and industrial Internet of Things (IoT) 4.0. These channels are often modeled as time-dispersive fading channels with the time-delay-domain representation given by
\begin{equation}
h(t, \tau)\new{=\sum_{i=1}^{P} h_{i}(t) \delta(\tau-\tau_{i}) }=\sum_{i=1}^{P} h_{i} \delta(\tau-\tau_{i}),
\label{eqyhr0414.3}
\end{equation}
which is irrelevant to time $t$. \new{Here, we assume that the time-variation of  $\tau_i(t)$ can be neglected during the channel stationary time, and the time-variation of  $\nu_i(t)$ can be further neglected during the coherence time, i.e., $\tau_i(t)=\tau_i$, $h_i(t)=h_i$.} Its corresponding TF domain response can be expressed as:
\begin{equation}
 H(t,f) \new{= \sum\limits_{i = 1}^P {h_i(t)}{e^{ - j2\pi f{\tau _i}}} }= \sum\limits_{i = 1}^P {h_i}{e^{ - j2\pi f{\tau _i}}}.
\end{equation}

The time-dispersive channels (TDC) have a smaller coherent bandwidth and a larger delay extension. By leveraging the orthogonality of subcarriers, OFDM serves as the primary waveform for TDC, widely adopted in 4G long-term evolution (LTE), 5G new radio (NR), and Wireless Fidelity (Wi-Fi) systems.

\subsubsection{Frequency-dispersive channels}
In some high-mobility NTN, such as unmanned aerial vehicle (UAV) and satellite communications, wireless channels may exhibit strong Doppler shift but little or no multipath. Such channels can be modeled as frequency-dispersive channels (FDC) with a time-delay domain representation given by
\begin{equation}
h(t, \tau) \new{= \sum_{i=1}^{P} h_{i}(t) \delta(\tau) }=\sum_{i=1}^{P} h_{i} \delta(\tau) e^{j 2 \pi \nu_{i} t}.
\label{eqyhr0414.5}
\end{equation}
\new{Without loss of generality, perfect time synchronization is assumed, such that the LoS propagation delay is compensated. Therefore, $\tau_i(t)$ corresponds to the relative delay with respect to the first-arriving path. Here, we assume that the time variation of $h_{i}(t)$ is mainly caused by the Doppler shift $\nu_i$, i.e., $h_{i}(t) = h_{i} e^{j 2 \pi \nu_i t}$, and the time-variation of $\tau_i(t)$ can be totally neglected during the channel stationary time with very few multipaths, i.e., $\tau_i(t)\approx 0$.}
Similarly, the corresponding TF domain response can be expressed as:
\begin{equation}
H(t,f) \new{ =\sum_{i=1}^{P} {h_i(t)} }=\sum_{i=1}^{P} {h_i}{e^{j2\pi \nu_{i} t}}.
\end{equation}

Note that FDCs are characterized by rapid changes in the channel impulse response over time, which leads to signal fading in the time domain. Conventional OFDM is weakly resistant to frequency-dispersive; therefore, the use of other domain modulations tends to be more prevalent, like OCDM or FrFT-OFDM. 

\new{To ground the unified framework in practice, Table~\ref{tab:usecases} summarizes representative 5G/6G use cases and maps them to typical operating bands, mobility ranges, channel selectivity, and modeling choices.}

\vspace{-5pt}
\subsection{Unified Signal Representation}
\label{sec2-2}
\begin{figure*}[tp]
    \centering
    \includegraphics[width=\textwidth]{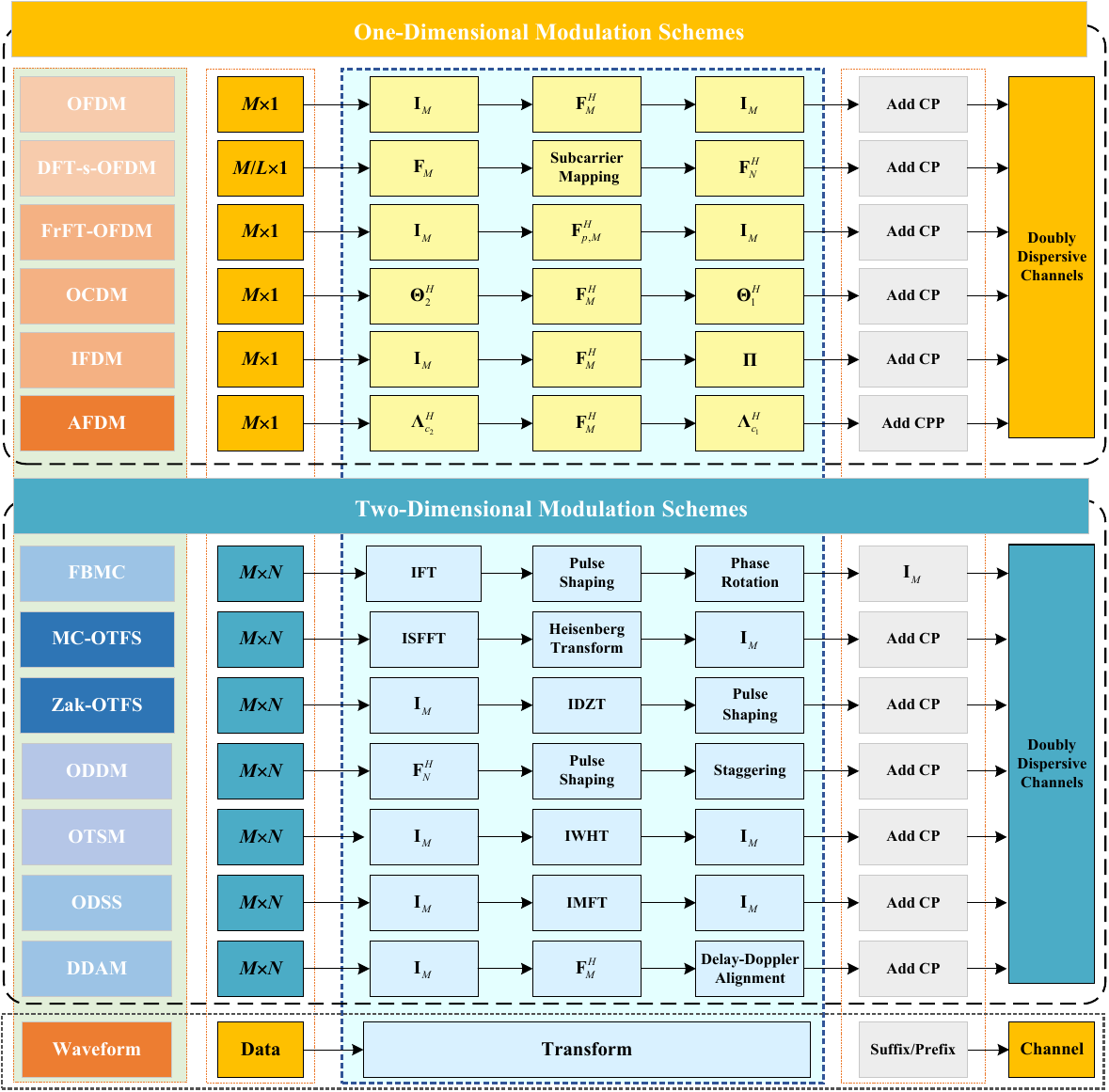}
    \caption{System diagram for the unified multicarrier waveform framework.}
    \label{fig2-1}
    \vspace{-8pt}
\end{figure*}

In this subsection, a unified signal representation is proposed to accommodate both 1D and 2D waveforms by selecting the matrix parameter and setting the basis function, as illustrated in Fig. \ref{fig2-1}.
The core of the unified representation lies in a parameterized signal model, where the transmitted signal is expressed as 
\begin{equation}
    \mathbf{s}=\mathbf{Ux}.
    \label{eq2-2-1}
\end{equation}

Here, $\mathbf{x}\in \mathbb{C}^{M \times N}$ denotes the information-bearing symbol matrix of the size $M\times N$, $\mathbf{s}\in \mathbb{C}^{M \times N}$ denotes \new{the discrete-time baseband signal}, and $\mathbf{U}\in \mathbb{C}^{M \times M}$ is a unitary operator matrix determined by specific modulation techniques. This framework unifies 1D and 2D waveforms by allowing seamless transitions between modulation domains through specific operational adjustments. For instance, setting $N=1$ reduces the model to a 1D structure suitable for OFDM or OCDM, while increasing $N (N > 1)$ and selecting a delay-Doppler basis adapts it to OTFS or ODDM. This classification not only reflects the structural differences in signal design but also highlights their suitability for different channel characteristics. For instance, 1D waveforms typically operate in a single domain (e.g., frequency or chirp), while 2D waveforms exploit joint TF or delay-Doppler representations, both offering enhanced resilience to delay-Doppler spreads in different ways.

\new{In practical wireless transceivers, the transmitted signal originates from a discrete-time symbol matrix but ultimately propagates in continuous time as a passband waveform. The continuous-time baseband signal generated by a transmit pulse shaping filter $g(t)$ can be denoted as}
\begin{equation}\new
    s(t) = \sum_{k} s[k]\; g(t-kT_s),
    \label{eq:cts_baseband}
\end{equation}
\new{where $T_s$ represents the time spacing of the basis functions, and $s[k]$ is the $k$-th element of the vectorized discrete-time modulated signal $\mathbf{\tilde{s}} \in \mathbb{C}^{MN \times 1}$, i.e., $\mathbf{\tilde{s}} =\operatorname{vec}(\mathbf{s})$. The function $g(t)$ determines the temporal support and spectral compactness of the waveform. In this work, unless otherwise specified, rectangular pulse shaping is used so that the intrinsic characteristics of different waveforms can be compared on a common basis.}

\new{The RF passband signal actually radiated by the transmitter is obtained through standard quadrature modulation as}
\begin{equation}\new
    s_{\mathrm{RF}}(t)=\Re\!\left\{ s(t)\,e^{j2\pi f_c t}\right\},
\end{equation}
\new{where the multiplication by $e^{j2\pi f_c t}$ represents the upconversion of the complex baseband waveform to the carrier frequency $f_c$. The up-converted signal is then passed through a bandpass filter (BPF) enforcing the allocated spectrum mask. At the receiver, the RF signal undergoes symmetric down-conversion and low-pass filtering (LPF), producing the received complex baseband waveform used for demodulation. These modules will affect the input-output relationship of the signal\cite{lin2023multicarrier,liu2025affine}. Since BPF is a linear time-invariant (LTI) operator, its effect can be equivalently absorbed into the overall end-to-end channel response. Therefore, the proposed unified framework in \eqref{eq2-2-1} is constructed upon a discrete-time baseband equivalent model with $s(t)$. This choice isolates waveform-intrinsic properties from RF implementation details for simplicity.} 

\begin{figure*}[thbp]
    \centering
    \includegraphics[width=\textwidth]{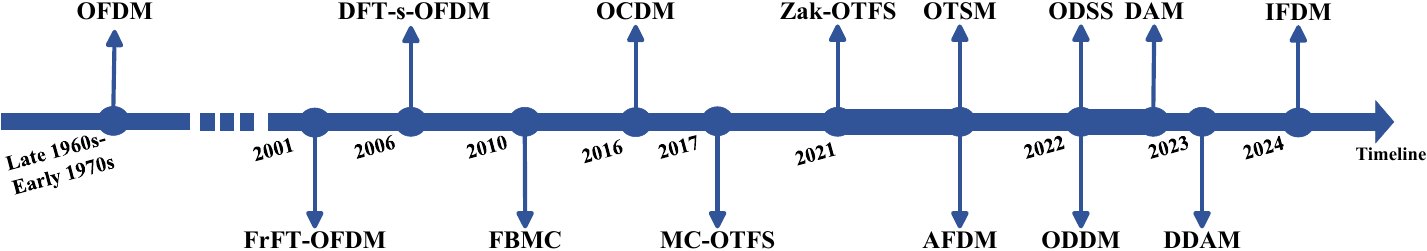}
    \caption{Timeline of existing multicarrier modulation schemes.}
    \label{fig2-2}
    \vspace{-6pt}
\end{figure*}

Fig. \ref{fig2-2} shows a timeline of the development of existing waveforms highlighted in this paper, and it can be seen that many new waveforms have been emerging in the last five years. As mentioned before, we divide them into 1D waveforms and 2D waveforms. Next, we will explain each type of waveform in detail.

\subsubsection{1D Modulation Waveforms}
1D modulation waveforms refer to schemes where information symbols are mapped onto $M\times 1$ vectors. This structure enables 1D waveforms to achieve high orthogonality and spectral efficiency with relatively low computational complexity. 

For instance, OFDM employs an inverse fast Fourier transform (IFFT) to map symbols into orthogonal subcarriers, offering robust performance in multi-path environments with minimal receiver complexity. Based on this, DFT-s-OFDM extends OFDM by using a precoded DFT, making its CP length flexible and adjustable. Similarly, FrFT-OFDM leverages the fractional Fourier transform in fractional domains for non-stationary channels, while OCDM utilizes linear chirp signals to maintain orthogonality, enhancing resilience to time-varying channels. The most typical 1D waveform that utilizes chirp domain modulation is the AFDM, which employs the DAFT to introduce a modulation scheme with two manageable chirp parameters, effectively combining backward compatibility and diversity capability. In addition, chirp-convolved data transmission (CCDT)\cite{berggren2021chirpconvolved} also multiplexes the chirp basis with three parameters and simplifies the waveform generation, which can equivalently be represented as DFT-s-OFDM with a frequency domain chirp filter. Another 1D modulation scheme uses a hybrid carrier approach, such as the hybrid carrier scheme based on weighted fractional Fourier transform\cite{song2024robust} or weighted affine Fourier transform\cite{li2024hybrid}. In addition, IFDM uses a 1D randomly interleaved inverse Fourier transform for modulation, which can enhance the statistical stability of the channel, enabling the equivalent channel matrix to satisfy the right-unitarily invariant assumption for MAMP detectors\cite{chi2024interleave}.

\subsubsection{2D Modulation Waveforms}
2D modulation waveforms extend the signal design to an $M\times N$ matrix, where information symbols are modulated across two resource dimensions. 

FBMC is a typical example of 2D modulation in the TF domain. The main principle of FBMC is to modulate offset quadrature amplitude modulation (OQAM) symbols across a set of subcarriers, each represented by a combination of a TF localized filter and a phase offset, with the real and imaginary parts of the information symbols separated to ensure orthogonality. Unlike the TF domain, the delay-Doppler domain waveforms map information symbols into a 2D grid of delay and Doppler shifts, which provides a more sparse and intuitive characterization of time-varying channels. A prominent example is OTFS, which modulates data in the delay-Doppler domain using an $M\times N$ symbol matrix $\mathbf{s}$. Currently, OTFS can be classified into two types according to the realization, one is the multicarrier version of OTFS (MC-OTFS), and the other is Zak-transform based OTFS (Zak-OTFS), which are implemented through the ISFFT and the Zak transform (ZT), respectively. Similar to OTFS, ODDM places information symbols on a 2D delay-Doppler grid to combat the effects of DDC. However, ODDM differs from OTFS in that it introduces a $T/m$ interval staggering, which can be regarded as a form of staggered multitone (SMT) modulation. \new{Additionally, their considerations regarding pulse shaping also differ. OTFS is based on TF domain pulses\cite{deng2025unifyingb}, while ODDM modulation is based on orthogonally realizable pulses in the DD plane\cite{lin2022orthogonal}, which means ODDM creatively shifts the pulse shaping domain from TF domain to DD domain.} \new{In \cite{bai2026orthogonal}, orthogonal chirp delay-Doppler division multiplexing (CDDM) modulation is proposed, which is formulated via the chirp-Zak transform (CZT) in the DD domain.} There are also waveforms that modulate signals in other 2D domains besides the DD domain, and in this paper, they are uniformly classified as 2D modulation waveforms, of which the representative ones are OTSM for the delay-sequence domain, ODSS for the delay-scale domain, and frequency-Doppler division multiplexing (FDDM)\cite{gong2024frequencydoppler} for the frequency-Doppler domain. In addition, DDAM achieves channel path separation in the spatial domain by actively compensating for delay and Doppler components, thus significantly mitigating delay-Doppler channel spreads.

\vspace{-5pt}
\subsection{Unified Time-Domain Input-Output Relationship}
\label{sec2-3}

\new{Let $s(t)$ represent the transmitted time-domain complex baseband equivalent signal. Then, the received time-domain complex baseband equivalent signal can be denoted as} 
\begin{equation}
r(t)=\int_{-\infty}^{\infty} h(t, \tau) s(t-\tau) \text{d} \tau + w(t),
\label{eqyhr0413.2}
\end{equation}
where $w(t)$ is the additive white Gaussian noise (AWGN). The expression \eqref{eqyhr0413.2} can be equivalently represented by
\begin{equation}
\mathbf{r}=\mathbf{Hs}+\mathbf{w}.
\label{eq2-3-1}
\end{equation}
Here, $\mathbf{r} \in \mathbb{C}^{M \times N}$ denotes the time-domain signal after passing through the channel, $\mathbf{H} \in \mathbb{C}^{M \times M}$ denotes the equivalent channel matrix, and $\mathbf{w} \in \mathbb{C}^{M \times N}$ denotes the AWGN signal, respectively.
Therefore, we can obtain the input-output relationship under different channel models by substituting different $h(t, \tau)$ into \eqref{eqyhr0413.2}.

Substituting (\ref{eqyhr0413.1}) into (\ref{eqyhr0413.2}) under the wideband DDC, we have
\begin{align}
r(t)&=\sum_{i=1}^{P} h_{i} \int_{-\infty}^{\infty}\delta\left(\tau-\left(\tau_{i}-\alpha_{i} t\right)\right) e^{j 2 \pi \nu_{i} t} s(t-\tau) \text{d} \notag \tau\\
&=\sum_{i=1}^{P} h_{i} s\left(t-\left(\tau_{i}-\alpha_{i} t\right)\right) e^{j 2 \pi \nu_{i} t}.
\label{eqyhr0413.3}
\end{align}

Substituting (\ref{eqyhr0414.1}) into (\ref{eqyhr0413.2}) under the narrowband DDC, we have the received noise-free time-domain signal as
\begin{align}
r(t)&=\sum_{i=1}^{P} h_{i} \int_{-\infty}^{\infty}\delta\left(\tau-\tau_{i}\right) e^{j 2 \pi \nu_{i}  t} s(t-\tau) \text{d} \notag \tau\\
&=\sum_{i=1}^{P} h_{i} s\left(t-\tau_{i}\right) e^{j 2 \pi \nu_{i} t},
\label{eqyhr0414.2}
\end{align}
which is the superposition of $P$ delay-Doppler-shifted replicas of the transmitted signal. Note that (\ref{eqyhr0414.1}) and (\ref{eqyhr0414.2}) can be obtained by letting $\tau_{i}(t)\approx  \tau_{i}, (i=1,\dots,P)$ in (\ref{eqyhr0413.1}) and (\ref{eqyhr0413.3}), respectively.

Substituting (\ref{eqyhr0414.3}) into (\ref{eqyhr0413.2}) under TDC, we have the received noise-free time-domain signal as
\begin{align}
r(t)&=\sum_{i=1}^{P} h_{i} \int_{-\infty}^{\infty}\delta\left(\tau-\tau_{i}\right) s(t-\tau) \text{d} \notag \tau\\
&=\sum_{i=1}^{P} h_{i} s\left(t-\tau_{i}\right) ,
\label{eqyhr0414.4}
\end{align}
which is the superposition of $P$ delay-shifted replicas of the transmitted signal. Note that (\ref{eqyhr0414.3}) and (\ref{eqyhr0414.4}) can be obtained by letting $\nu_{i}=0, (i=1,\dots,P)$ in (\ref{eqyhr0414.1}) and (\ref{eqyhr0414.2}), respectively.

Substituting (\ref{eqyhr0414.5}) into (\ref{eqyhr0413.2}) under FDC, we have the received noise-free time-domain signal as
\begin{align}
r(t)&=\sum_{i=1}^{P} h_{i} \int_{-\infty}^{\infty}\delta\left(\tau\right) e^{j 2 \pi \nu_{i}  t} s(t-\tau) \text{d} \notag \tau\\
&= \sum_{i=1}^{P} h_{i} s\left(t\right) e^{j 2 \pi \nu_{i} t}.
\label{eqyhr0414.6}
\end{align}
Similarly, (\ref{eqyhr0414.5}) and (\ref{eqyhr0414.6}) can be obtained by letting $\tau_i=0, (i=1,\dots,P)$ in (\ref{eqyhr0414.1}) and (\ref{eqyhr0414.2}), respectively.

Finally, a unified representation of the time-domain input-output relationship can be summarized as shown in Table \ref{tab2-1}.

\begin{table*}[tbhp]
    \caption{Time-domain input-output relationship for different channel conditions}
    \renewcommand\arraystretch{1.2}
    \centering
\begin{tabular}{|m{2.8cm}|m{8.2cm}|m{4.9cm}|}
\hline
\textbf{Channel types}                & \textbf{Input-output relationship in time domain}                                                                               & \textbf{Examples of channel models\cite{kumar2024review}}                                                                        \\ \hline
Wideband doubly-dispersive channels   & $r(t)=\sum_{i=1}^{P} h_{i} s\left(t-\left(\tau_{i}-\alpha_{i} t\right)\right) e^{j 2 \pi \nu_{i} t} +w(t)$                   & \multirow{4}{*}{\begin{tabular}[c]{@{}l@{}}$\bullet$ Extended pedestrian A model (EPA): \\$P=7, \tau_{\max}=410ns, \nu_{\max} =5Hz$;\\ $\bullet$ Extended vehicular A model (EVA): \\$P=9, \tau_{\max}=2510ns, \nu_{\max} =70Hz$; \\ $\bullet$ Extended typical urban model (ETU): \\$P=9, \tau_{\max}=5000ns, \nu_{\max} =300Hz$.\end{tabular}} \\ \cline{1-2}
Narrowband doubly-dispersive channels & $r(t)=\sum_{i=1}^{P} h_{i} s\left(t-\tau_{i}\right) e^{j 2 \pi \nu_{i} t} +w(t), \tau_{i}(t)\approx \tau_{i}, (i=1,\dots,P)$ &                                                                                                            \\ \cline{1-2}
Time-dispersive channels              & $r(t)=\sum_{i=1}^{P} h_{i} s\left(t-\tau_{i}\right) +w(t), \nu_{i}=0, (i=1,\dots,P)$                                         &                                                                                                            \\ \cline{1-2}
Frequency-dispersive channels         & $r(t)= \sum_{i=1}^{P} h_{i} s\left(t\right) e^{j 2 \pi \nu_{i} t} +w(t), \tau_{i}=0, (i=1,\dots,P)$                                                              &                                                                                                            \\ \hline
\end{tabular}
\label{tab2-1}
\vspace{-6pt}
\end{table*}

\begin{figure*}[bhp]
    \centering
    \includegraphics[width=0.96\textwidth]{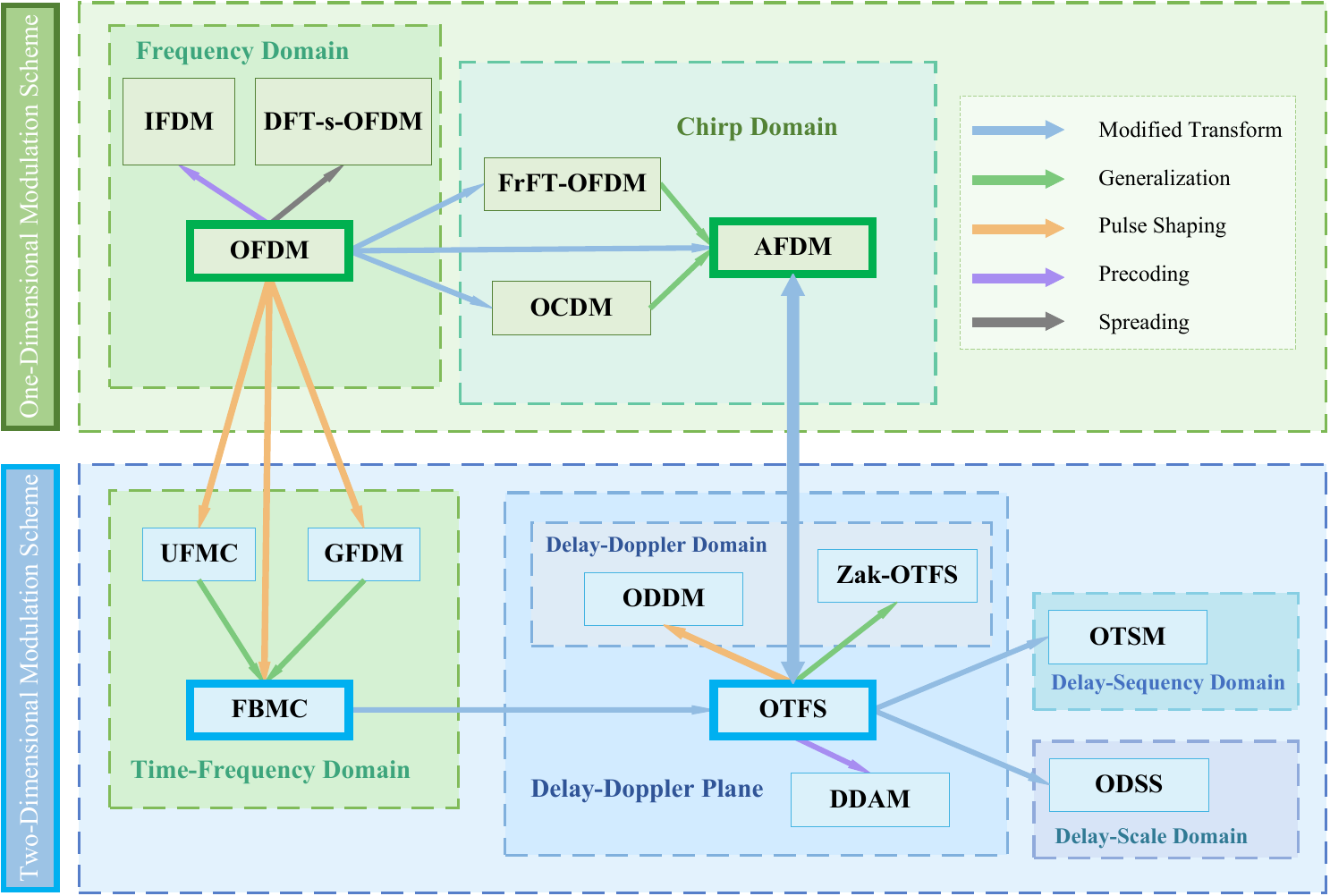}
    \caption{\color{b}Relationships between different waveforms and their signal domains.}
    \label{fig3-1}
\end{figure*}

\vspace{-5pt}
\section{Basics and Principles of Existing Multicarrier Waveforms}
\label{sec3}
Following the classification criteria of 1D waveforms and 2D waveforms introduced in Section \ref{sec2-2}, we proceed to give details of the two types of waveforms in this section to sort out the basics and principles, each of which considers both the transmitter and receiver processes.
Fig. \ref{fig3-1} illustrates the relationship between the two types of multicarrier schemes in the signal domain. The 1D waveforms progress from OFDM to AFDM, and the 2D waveforms progress from FBMC to OTFS, presenting a scattering-like correlation to the other modulation schemes. There are also close connections between different waveforms, which are marked with arrows of different colors. \new{It is worth noting that the bidirectional arrow between AFDM and OTFS in Fig.~\ref{fig3-1} indicates their bijective relation\cite{yinCDDS}. Specifically, OTFS capitalizes on the sparse representation of the channel in the DD domain, enabling the distinct resolution of multipath components in terms of delay and Doppler shifts. This property can be equivalently realized in AFDM, whose equivalent channel matrix exhibits a one-to-one mapping with the DD domain, allowing similar channel decoupling in the DAFT domain.}

\vspace{-5pt}
\subsection{1D Modulation Waveforms}
\label{sec3-1}
\subsubsection{OFDM}

\begin{figure}[H]
	\centering
\includegraphics[width=0.36\textwidth]{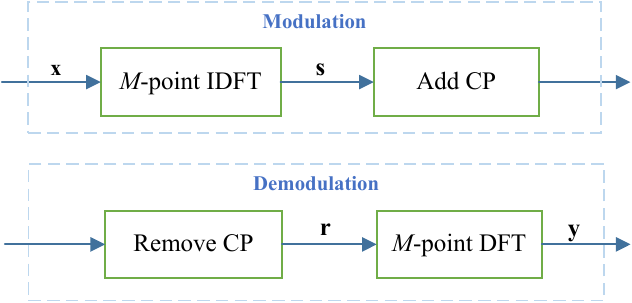}
	\caption{Diagram of OFDM.}
	\label{block1OFDM}
    \vspace{-8pt}
\end{figure}

OFDM leverages the orthogonality of subcarriers to multiplex information symbols across the frequency domain, converting a wideband frequency-selective fading channel into a set of parallel narrowband flat-fading subchannels\cite{farhang-boroujeny2016ofdm,shahriar2015phylayer}. As shown in Fig. \ref{block1OFDM}, let $\mathbf{x}^{\text{OFDM}} \in \mathbb{C}^{M \times 1}$ denote a vector of quadrature amplitude modulation (QAM) symbols, then the $M$-points inverse discrete Fourier transform (IDFT) operation is performed to map $\mathbf{x}^{\text{OFDM}}$ to the time domain as

\begin{equation}
    s^{\text{OFDM}}[l]=\frac{1}{\sqrt{M}} \sum_{k=0}^{M-1} x^{\text{OFDM}}[k] e^{j 2 \pi \frac{k l}{M}},
    \label{eqOFDM-1}
\end{equation}
or, equivalently, the matrix form is given by

\begin{equation}
    \mathbf{s}^{\text{OFDM}}=\mathbf{F}_M^{H} \mathbf{x}^{\text{OFDM}},
    \label{eqOFDM-2}
\end{equation}
where $\mathbf{F}_M \in \mathbb{C}^{M \times M}$ is the DFT matrix with entries $e^{-j2 \pi kl/M }/\sqrt{M}$. Then, a CP of length $L_{\text{cp}}$ is added, noted that $L_{\text{cp}}$ should be any integer greater than or equal to the value in samples of the maximum delay spread of the channel. Thus, the transmitted signal can be written as
\begin{equation}
\mathbf{s}^{\text{OFDM}}_{\text{cp}}=\left[s^{\text{OFDM}}\left[M-L_{\text{cp}}\right]: s^{\text{OFDM}}[M-1] ; \mathbf{s}^{\text{OFDM}}\right] .
\label{eqOFDM-CP}
\end{equation}

At the receiver, considering the time domain signal $\mathbf{r}^{\text{OFDM}}$ after removing the CP, the DFT operation is performed to obtain the received signal $\mathbf{y}^{\text{OFDM}}$, i.e.,
\begin{equation}
    \mathbf{y}^{\text{OFDM}} = \mathbf{F}_M \mathbf{r}^{\text{OFDM}}.
\end{equation}

\subsubsection{DFT-s-OFDM}
\begin{figure}[H]
	\centering
\includegraphics[width=0.36\textwidth]{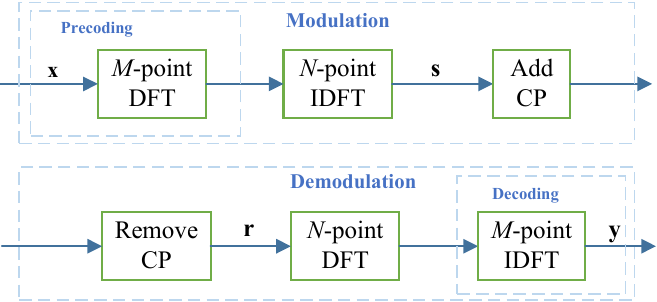}
	\caption{Diagram of DFT-s-OFDM.}
	\label{block2DFT-s-OFDM}
    \vspace{-5pt}
\end{figure}
DFT-s-OFDM employs a DFT precoding step before OFDM modulation to spread information symbols across subcarriers in the frequency domain, which can reduce the PAPR and maintain compatibility with OFDM systems\cite{berardinelli2016generalized,sahin2016flexible}. As shown in Fig. \ref{block2DFT-s-OFDM}, let $\mathbf{x}^\text{DFT-s-OFDM} \in \mathbb{C}^{M \times 1}$ denote a vector of QAM symbols, then, with DFT precoding, the frequency domain samples are obtained as
\begin{equation}
X\!^\text{DFT-s-OFDM}[n]\!=\!\frac{1}{\sqrt{M}}\!\sum_{k=0}^{M-1} \!\!x^\text{DFT-s-OFDM}\![k] e^{-j 2 \pi kn / M}.
\end{equation}

After zero padding and subcarrier mapping, DFT-s-OFDM performs conventional OFDM modulation, i.e., the sample $X^\text{DFT-s-OFDM}$ is converted into a time-domain transmit signal using $N$-point ($N\geq M$) IDFT as
\begin{equation}
s^\text{DFT-s-OFDM}[l]=\frac{1}{\sqrt{N}}\!\!\! \sum_{n=-N / 2}^{N / 2-1} \!\!\!\!X^\text{DFT-s-OFDM}[n] e^{j 2 \pi  nl / N}.
\label{eqDFT-S-OFDM2}
\end{equation}
If the active data subcarrier is assumed to be placed at the center of the channel, then \eqref{eqDFT-S-OFDM2} can also be expressed as\cite{deng2024bit}
\begin{equation}
    \begin{aligned}
    &s^\text{DFT-s-OFDM}{[l]}\\=&\frac{1}{\sqrt{MN}} \sum_{n=-N / 2}^{N / 2-1} \sum_{k=0}^{M-1}x^\text{DFT-s-OFDM}{[k]} e^{-j 2 \pi kn / M}e^{j 2 \pi  nl / N}\\
    =&\frac{1}{\sqrt{M N}} \sum_{k=0}^{M-1} x^\text{DFT-s-OFDM}{[k]} g\left[l-k N / M\right],
    \end{aligned}
    \label{eqDFT-S-OFDM3}
\end{equation} 
where the effective pulse-shaping function $g[\nu]$ is\cite{gokceli2021novel}
\begin{equation}
g[\nu]=e^{-j(\pi \nu / N)} \frac{\sin \left(\pi M \nu / N\right)}{\sin (\pi \nu / N)}.
\end{equation}

In matrix form, \eqref{eqDFT-S-OFDM3} can be rewritten as
\begin{equation}
\mathbf{s}^{\text{DFT-s-OFDM}}=\mathbf{F}_N^{H} \mathbf{P} \mathbf{F}_M \mathbf{x}^{\text{DFT-s-OFDM}},
\end{equation}
where $\mathbf{F}_N \in \mathbb{C}^{ N \times N}$ and $\mathbf{F}_M \in \mathbb{C}^{ M \times M}$ represent the DFT matrices. The $\mathbf{P}\in \mathbb{C}^{ N \times M}$ is the localized subcarrier mapping matrix.

After adding CP, the DFT-s-OFDM signal is transmitted through the channel. At the receiver, the CP is discarded, and then the received signal $\mathbf{r}^{\text{DFT-s-OFDM}}$ is subjected to $N$-point DFT, subcarrier demapping, and $M$-point IDFT de-spreading, which yields
\begin{equation}
\mathbf{y}^{\text{DFT-s-OFDM}}=\mathbf{F}_M^{H} \mathbf{P}^{H} \mathbf{F}_N \mathbf{r}^{\text{DFT-s-OFDM}}.
\end{equation}

\subsubsection{FrFT-OFDM}
FrFT-OFDM integrates the fractional Fourier transform into the OFDM framework to modulate symbols in a rotated frequency domain with rotation $\alpha $, which can enhance the resilience to non-stationary and time-varying channels\cite{martone2001multicarrier,soo-changpei2000closedform}. Here, let $\mathbf{x}^{\text{FrFT-OFDM}} \in \mathbb{C}^{M \times 1}$ denote a vector of QAM symbols. The transmit signal of FrFT-OFDM after the inverse discrete fractional Fourier transform (IDFrFT) operation can be represented as
\begin{equation}
    s^{\text{FrFT-OFDM}}[l]=\frac{1}{\sqrt{M}} \sum_{k=0}^{M-1} x^{\text{FrFT-OFDM}}[k] F_{-\alpha}[l, k].
    \label{eqFrFTOFDM-1}
\end{equation}
Here, $F_\alpha$ denotes the kernel of DFrFT, which can be given by\cite{zheng2010ici,mokhtari2016nearoptimal,soo-changpei2000closedform}
\begin{equation}
    \begin{aligned}
    F_\alpha[l, k]= & \sqrt{\frac{\sin \alpha-j * \cos \alpha}{M}} \exp \left(\frac{j}{2} l^2 \cot \alpha(\Delta u)^2\right) \\
    & \times \exp \left(\frac{j}{2} k^2 \cot \alpha\left(T_s\right)^2\right) e^{-j 2 \pi \frac{k l}{M}}
    \end{aligned},
    \label{eqFrFTOFDM-2}
\end{equation}
where $\alpha=p * \pi / 2(\operatorname{set} 0<\alpha<\pi)$, $p$ is the fractional factor of the transform, $\Delta u$ is the sampling space in fractional Fourier domain, and $\Delta u T_s=2 \pi|\sin \alpha| / M$ with $T_s$ representing the sampling interval in the time domain. When $\alpha=\pi / 2$, the system is equivalent to the OFDM system.

\begin{figure}[H]
	\centering
\includegraphics[width=0.4\textwidth]{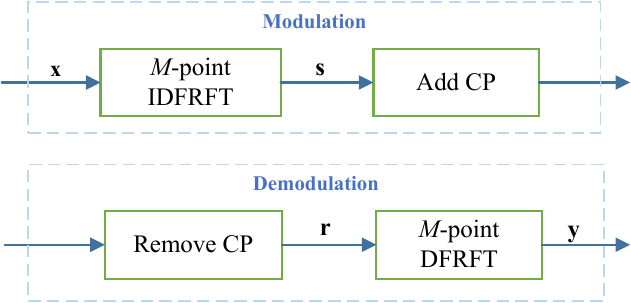}
	\caption{Diagram of FrFT-OFDM.}
	\label{block3FrFT-OFDM}
    \vspace{-8pt}
\end{figure}

In matrix form, \eqref{eqFrFTOFDM-1} can be denoted as
\begin{equation}
    \mathbf{s}^{\text{FrFT-OFDM}}\!=\!\mathbf{F}_{-p,M} \mathbf{x}^{\text{FrFT-OFDM}}\! =\!\mathbf{F}_{p,M}^H \mathbf{x}^{\text{FrFT-OFDM}},
\end{equation}
where $\mathbf{F}_{p,M} \in \mathbb{C}^{ M \times M}$ represent the DFrFT matrix with fractional factor $p$.
Similar to \eqref{eqOFDM-CP} in the OFDM system, CP is added at the beginning of the FrFT-OFDM block and is discarded at the receiver to get the time domain signal $\mathbf{r}^{\text{FrFT-OFDM}}$ as shown in Fig. \ref{block3FrFT-OFDM}. The DFrFT is then applied to obtain the received signal $\mathbf{y}^{\text{FrFT-OFDM}}$ in fractional Fourier domain, i.e., 
\begin{equation}
    \mathbf{y}^{\text{FrFT-OFDM}}=\mathbf{F}_{p,M} \mathbf{r}^{\text{FrFT-OFDM}}.
\end{equation}

\subsubsection{OCDM}
\begin{figure}[H]
	\centering
\includegraphics[width=0.4\textwidth]{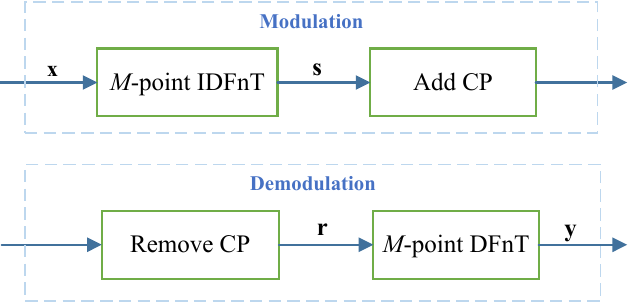}
	\caption{Diagram of OCDM.}
	\label{block4OCDM}
    \vspace{-8pt}
\end{figure}

OCDM modulates signals in the chirp domain using the DFnT. It outperforms OFDM in TDC and has been widely used in radar and acoustic communications\cite{ouyang2016orthogonal,omar2021performance}. As shown in Fig. \ref{block4OCDM}, let $\mathbf{x}^{\text{OCDM}} \in \mathbb{C}^{M \times 1}$ denote a vector of QAM symbols, the $M$-point inverse DFnT (IDFnT) is performed to map $\mathbf{x}^{\text{OCDM}}$ to the time domain as
\begin{equation}
    \begin{aligned}
    s^{\text{OCDM}}[l]=&\frac{1}{\sqrt{M}} e^{j \frac{\pi}{4}} \sum_{k=0}^{M-1} x^{\text{OCDM}}[k] \\
    &\times\left\{\begin{array}{l}e^{-j \frac{\pi}{M}(l-k)^2} \quad M \equiv 0(\bmod 2) \\ e^{-j \frac{\pi}{M}\left(l-k-\frac{1}{2}\right)^2} M \equiv 1(\bmod 2) .\end{array}\right.
    \label{OCDM}
    \end{aligned}
\end{equation}

The DFnT matrix can be decomposed by using the equation $\boldsymbol{\Phi}=\boldsymbol{\Theta}_2 \mathbf{F} \boldsymbol{\Theta}_1$, where $\boldsymbol{\Theta}_1$ and $\boldsymbol{\Theta}_2$ are two diagonal matrices given by
\begin{equation}
    \Theta_1(k)=e^{-j \frac{\pi}{4}} \times\left\{\begin{array}{ll}e^{j\frac{\pi}{M} k^2} & M \equiv 0(\bmod 2) \\e^{j \frac{\pi}{4M}} e^{j \frac{\pi}{M}\left(k^2+k\right)} & M \equiv 1(\bmod 2)\end{array} \right.
\end{equation}
and
\begin{equation}
     \Theta_2(l)=\left\{\begin{array}{ll}e^{j \frac{\pi}{M} l^2} & M \equiv 0(\bmod 2) \\e^{j \frac{\pi}{M} \left(l^2-l\right)} & M\equiv 1(\bmod 2).
    \end{array} \right.
\end{equation}

\eqref{OCDM} can be rewritten in matrix form as
\begin{equation}
    \mathbf{s}^{\text{OCDM}}=\boldsymbol{\Theta}_1^H \mathbf{F}^H_M \boldsymbol{\Theta}_2^H \mathbf{x}^{\text{OCDM}}=\boldsymbol{\Phi}^H \mathbf{x}^{\text{OCDM}}.
    \vspace{-8pt}
\end{equation}
After adding CP, the OCDM signal is transmitted through the channel. Therefore, similar to OFDM, the receiver demodulates the time domain signal $\mathbf{r}^{\text{OCDM}}$ after removing the CP to produce the received signal $\mathbf{y}^{\text{OCDM}}$ in the chirp domain:
\begin{equation}
    \mathbf{y}^{\text{OCDM}}=\boldsymbol{\Phi} \mathbf{r}^{\text{OCDM}}=\boldsymbol{\Theta}_2 \mathbf{F}_M \boldsymbol{\Theta}_1 \mathbf{r}^{\text{OCDM}}.
\end{equation}

\subsubsection{IFDM}
\begin{figure}[H]
	\centering
\includegraphics[width=0.4\textwidth]{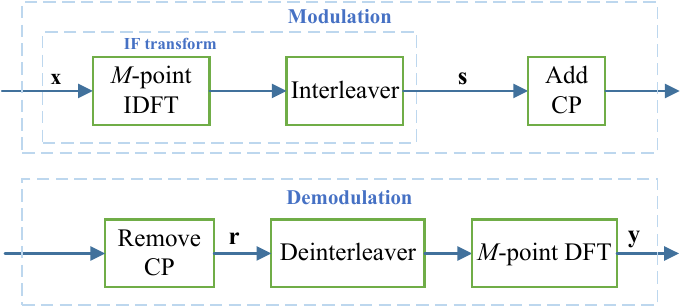}
	\caption{Diagram of IFDM.}
	\label{block5IFDM}
    \vspace{-8pt}
\end{figure}

IFDM is implemented by adding a random interleaving matrix to the OFDM signal via the interleave Fourier (IF) transform to ensure that the signal undergoes sufficient statistical channel fading to make the equivalent channel matrix more numerically stable\cite{chi2024interleave}.
Let $\mathbf{x}^{\text{IFDM}} \in \mathbb{C}^{M \times 1}$ denote a vector of QAM symbols in the IF domain. By interleaving the $M$-points IDFT matrix, the inverse IF transform is produced as shown in Fig. \ref{block5IFDM}.

After the inverse IF transform, the obtained signal in the time domain is
\begin{equation}
\label{eqIFDM}
    s^{\text{IFDM}}[l] = \frac{1}{\sqrt{M}} \sum_{k=0}^{M-1} x^{\text{IFDM}}[k] \cdot e^{j2\pi \frac{\alpha(l) k}{M}}, 
\end{equation}
where $\alpha(l)$ is the interleaved index corresponding to the 
$l$-th output.

\eqref{eqIFDM} can be rewritten in matrix form as
\begin{equation}
    \mathbf{s}^{\text{IFDM}}=\mathbf{\Pi} \mathbf{F}_M^H\mathbf{x}^{\text{IFDM}},
\end{equation}
where $\mathbf{\Pi} \mathbf{F}_M^H$ denotes the inverse IF transform with the random interleave matrix $\mathbf{\Pi} \in \mathbb{C}^{M \times M}$. 

Similarly, CP is added to the front of the transmitted signal for ISI mitigation, and the time domain signal $\mathbf{r}^{\text{IFDM}}$ with CP removed is obtained at the receiver. During demodulation, the opposite operation is performed on $\mathbf{r}^{\text{IFDM}}$ as 
\begin{equation}
\mathbf{y}^{\text{IFDM}}=\mathbf{F}_M\mathbf{\Pi}^{-1} \mathbf{r}^{\text{IFDM}},
\end{equation}
i.e., the deinterleaving step is performed, followed by the DFT, to yield $\mathbf{y}^{\text{IFDM}}$ in the IF domain.

\subsubsection{AFDM}
\begin{figure}[H]
	\centering
\includegraphics[width=0.4\textwidth]{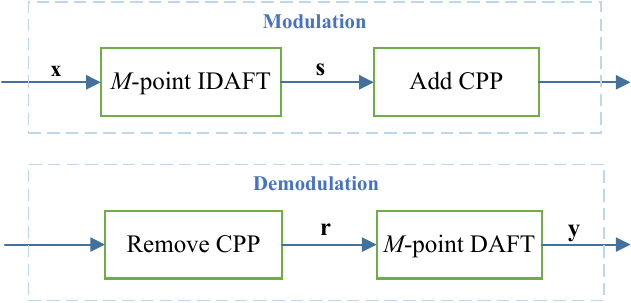}
	\caption{Diagram of AFDM.}
	\label{block6AFDM}
    \vspace{-8pt}
\end{figure}

Similar to OCDM, AFDM modulates the signal in the chirp domain (DAFT domain), utilizing two parameters inherent in DAFT to achieve full diversity performance and robustness to high mobility scenarios\cite{bemani2021afdm,bemani2021affine}.
Let $\mathbf{x}^{\text{AFDM}} \in \mathbb{C}^{M \times 1}$ denote a vector of QAM symbols in the DAFT domain. As shown in Fig. \ref{block6AFDM}, the $M$-points IDAFT is performed to map $\mathbf{x}^{\text{AFDM}}$ to the time domain as\cite{bemani2023affinea}
\begin{equation}
s^{\text{AFDM}}[l]=\frac{1}{\sqrt{M}} \sum_{k=0}^{M-1} x^{\text{AFDM}}[k] e^{j 2 \pi\left(c_{2} k^{2}+\frac{1}{M} kl+c_{1} l^{2}\right)},
\label{eqAFDM-1}
\end{equation}
where $M$ denotes the number of subcarriers, $c_1$ and $c_2$ are two fundamental parameters of AFDM.
Hence, \eqref{eqAFDM-1} can be written in a matrix form as
\begin{equation}
   \mathbf{s}^{\text{AFDM}}=\boldsymbol{\Lambda}_{c_1}^H \mathbf{F}_M^H \boldsymbol{\Lambda}_{c_2}^H \mathbf{x}^{\text{AFDM}},
\end{equation}
where
\begin{equation}
\boldsymbol{\Lambda}_{c_i}=\operatorname{diag}\left(e^{-j 2 \pi c_i l^2}, l=0,1, \ldots, M-1, i=1,2\right) .
\vspace{-8pt}
\end{equation}

Before transmitting $\mathbf{s}^{\text{AFDM}}$ into the channel, a chirp-periodic prefix (CPP) should be added with a length of $L_{\text{cpp}}$, which is an integer greater than or equal to the value in samples of the maximum delay spread of the channel. The prefix is
\begin{equation}
   s^{\text{AFDM}}_{\text{cpp}}[l]=s^{\text{AFDM}}[M+l] e^{-j 2 \pi c_1\left(M^2+2 M l\right)}, l=-L_{\text{cpp}}, \cdots,-1 . 
\end{equation}
Thus, the transmitted signal can be written as
\begin{equation}
\mathbf{s}^{\text{AFDM}}_{\text{cpp}}=\left[s^{\text{AFDM}}_{\text{cpp}}\left[-L_{\text{cpp}}\right]: s^{\text{AFDM}}_{\text{cpp}}[-1] ; \mathbf{s}^{\text{AFDM}}\right].
\end{equation}

After passing through the channel, the CPP is first removed to obtain the received time domain signal $\mathbf{r}^{\text{AFDM}}$, which is then demodulated to yield the DAFT domain signal $\mathbf{y}^{\text{AFDM}}$, given by
\begin{equation}
\mathbf{y}^{\text{AFDM}}=\boldsymbol{\Lambda}_{c_2} \mathbf{F}_M \boldsymbol{\Lambda}_{c_1} \mathbf{r}^{\text{AFDM}}.
\end{equation}

\vspace{-5pt}
\subsection{2D Modulation Waveforms}
\label{sec3-2}
\subsubsection{FBMC}
\begin{figure}[H]
	\centering
\includegraphics[width=0.4\textwidth]{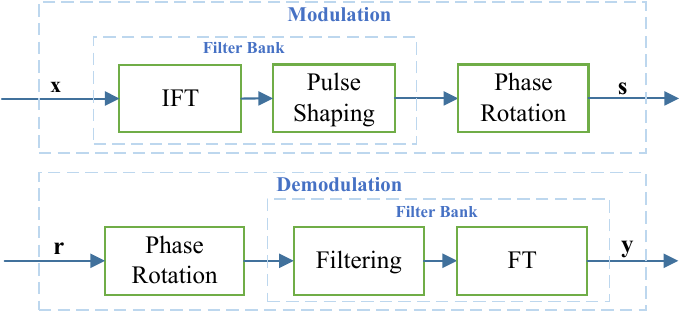}
	\caption{Diagram of FBMC.}
	\label{block7FBMC}
    \vspace{-8pt}
\end{figure}

Unlike OFDM, instead of a simple rectangular window function, the subcarriers of FBMC have a smooth filter shape to reduce OOBE\cite{farhang-boroujeny2011ofdm}. This design effectively reduces the interference between subcarriers, while CP can be eliminated. There are many types of FBMC, and the most common variant is FBMC/OQAM\cite{ihalainen2011channel,qu2013multiblock,nissel2016pilotsymbol}, which achieves orthogonality by interleaving the real and imaginary parts of the transmission (offset by half a symbol period) to improve spectral efficiency as shown in Fig. \ref{block7FBMC}. Let $\mathbf{x}^{\text{FBMC}} \in \mathbb{C}^{M \times N}$ denote the matrix of QAM symbols in the TF domain. The transmitted signal in the time domain is

\begin{equation}
s^{\text{FBMC}}(t)= \sum_{k=0}^{N-1} \sum_{l=0}^{M-1}  g_{l, k}(t) x^{\text{FBMC}}_{[l, k]},
\label{eqFBMC-1}
\end{equation}
where $x^{\text{FBMC}}_{[l, k]}$ represents the transmitted QAM symbol at subcarrier position $l$ and time position $k$ with\cite{nissel2017filter}

\begin{equation}
g_{l, k}(t)=p(t-k T) \text{e}^{j 2 \pi l F(t-k T)} \text{e}^{j \frac{\pi}{2}(l+k)}.
\label{eq:g_lk(t)}
\end{equation}
Here, the basis pulse $g_{l, k}(t)$ is a time and frequency-shifted version of the prototype filter $p(t)$, with $T$ denoting the time spacing and $F$ denoting the frequency spacing (subcarrier spacing). In FBMC, the prototype filter must be a real-valued even function, i.e., $p(t)=p(-t)$, and orthogonal for a TF spacing of $T \times F = T F=2$.
Typically, prototype filter $p(t)$ can be generated by Hermite polynomials $H(\cdot)$\cite{haas1997timefrequency}:

\begin{equation}
p(t)=\frac{1}{\sqrt{T_0}} \text{e}^{-2 \pi\left(\frac{t}{T_0}\right)^2} \!\!\!\!\!\sum_{\substack{i=\{0,4,8, \\12,16,20\}}} \!\!\!a_i H_i\left(2 \sqrt{\pi} \frac{t}{T_0}\right),
\end{equation}
the coefficients of $a_i$ are defined as follows\cite{nissel2017filter}:
$$
\begin{array}{ll}
a_0=1.412692577 & a_{12}=-2.2611 \cdot 10^{-9} \\
a_4=-3.0145 \cdot 10^{-3} & a_{16}=-4.4570 \cdot 10^{-15} \\
a_8=-8.8041 \cdot 10^{-6} & a_{20}=1.8633 \cdot 10^{-16}
\end{array}
$$
The prototype filter $p(t)$ guarantees orthogonality of the basis pulses for a time spacing of $T=T_0$ and a frequency spacing of $F=\frac{2}{T_0}$. 

To improve the spectral efficiency in FBMC, the time-spacing as well as the frequency spacing are both reduced by a factor of two, leading to $T F=0.5$. That is, in FBMC/OQAM, only the real part of the data is transmitted, causing the original orthogonality to become orthogonality in the real number domain.

Then, \eqref{eqFBMC-1} can be expressed as
\begin{equation}
    \mathbf{s}^{\text{FBMC}}=\mathbf{G} \mathbf{x}^{\text{FBMC}},
\end{equation}
where $\mathbf{G} \in \mathbb{C}^{\left(\frac{(M-1) T+6 T_0}{\Delta t}+1\right) \times M N}$ is given by\cite{nissel2016pilotsymbol}:
\begin{equation}
[\mathbf{G}]_{i, l+k M}=\left.\sqrt{\Delta t} g_{l, k}(t)\right|_{t=\Delta t i-3 T_0},
\end{equation}
with sampling interval $\Delta t$.

At the receiver, the received time domain signal $\mathbf{r}^{\text{FBMC}}$ is transformed back to the frequency domain through the corresponding receive matrix $\mathbf{G}^H$(including filtering and FT), i.e.,
\begin{equation}
    \mathbf{y}^{\text{FBMC}}=\mathbf{G}^H \mathbf{r}^{\text{FBMC}}.
\end{equation}

\subsubsection{OTFS}
OTFS\footnote{Before OTFS was proposed, there were some waveforms that were equivalent to OTFS in terms of transmit and receive processing, such as vector OFDM (V-OFDM)\cite{xia2000precoded} and orthogonal signal-division multiplexing (OSDM)\cite{ebihara2014underwater}, although they were not proposed for DDC\cite {xia2022comments,vanderwerf2024equivalence,deng2025unifyingb}.}, as a representative of 2D modulation, multiplexes information symbols in the DD domain. There are two ways to realize OTFS, one is based on multicarrier (i.e., MC-OTFS)\cite{hadani2017orthogonal,wei2021orthogonal,ge2021receiver} and the other is based on Zak-transform (i.e., Zak-OTFS)\cite{jayachandran2024zakotfs,mohammed2024otfs}. MC-OTFS with two steps will be introduced first, followed by Zak-OTFS with one step.

\begin{figure}[H]
	\centering
\includegraphics[width=0.4\textwidth]{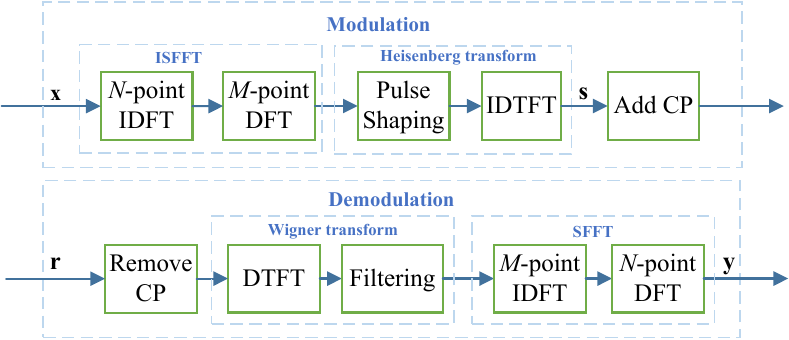}
	\caption{Diagram of MC-OTFS.}
	\label{block8MC-OTFS}
    \vspace{-8pt}
\end{figure}

Different from the TF domain, DD domain modulation requires the signal to be placed on the DD lattice first, which is denoted as\cite{murali2018otfs}
\begin{equation}
    L\!=\!\left\{\!\!\left(\frac{l}{M \Delta f}, \frac{k}{N T}\right)\!,
     \!l=0, \!\cdots,\! M-1,\! k=0, \!\cdots,\! N-1\!\right\},
\end{equation}
where $T$ is the symbol period which is determined to be greater than the maximum multipath delay spread; $\Delta f=\frac{1}{T}$ is the subcarrier spacing which should be larger than the maximum Doppler spread; $M$ and $N$ are the number of subcarriers and time slots, also representing the number of delay and Doppler bins, respectively.

Let $\mathbf{\tilde{x}}^{\text{MC-OTFS}} \in \mathbb{C}^{M \times N}$ denote the matrix of QAM symbols in the delay-Doppler domain. The ISFFT, which involves $N$-point inverse DFT along the Doppler axis and $M$-point DFT along the delay axis, is performed to map $\mathbf{\tilde{x}}^{\text{MC-OTFS}}$ to the TF domain as
\begin{equation}
  X^{\text{MC-OTFS}}_{[m,n]} = \frac{1}{\sqrt{NM}} \sum_{k=0}^{N-1} \sum_{l=0}^{M-1} x^{\text{MC-OTFS}}_{[l,k]} e^{j2\pi\left(\frac{kn}{N} - \frac{lm}{M} \right)}, 
  \label{eqMCOTFS-1}
\end{equation}
where \( X^{\text{MC-OTFS}}_{[m,n]} \) represents the TF domain signal, \( n \) and \( m \) index the time and frequency dimensions.

Then, the Heisenberg transform is applied to convert \( X^{\text{MC-OTFS}}_{[m,n]} \) into the time domain signal, which can be expressed as
\begin{equation}
  s(t)^\text{MC-OTFS}\! =\!\!\! \sum_{n=0}^{N-1}\! \sum_{m=0}^{M-1} \! X^{\text{MC-OTFS}}_{[m,n]}  \!\!g_\text{tx}(t \!- \!nT) e^{j2\pi m \Delta f (t\! -\! nT)},
  \label{eqMCOTFS-2}
\end{equation}
  where \( g_\text{tx}(t) \) represents the transmit pulse.

We assume that the total duration of the transmitted signal is $NT$ and the sampling interval is $T/M$. In matrix form, \eqref{eqMCOTFS-1} and \eqref{eqMCOTFS-2} can be expressed as\cite{deng2025unifyingb}
\begin{equation}
    \begin{split}
        \mathbf{\tilde{s}}^{\text{MC-OTFS}}=&\mathbf{G}_\text{tx} \mathbf{F}_M^{\text{H}}\left(\mathbf{F}_M \mathbf{\tilde{x}}^{\text{MC-OTFS}} \mathbf{F}_N^H\right)\\=&\mathbf{G}_\text{tx} \mathbf{\tilde{x}}^{\text{MC-OTFS}} \mathbf{F}_N^H,
    \end{split}
\end{equation}
where $\mathbf{G}_\text{tx}  \in \mathbb{C}^{M \times M}$ is the diagonal matrix with the samples of $g_\text{tx}(t)$, i.e.,
\begin{equation}
\mathbf{G}_\text{tx}=\operatorname{diag}[g_\text{tx}(0), g_\text{tx}\left(\frac{T}{M}\right), \cdots, g_\text{tx}\left(\frac{(M-1) T}{M}\right)];
\end{equation}
$\mathbf{\tilde{s}}^{\text{MC-OTFS}} \in \mathbb{C}^{M \times N}$ is the discrete transmit signal matrix, which can be given in $MN \times 1$ vector form as 
\begin{equation}
\mathbf{s}^{\text{MC-OTFS}}=\operatorname{vec}(\mathbf{\tilde{s}}^{\text{MC-OTFS}})=\left(\mathbf{F}_N^{\text{H}} \otimes \mathbf{G}_{\text{tx}}\right) \mathbf{x}^{\text{MC-OTFS}}
\end{equation}
with $\mathbf{x}^{\text{MC-OTFS}}=\operatorname{vec}(\mathbf{\tilde{x}}^{\text{MC-OTFS}})$.
After the vectorization of $\mathbf{\tilde{s}}^{\text{MC-OTFS}}$, the CP is added to $\mathbf{s}$ for transmission as shown in Fig. \ref{block8MC-OTFS}. 

At the receiver, we obtain the received signal $\mathbf{r}^{\text{MC-OTFS}}$ in the time domain with the CP removed, and convert $\mathbf{r}^{\text{MC-OTFS}}$ to the TF domain $\mathbf{\tilde{y}}^{\text{MC-OTFS}}_\text{TF}$ by the Wigner transform, i.e., 
\begin{equation}
\begin{split}
    &Y^{\text{MC-OTFS}}(t, f)\\
    =&\int g_\text{rx}^*\left(t^{\prime}-t\right) r^{\text{MC-OTFS}}\left(t^{\prime}\right) e^{-j 2 \pi f\left(t^{\prime}-t\right)} d t^{\prime}.
\end{split}
\label{eqMCOTFS-3}
\end{equation}

Equivalently, \eqref{eqMCOTFS-3} can be expressed in matrix form as
\begin{equation}
\mathbf{\tilde{y}}^{\text{MC-OTFS}}_\text{TF}=\mathbf{F}_M \mathbf{G}_{\text{rx}} \mathbf{\tilde{r}}^{\text{MC-OTFS}},
\end{equation}
with $\mathbf{\tilde{r}}^{\text{MC-OTFS}}\in \mathbb{C}^{M \times N}=\operatorname{vec}^{-1}(\mathbf{r}^{\text{MC-OTFS}})$, $\mathbf{G}_\text{rx}\in \mathbb{C}^{M \times M}$ is the diagonal matrix with the samples of $g_\text{rx}(t)$, i.e., $\mathbf{G}_\text{rx}=\operatorname{diag}[g_\text{rx}(0), g_\text{rx}\left(\frac{T}{M}\right), \cdots, g_\text{rx}\left(\frac{(M-1) T}{M}\right)]$ and then $\mathbf{\tilde{y}}^{\text{MC-OTFS}}_\text{DD}=\operatorname{vec}^{-1}(\mathbf{y}^{\text{MC-OTFS}})$ becomes
\begin{equation}
    \begin{split}
        \mathbf{\tilde{y}}^{\text{MC-OTFS}}_\text{DD}=& \mathbf{F}_M^H \mathbf{\tilde{y}}^{\text{MC-OTFS}}_\text{TF}\mathbf{F}_N 
        \\=&\mathbf{F}_M^H\left(\mathbf{F}_M \mathbf{G}_{\text{rx}} \mathbf{\tilde{r}}^{\text{MC-OTFS}}\right) \mathbf{F}_N .
    \end{split}
\end{equation}

Finally, $\mathbf{\tilde{y}}^{\text{MC-OTFS}}_\text{TF}$ in the TF domain is transformed into $\mathbf{\tilde{y}}^{\text{MC-OTFS}}_\text{DD}$ in the DD domain. Then, the whole process at the receiver can be represented by a vectorized form as
\begin{equation}
\mathbf{y}^{\text{MC-OTFS}}=\left(\mathbf{F}_N \otimes \mathbf{G}_\text{rx}\right) \mathbf{r}^{\text{MC-OTFS}}.
\end{equation}

\begin{figure}[H]
	\centering
\includegraphics[width=0.4\textwidth]{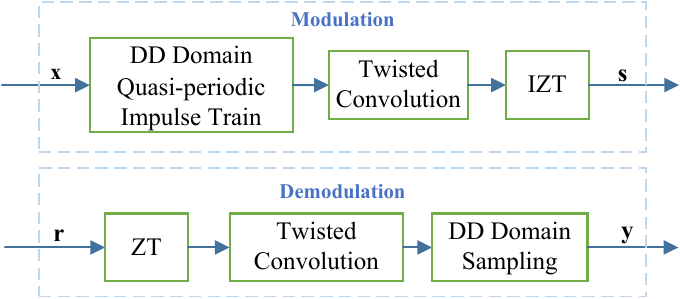}
	\caption{Diagram of Zak-OTFS.}
	\label{block8Zak-OTFS1}
    \vspace{-8pt}
\end{figure}

Unlike the two-step MC-OTFS, Zak-OTFS cleverly utilizes the ZT to implement the OTFS modulation process in just one step, as shown in Fig. \ref{block8Zak-OTFS1}. In Zak-OTFS, a pulse in the DD domain, which is a quasi-periodic localized function defined by a delay period $\tau_{\text{p}}$ and a Doppler period $\nu_{\text{p}} = 1/\tau_{\text{p}}$, serves as the basic information carrier\cite{jayachandran2024zakotfs}. The fundamental period in the DD domain is defined as
\begin{equation}
\mathcal{D} = \{(\tau, \nu) : 0 \le \tau < \tau_{\text{p}}, 0 \le \nu < \nu_{\text{p}}\},
\end{equation}
where $\tau$ and $\nu$ represent the delay and Doppler variables, respectively. The fundamental period is discretized into $M$ bins on the delay axis and $N$ bins on the Doppler axis, represented as the lattice $\Lambda = \{(l \frac{\tau_{\text{p}}}{M}, k \frac{\nu_{\text{p}}}{N}) \mid l = 0, \dots, M-1, k = 0, \dots, N-1\}$. 

The time-domain Zak-OTFS frame is limited to a duration $T = N\tau_{\text{p}}$ and bandwidth $B = M\nu_{\text{p}}$. In each frame, $MN$ information symbols drawn from a modulation alphabet $\mathbb{A}$, denoted as $x^{\text{Zak-OTFS}}[l, k] \in \mathbb{A}$ (where $l$ and $k$ index the delay and Doppler bins, respectively), are multiplexed in the DD domain. The information symbol $x^{\text{Zak-OTFS}}[l, k]$ is carried by a DD domain pulse $x^{\text{Zak-OTFS}}_{\text{DD}}[l, k]$, which is a quasi-periodic function with period $M$ along the delay axis and period $N$ along the Doppler axis. Specifically, for any integers $n, m \in \mathbb{Z}$, the quasi-periodicity is given by
\begin{equation}
x^{\text{Zak-OTFS}}_{\text{DD}}[l + nM, k + mN] = x^{\text{Zak-OTFS}}[l, k] e^{j 2 \pi n \frac{k}{N}}.
\end{equation}

These discrete DD domain signals are supported on the information lattice $\Lambda_{\text{DD}} = \{ ( l \frac{\tau_{\text{p}}}{M}, k \frac{\nu_{\text{p}}}{N} ) \mid l, k \in \mathbb{Z} \}$. The continuous DD domain information symbol $x^{\text{Zak-OTFS}}_{\text{DD}}(\tau, \nu)$ is then constructed as
\begin{equation}
\begin{split}
x^{\text{Zak-OTFS}}_{\text{DD}}(\tau, \nu) = \sum_{l, k \in \mathbb{Z}} & x^{\text{Zak-OTFS}}_{\text{DD}}[l, k] \delta \left( \tau - \frac{l\tau_{\text{p}}}{M} \right) \\
&\times \delta \left( \nu - \frac{k\nu_{\text{p}}}{N} \right),
\end{split}
\end{equation}
where $\delta(\cdot)$ denotes the Dirac-delta impulse function. This continuous signal satisfies the quasi-periodicity condition:
\begin{equation}
x^{\text{Zak-OTFS}}_{\text{DD}}(\tau + n\tau_{\text{p}}, \nu + m\nu_{\text{p}}) = e^{j2\pi n \nu \tau_{\text{p}}} x^{\text{Zak-OTFS}}_{\text{DD}}(\tau, \nu).
\end{equation}

The actual DD domain transmit signal, denoted as $x^{\text{Zak-OTFS}}_{w}(\tau, \nu)$, is obtained by the twisted convolution of the transmit pulse shaping filter $w_{\text{tx}}(\tau, \nu)$ with $x^{\text{Zak-OTFS}}_{\text{DD}}(\tau, \nu)$. To accommodate the column width, the operation is expressed as
\begin{equation}
\begin{split}
x^{\text{Zak-OTFS}}_{w}(\tau, \nu) &= w_{\text{tx}}(\tau, \nu) *_{\sigma} x^{\text{Zak-OTFS}}_{\text{DD}}(\tau, \nu) \\
&= \iint w_{\text{tx}}(\tau', \nu') x^{\text{Zak-OTFS}}_{\text{DD}}(\tau - \tau', \nu - \nu') \\
&\quad \times e^{j2\pi \nu'(\tau - \tau')} d\tau' d\nu',
\end{split}
\end{equation}
where $*_{\sigma}$ denotes the twisted convolution. 

Finally, the transmitted time-domain signal $s^{\text{Zak-OTFS}}(t)$ is the inverse time-Zak transform (IZT) of $x^{\text{Zak-OTFS}}_{w}(\tau, \nu)$, given by
\begin{equation}
s^{\text{Zak-OTFS}}(t) = \mathcal{Z}_t^{-1} \{x^{\text{Zak-OTFS}}_{w}(\tau, \nu)\}.
\end{equation}
Here, the pulse-shaping filter $w_{\text{tx}}(\tau, \nu)$ is essential to limit the time duration and bandwidth. The IZT of a generic DD function $a(\tau, \nu)$ is defined as
\begin{equation}
\mathcal{Z}_t^{-1}\{a(\tau, \nu)\} \triangleq \sqrt{\tau_{\text{p}}} \int_0^{\nu_{\text{p}}} a(t, \nu) d\nu,
\end{equation}
and the ZT of a continuous time-domain signal $a(t)$ is defined as
\begin{equation}
\mathcal{Z}_t\{a(t)\} \triangleq \sqrt{\tau_{\text{p}}} \sum_{k \in \mathbb{Z}} a(t + k\tau_{\text{p}}) e^{-j2\pi \nu k \tau_{\text{p}}}.
\end{equation}

At the receiver, the received signal $r^{\text{Zak-OTFS}}(t)$ after passing through the channel is transformed back to the DD domain via the ZT, denoted as 
\begin{equation}
y^{\text{Zak-OTFS}}_{\text{DD}}(\tau, \nu) = \mathcal{Z}_t \{r^{\text{Zak-OTFS}}(t)\}.
\end{equation}
Then, the demodulated symbol $y^{\text{Zak-OTFS}}[l, k]$ is obtained by sampling the twisted convolution of $y^{\text{Zak-OTFS}}_{\text{DD}}(\tau, \nu)$ and a receive pulse $w_{\text{rx}}(\tau, \nu)$ on the information lattice $\Lambda_{\text{DD}}$, which can be expressed as
\begin{equation}
\begin{split}
y^{\text{Zak-OTFS}}[l, k] &= \left. \left( w_{\text{rx}}(\tau, \nu) *_{\sigma} y^{\text{Zak-OTFS}}_{\text{DD}}(\tau, \nu) \right) \right|_{(\frac{l\tau_{\text{p}}}{M}, \frac{k\nu_{\text{p}}}{N})}.
\end{split}
\end{equation}

\begin{figure}[H]
	\centering
\includegraphics[width=0.4\textwidth]{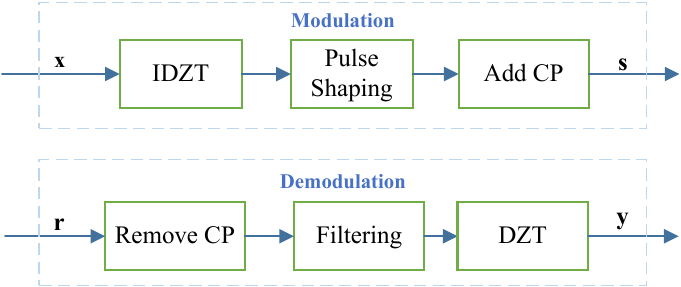}
	\caption{Diagram of Zak-OTFS using DZT.}
	\label{block8Zak-OTFS}
    \vspace{-8pt}
\end{figure}

While the aforementioned continuous Zak transform framework offers a theoretical basis, practical digital implementations typically rely on discretized transformations. As shown in Fig. \ref{block8Zak-OTFS}, OTFS based on the discrete Zak transform (DZT) has also been proposed in \cite{lampel2022otfs}.

At the transmitter, we also consider QAM symbols, denoted by $x^{\text{Zak-OTFS}}_{[l,k]}$. They are multiplexed over an $M \times N$ DD grid, which is converted to the time domain using the inverse discrete Zak transform (IDZT) as\cite{lampel2022otfs}
\begin{equation}
s_{[n]}^{\text{Zak-OTFS}}=\frac{1}{\sqrt{N}} \sum_{k=0}^{N-1} x^{\text{Zak-OTFS}}_{[(n)_L,k]} e^{j 2 \pi \frac{\lfloor n / L\rfloor}{N} k},
\label{eqZakOTFS-1}
\end{equation}
where $(\cdot)_L$ denotes modulo-L operation and $\lfloor\cdot\rfloor$ denotes floor operation.

Similar to MC-OTFS, CP is added to the transmitted signal $\mathbf{s}^{\text{Zak-OTFS}}$ and then the discrete sequence is mounted on a time-shifted transmit pulse $g_\text{tx}(t)$. Assuming that $\mathbf{Z}$ denotes the Zak-transform, the matrix form in \eqref{eqZakOTFS-1} can be expressed as
\begin{equation}
    \mathbf{s}^{\text{Zak-OTFS}} = \mathbf{G}_\text{tx} \mathbf{Z}^H \mathbf{x}^{\text{Zak-OTFS}}.
\end{equation}

At the receiver, the $\mathbf{r}^{\text{Zak-OTFS}}$ in the time domain is transformed back to the DD domain $\mathbf{y}^{\text{Zak-OTFS}}$ via the DZT with matched filter $g_\text{rx}(t)$ as 
\begin{equation}
    \mathbf{y}^{\text{Zak-OTFS}} = \mathbf{Z} \mathbf{G}_\text{rx} \mathbf{r}^{\text{Zak-OTFS}}.
\end{equation}

\subsubsection{ODDM}
Similar to OTFS, ODDM also modulates the signal in the DD domain\cite{lin2022orthogonal}. By utilizing delay-Doppler plane orthogonal pulses (DDOP)\cite{lin2022delaydoppler}, which uses a square-root Nyquist pulse train, ODDM achieves sufficient orthogonality on the DD plane with fine resolutions\footnote{\newer{Although rectangular pulse shaping is primarily adopted in this unified framework for ease of exposition and fair comparison, it is known to exhibit high OOBE. In contrast, the square-root Nyquist pulses utilized in ODDM provide superior OOBE performance. As investigated in \cite{shen2022error}, rectangular pulse-shaped OTFS may suffer from severe interference when practical BPFs are employed at the receiver, whereas the pulse shaping in ODDM helps mitigate this issue.}}.

\begin{figure}[H]
	\centering
\includegraphics[width=0.4\textwidth]{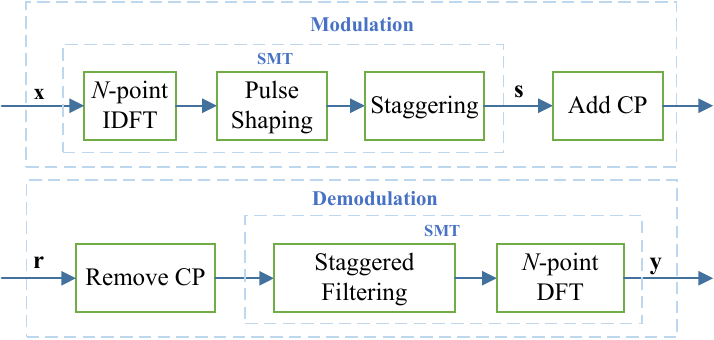}
	\caption{Diagram of ODDM.}
	\label{block9ODDM}
    \vspace{-8pt}
\end{figure}

As shown in Fig. \ref{block9ODDM}, let $\mathbf{X}^{\text{ODDM}} \in \mathbb{C}^{M \times N}$ denote the matrix of QAM symbols in the delay-Doppler domain for ODDM. Assume that the subcarrier spacing is $1/NT$, and the symbol length is $NT$. Then, the QAM symbols $\mathbf{x}^{\text{ODDM}}\in \mathbb{C}^{MN \times 1}=\operatorname{vec}(\mathbf{X}^{\text{ODDM}})$ are spaced by a short interval $ \frac{T}{M} $, which results in a staggered signal structure. The ODDM signal in the continuous-time domain is thus given by
\begin{equation}
s^{\text{ODDM}}(t)=\sum_{l=0}^{M-1} \sum_{k=0}^{N-1} x^{\text{ODDM}}_{[l,k]} \check{g}_{t x}\left(t-l \frac{T}{M}\right) e^{j 2 \pi \frac{k}{N T}\left(t-l \frac{T}{M}\right)},
\label{eqODDM-1}
\end{equation}
where $\check{g}_{t x}(t)$ is the DDOP with square-root Nyquist pulse $a(t)$ given by
\begin{equation}
\check{g}(t)=\sum_{\dot{k}=0}^{N-1} a(t-\dot{k} T),
\end{equation}
and usually $a(t)$ is struncated support of  $\left(-Q \frac{T}{M}, Q \frac{T}{M}\right)$, where $Q$ is an integer and $2 Q \ll M$. Therefore, the range of the time-domain transmitted signal is $-Q \frac{T}{M} \leq t \leq N T+(Q-1) \frac{T}{M}$.

In matrix form, \eqref{eqODDM-1} can be expressed as\cite{tong2024orthogonal}
\begin{equation}
\mathbf{s}^{\text{ODDM}}=\mathbf{G}_\text{tx}\boldsymbol{\Pi}\left(\mathbf{I}_M \otimes \mathbf{F}_N^H\right) \mathbf{x}^{\text{ODDM}},
\end{equation}
where the permutation matrix $\boldsymbol{\Pi}  \in \mathbb{C}^{MN \times MN}$ is 
\begin{equation}
[\boldsymbol{\Pi}]_{k, k^{\prime}}=\left\{\begin{array}{ll}
1, k^{\prime}= & (k)_M N+\left\lfloor\frac{k}{M}\right\rfloor \\
0, & \text { otherwise }
\end{array} .\right.
\end{equation}

At the receiver, the received time-domain signals $\mathbf{r}^{\text{ODDM}}$ after CP removal is matched filtered by a receive pulse $g_{rx}(t)$ and converted back to the DD domain to yield $\mathbf{y}^{\text{ODDM}}$, i.e.,
\begin{equation}
\mathbf{y}^{\text{ODDM}}=\mathbf{G}_\text{rx}\left(\mathbf{I}_M \otimes \mathbf{F}_N^H\right)^H \boldsymbol{\Pi}^H \mathbf{r}^{\text{ODDM}}.
\end{equation}

\subsubsection{OTSM}
Apart from OTFS, which places its multiplex information symbols in the DD domain, OTSM places them in the delay-sequence domain\cite{thaj2021orthogonal}. It is realized by using the inverse Walsh-Hadamard transform (IWHT) along the sequency domain. In this manner, the OTSM transfers the channel delay spread and Doppler spread to the delay dimension and sequence dimension, respectively, while ensuring that they are separable at the receiver\cite{neelam2023jointa,sui2024ber,sui2024performance}. 

\begin{figure}[H]
	\centering
\includegraphics[width=0.4\textwidth]{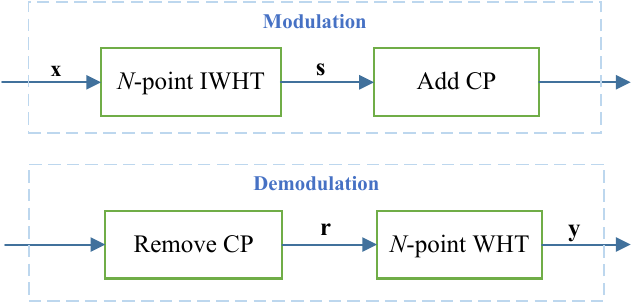}
	\caption{Diagram of OTSM.}
	\label{block10OTSM}
    \vspace{-8pt}
\end{figure}
As shown in Fig. \ref{block10OTSM}, the input information symbols \( \mathbf{x}^{\text{OTSM}} = {[\mathbf{x}_0^T, \dots, \mathbf{x}_{M-1}^T]}^T \in \mathbb{C}^{MN \times 1}\) in the delay-sequency domain are first decomposed into symbol vectors \( \mathbf{x}_m \in \mathbb{C}^{N \times 1}\) at the transmitter. Let $\mathbf{X}^{\text{OTSM}}\in \mathbb{C}^{M \times N}$ denote the matrix of QAM symbols in the delay-sequency domain. The symbol vectors are then rearranged into $\mathbf{X}^{\text{OTSM}} = [\mathbf{x}_0, \mathbf{x}_1, \dots, \mathbf{x}_{M-1}]^T$. Then, IWHT is applied to
every row of $\mathbf{X}^{\text{OTSM}}$ to obtain the delay-time domain transmitted matrix, given by

\begin{equation}
\mathbf{S}^{\text{OTSM}}=\mathbf{X}^{\text{OTSM}} \mathbf{W}_N . 
\end{equation}

The elements of the Walsh-Hadamard transform (WHT) matrix $\mathbf{W}_N\in \mathbb{C}^{N \times M}$ is given by $\mathbf{W}_N[n, m]=\tilde{\mathbf{W}}[n, m / N+0.5 / N] / \sqrt{N}$, where $\tilde{\mathbf{W}}[n, \lambda]$ represents the continuous Walsh functions within $0 \leq \lambda \leq 1$\cite{thaj2021orthogonal}. The matrix $\mathbf{S}^{\text{OTSM}}$ contains the delay-time samples, which are column-wise vectorized to obtain the time domain samples $\mathbf{s}^{\text{OTSM}}\in \mathbb{C}^{MN \times 1}$, given by

\begin{equation}
    \mathbf{s}^{\text{OTSM}}=\operatorname{vec} (\mathbf{S}^{\text{OTSM}})
\end{equation}

Hence, the transmitted signal can be obtained as:
\begin{equation}
    \mathbf{s}^{\text{OTSM}} = (\mathbf{W}_N \otimes \mathbf{I}_M) \cdot (\mathbf{P} \cdot \mathbf{x}^{\text{OTSM}}),
\end{equation}
where
\begin{equation}
\boldsymbol{P}=\left[\begin{array}{c}
\boldsymbol{I}_N \otimes \boldsymbol{e}_M^T(0) \\
\boldsymbol{I}_N \otimes \boldsymbol{e}_M^T(1) \\
\vdots \\
\boldsymbol{I}_N \otimes \boldsymbol{e}_M^T(M-1)
\end{array}\right] .
\end{equation}
\( \mathbf{P}  \in \mathbb{C}^{MN \times MN} \) is the row-column permutation matrix, which is also known as the perfect shuffle matrix. After adding the CP, the resulting time-domain samples are pulse-shaped and converted from digital to analog, and then transmitted over the channel.

At the receiver, $\mathbf{r}^{\text{OTSM}} \in \mathbb{C}^{MN \times 1}$ is processed via analog to digital conversion and CP removal, yielding a time domain vector of size ${MN\times 1}$. The received delay-sequency information is obtained by taking an $N$-point WHT, i.e.,
\begin{equation}\mathbf{y}^{\text{OTSM}}=(\mathbf{I}_{M}\otimes\mathbf{W}_{N})\cdot(\mathbf{P}^{\text{T}}\cdot\mathbf{r}^{\text{OTSM}}).
\end{equation}

\subsubsection{ODSS}
Unlike OTFS, ODSS can improve BER performance in wideband time-varying channels\cite{k.p.2022orthogonal}, which introduces a preprocessing 2D transform from the Mellin-Fourier domain to the delay-scale domain. This transform plays the role of ISFFT in OTFS; thus, ODSS is a typical type of 2D modulation.
\begin{figure}[H]
	\centering
\includegraphics[width=0.4\textwidth]{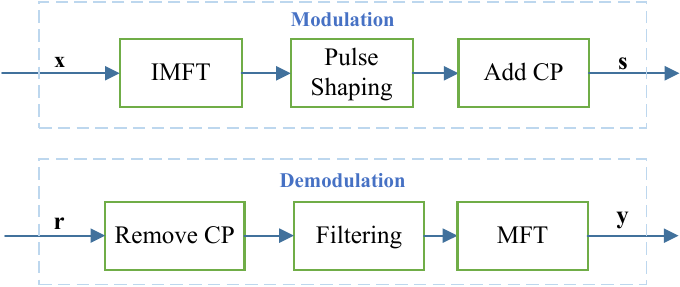}
	\caption{Diagram of ODSS.}
	\label{block11ODSS}
    \vspace{-8pt}
\end{figure}
As shown in Fig. \ref{block11ODSS}, let $\mathbf{x}^{\text{ODSS}} \in \mathbb{C}^{M_{\text{tot}} \times 1}$ denote the QAM symbols in the discrete Mellin-Fourier domain of size $M_{\text{tot}}=\sum_{n=0}^{N-1} M(n)$, where $M(n)=\left\lfloor q^n\right\rfloor$, $q$ is the sampling ratio. 

In ODSS, the lattice of the Mellin-Fourier domain can be denoted as
\begin{equation}
\Lambda^{\perp}=\{( l \Delta f, k \Delta \beta),l=0, \cdots, M(k)-1, k=0, \cdots, N-1\}.
\end{equation}
Here, $\Delta \beta=\frac{1}{N \ln q}$ and $\Delta f$ are the spacings on the Mellin and Fourier axis, respectively, where bandwidths $B = M \Delta f$.

Correspondingly, the lattice of the delay-scale domain can be denoted as
\begin{equation}
\Lambda=\{( m \Delta \tau, q^n),m=0, \cdots, M(n)-1, n=0, \cdots, N-1\},
\end{equation}
with $\Delta \tau = 1/B$ denoting delay axis and $q$ is the geometric sampling ratio on the scale axis.

After taking the inverse Mellin-Fourier transform (IMFT), i.e., an $N$-point inverse discrete Mellin transform along the scale axis, and an $M$-point discrete Fourier transform along the delay axis, the transmit signal in the delay-scale domain can be written as
\begin{equation}
     X_{[m,n]}^{\text{ODSS}} = \frac{q^{-\frac{n}{2}}}{N} \sum_{k=0}^{N-1} \frac{\sum_{l=0}^{M(k)-1} x^{\text{ODSS}}_{[l,k]} e^{j2\pi \left( \frac{m l}{M(k)} - \frac{n k}{N} \right)}}{M(k)} .
\end{equation}

Then, the ODSS modulator converts the 2D signal to a 1D continuous signal, given by:
\begin{equation}
    s^{\text{ODSS}}(t) = \sum_{n=0}^{N-1}\sum_{m=0}^{M(n)-1} X_{[m,n]}^{\text{ODSS}} \, q^{\frac{n}{2}} \, g_{\text{tx}} \left( q^n \left( t - \frac{m}{q^n W} \right) \right)
\end{equation}
where $g_{\text{tx}(t)}$ is the transmit pulse shaping function with duration $T =1/B$.

Thus, the ODSS modulation can be expressed in the matrix form as
\begin{equation}
\mathbf{s}^{\text{ODSS}}=\mathbf{G}_{\text{tx}}\mathcal{T}_{\text{iMF}} \mathbf{x}^{\text{ODSS}},
\end{equation}
where $\mathcal{T}_{\text{iMF}} \in \mathbb{C}^{M_{\text{tot}} \times M_{\text{tot}}}$ is the matrix representing the 2D ODSS transform. After adding the CP, the $\mathbf{s}^{\text{ODSS}}$ is transmitted through the channel.

At the receiver, the signal is extracted by sampling the cross-ambiguity function between the received signal $\mathbf{r}^{\text{ODSS}}$ and the receive pulse shaping function $\mathbf{G}_\text{rx}$. Then, the discrete Mellin-Fourier transform (MFT) is applied, which yields $\mathbf{y}^{\text{ODSS}}$ in the Mellin-Fourier domain:
\begin{equation}
\mathbf{y}^{\text{ODSS}}=\mathcal{T}_{\text{MF}}\mathbf{G}_{\text{rx}} \mathbf{r}^{\text{ODSS}}.
\end{equation}

\subsubsection{DDAM}
\begin{figure}[H]
	\centering
\includegraphics[width=0.4\textwidth]{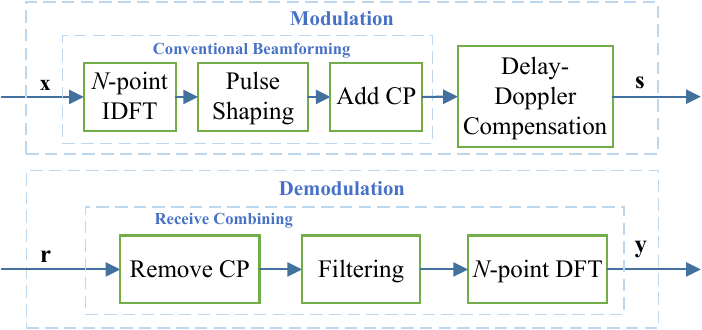}
	\caption{Diagram of DDAM.}
	\label{block12DDAM}
    \vspace{-8pt}
\end{figure}

DDAM is an extension of delay alignment modulation (DAM)\cite{lu2022delay}, and is also a 2D multicarrier modulation scheme. DDAM is usually employed in massive MIMO systems\cite{lu2024delaydoppler}, where the base station (BS) and user equipment (UE) are equipped with $M_t \gg 1$ and $M_r \geq 1$ antennas, respectively. By leveraging the high spatial resolution provided by large antenna arrays, individual multipath components can be distinguished and resolved in the spatial domain\cite{xiao2023integrated}. When multi-path components exhibit spatial orthogonality, it becomes possible to apply delay-Doppler compensation independently to each path without influencing others, meaning that each path can be manipulated separately with beamforming and delay-Doppler compensation, as shown in Fig. \ref{block12DDAM}.

Considering the narrowband DDC model demonstrated in \eqref{eqyhr0414.1}, let the minimum delay and Doppler shift among all $P$ multi-paths be denoted by ${\tau_{\min }} \buildrel \Delta \over = \mathop {\min }\limits_{1 \le i \le P} {\tau_i}$ and ${\nu_{\min }} \buildrel \Delta \over = \mathop {\min }\limits_{1 \le i \le P} {\nu_i}$, and the maximum delay and Doppler shift be denoted by ${\tau_{\max }} \buildrel \Delta \over = \mathop {\max }\limits_{1 \le i \le P} {\tau_i}$ and ${\nu_{\max }} \buildrel \Delta \over = \mathop {\max }\limits_{1 \le i \le P} {\nu_i}$. Thus, ${\tau_{{\rm{span}}}} = {\tau_{\max }} - {\tau_{\min }}$ and ${\nu_{{\rm{span}}}} = {\nu_{\max }} - {\nu_{\min }}$ represent the normalized channel delay spread and channel Doppler spread, respectively.
To perform compensation on each path, we can adjust the delay of each path by \( \kappa_i = \tau_{\text{max}} - \tau_i \), and the Doppler shift by \( -\nu_i \). Then, the transmitted signal is given by
\begin{equation}
	{\mathbf{s}}^{\text{DDAM}}_{[n]} = \sum\limits_{i = 1}^P {{{\mathbf{F}}_{i}}\mathbf{x}^{\text{DDAM}}_{[n - {\kappa _i}]}e^{-j2\pi \nu_i nT_s}},
	\label{eqDDAM-s}
\end{equation}
where the $\mathbf{x}^{\text{DDAM}}\in \mathbb{C}^{{N_\text{s}} \times 1}$ denotes the independent and identically (i.i.d.) information-bearing symbol vectors, with $\mathbb{E}[\mathbf{x}[n]\mathbf{x}^H[n]]=\mathbf{I}_{N_\text{s}}$; ${\mathbf{F}}_i\in \mathbb{C}^{M_\text{t} \times N_\text{s}}$ is the matrix denoting the transmit beamforming vector associated with path $i$, which can be achieved by maximal ratio transmission (MRT), zero-forcing (ZF), or minimum mean square error (MMSE).

After going through the channel, the received time domain signal $\mathbf{r}^{\text{DDAM}}$ is obtained. The signal for DDAM after the receive combining is
\begin{equation}
\mathbf{y}^{\text{DDAM}}=\mathbf{W}^H \mathbf{r}^{\text{DDAM}},
\end{equation}
with the receive combining matrix $\mathbf{W}\in \mathbb{C}^{{M_\text{r}} \times {N_\text{s}}}$. 

Specifically, considering the multiple-input single-output (MISO) channel with channel vector ${\mathbf{h}}_i \in \mathbb{C}^{{M_\text{t}} \times 1}$ for the $i$th multi-path, where $M_\text{t}$ is the number of transmit antennas, the channel can be expressed as
\begin{equation}
	\mathbf{h}^H[n,\tau] = \sum\limits_{i = 1}^P {\mathbf{h}}^H_i e^{j2\pi \nu_inT_s} \delta [\tau - {\tau_i}].
	\label{eqDDAM-ChannelMI}
\end{equation}

By first performing delay-Doppler compensation and then further applying path-based beamforming, the processed equivalent channel can be expressed as \eqref{eqDDAM-ChannelMI2}, shown at the top of the next page.
\begin{figure*}[thpb]
\centering
\begin{equation}
\begin{split}
        {\mathbf{\tilde h}}^H[n,\tau] &= \sum\limits_{i = 1}^P \mathbf{h}^H[n,\tau-\kappa_i]e^{-j2\pi \nu_inT_s}\mathbf{F}_i 
		= \sum\limits_{i = 1}^P\sum\limits_{i^\prime = 1}^P \mathbf{h}^H_{i^\prime}\mathbf{F}_i e^{j2\pi (\nu_{i^\prime}- \nu_i)nT_s} \delta[\tau - {\tau_{i^\prime}} - \kappa_i]\\
		&= \underbrace{\sum\limits_{i = 1}^P \mathbf{h}^H_{i}\mathbf{F}_i \delta [\tau - \tau_{\max }]}_{\text{compensated multi-paths}} 
		+\underbrace{\sum\limits_{i = 1}^P\sum\limits_{i^\prime \neq i}^P \mathbf{h}^H_{i^\prime}\mathbf{F}_i e^{j2\pi (\nu_{i^\prime}- \nu_i)nT_s} \delta [\tau - \tau_{\max } +({\tau_i}- {\tau_{i^\prime}})]}_{\text{undesired components}}
	\end{split}  
     \label{eqDDAM-ChannelMI2}
\end{equation}
\vspace{-16pt}
\end{figure*}

While undesired components exist in~\eqref{eqDDAM-ChannelMI2}, they can be mitigated by properly designing $\mathbf{F}_i$ in the MISO system.
Specifically, if $\{\mathbf{F}_i\}_{i=1}^P$ is designed so that
\begin{equation}
	\mathbf{h}^H_{i^\prime}\mathbf{F}_i = \mathbf{0}_{1 \times N_\text s}, \forall i \neq i^\prime,
	\label{eqDDAM-ISI-ZF}
\end{equation}
then the channel in~\eqref{eqDDAM-ChannelMI2} reduces to
\begin{equation}
	\begin{split}
		{\mathbf{\tilde h}}^H[\tau] =& \sum\limits_{i = 1}^P \mathbf{h}^H_{i} \mathbf{F}_i\delta [\tau - \tau_{\max }].
		\label{eqDDAM-ChannelMIZF}
	\end{split}
\end{equation}

Note that the manipulation of the channel in~\eqref{eqDDAM-ChannelMI2} can be equivalent to the processing of the transmitted signal or received signal. Thus, DDAM can be implemented either at the transmitter, at the receiver, or simultaneously at both ends~\cite{wang2025doubleside}. Moreover, ${\bf{x}}[n]$ can be signals modulated using other methods, such as OFDM or OTFS. This implies that DDAM and other waveforms are not exclusive. Combining DDAM with other waveforms is referred to as $\textit{DDAM} + \textit{``X''}$, where $\textit{``X''}$ could be any existing single- or multicarrier waveforms~\cite{xiao2025rethinking}. 

\begin{figure*}[bhp]
    \centering
    \includegraphics[width=\textwidth]{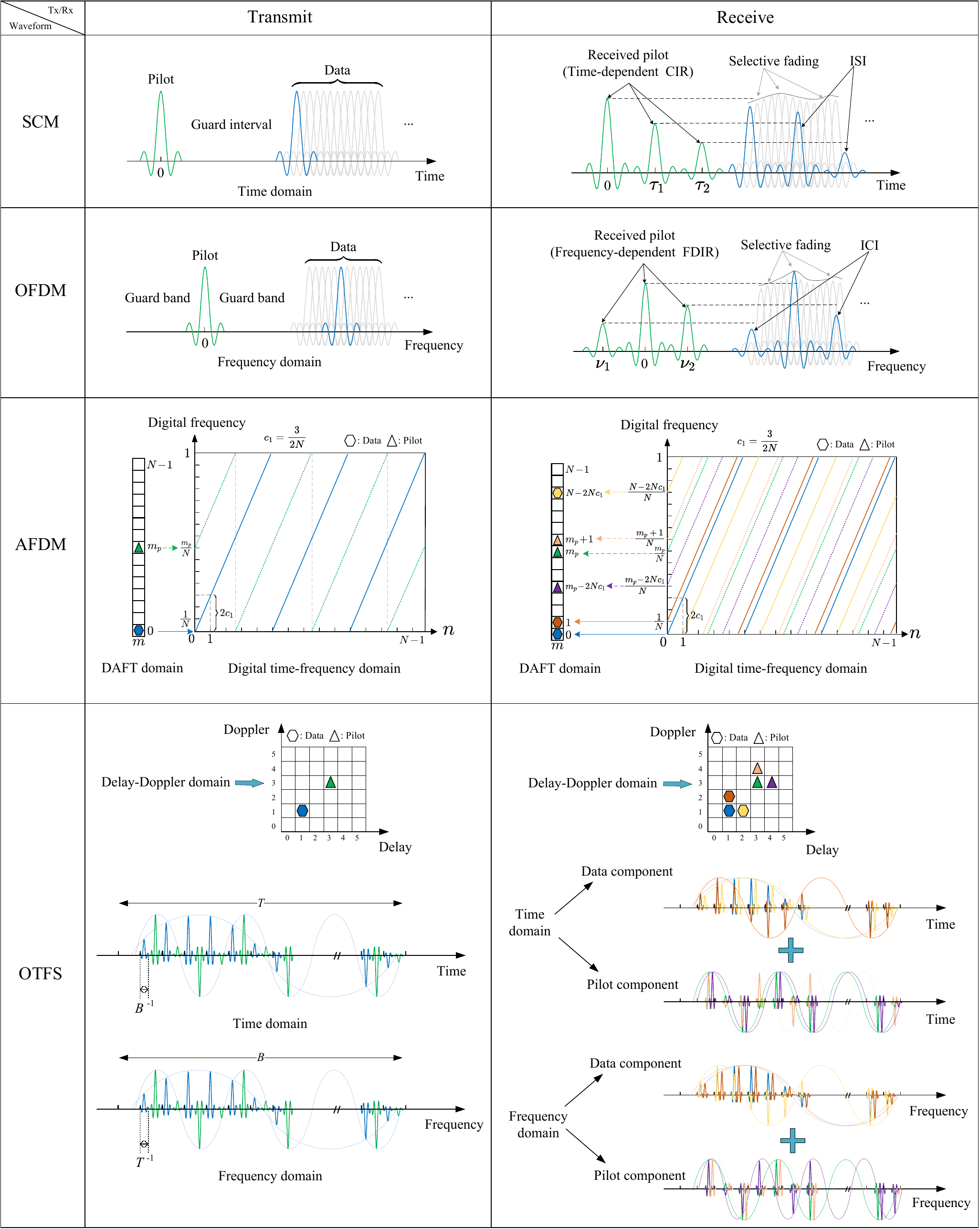}
    \caption{Influence of different dispersions on SCM, OFDM, AFDM, and OTFS.}
    \label{fig4-2}
\end{figure*}

\section{Generalized Modulation-domain Interference Analysis}
\label{sec4}
In Section \ref{sec3}, we have clarified the modulation methods for various waveforms. When new waveforms are introduced, using different domains for signal processing can lead to different perceptions of channels and their effects. \new{Depending on the signal representation domain, this difference can be either beneficial or detrimental. In particular, operating in certain modulation domains (e.g., DD or chirp) can render the channel response more sparse and separable, thereby facilitating channel estimation and equalization. However, due to the resolution limitation imposed by the uncertainty principle, the increased domain diversity may also introduce higher receiver complexity and potential interference coupling.} The most critical consideration is how different waveforms handle the channel-introduced interference. In this section, we will take a broader look at interference suppression, discussing its generation, classification, and solutions in OFDM and several other representative waveforms, aiming to provide a deeper understanding of the essence of waveform design.
\vspace{-5pt}
\subsection{Modulation-domain ISI}
\label{sec4-1}
\subsubsection{ISI}
ISI arises as a fundamental impairment in wireless communication systems when consecutive symbols overlap in the time domain, disrupting their independent detection at the receiver. In single-carrier modulation (SCM) systems, it manifests as multiple copies of the same symbol scattered across the time axis (see Fig. \ref{fig4-2}). When the channel delay spread $\tau_{\max}$ exceeds the symbol duration $T_s$ (i.e., $\tau_{\max}$ is comparable to $T_s$ or is not much smaller than $T_s$), the tail of one symbol's response extends into the time slot of the next symbol, causing ISI. This overlap destroys the temporal boundaries between symbols, complicating their separation at the receiver.

The generation of ISI is further exacerbated in DDC, where Doppler effects introduce time variations in \eqref{eqyhr0414.1}, causing the channel response to shift dynamically within a symbol duration. This not only broadens the effective delay spread but also introduces additional interference between symbols, as their relative timing is no longer fixed. Consequently, ISI emerges as a combined result of the transmitted signal’s temporal structure and the channel’s dispersive nature, necessitating waveform designs, such as the inclusion of a CP in OFDM, to mitigate its impact.

\subsubsection{ICI}
ICI emerges in multicarrier systems when the orthogonality between subcarriers is compromised, leading to unexpected energy leakage across frequency bands and degrading signal detection. 
Consider a multicarrier waveform like OFDM, where the transmitted signal is constructed by modulating $M$ information symbols onto $M$ orthogonal subcarriers via an IDFT, expressed as \eqref{eqOFDM-1}, with subcarrier spacing $\Delta f=1 / T_s$. Under ideal conditions, the orthogonality condition:
\begin{equation}
    \int_0^{T_s} e^{j 2 \pi k \Delta f t} e^{-j 2 \pi m \Delta f t} \text{d} t=0 ( k \neq m ) 
\end{equation}
ensures that subcarriers remain orthogonal, preventing interference between them at the receiver. However, under the DDC, the Doppler-induced frequency shift $f_D$ causes the subcarrier frequencies to deviate from their nominal values (e.g., $k \Delta f+f_D$). This shift results in a loss of orthogonality, which means that
\begin{equation}
    \int_0^{T_s} e^{j 2 \pi\left(k \Delta f+f_D\right) t} e^{-j 2 \pi m \Delta f t} \text{d} t \neq 0 ( k \neq m ). 
\end{equation}
This manifests itself as spectrum leakage, where energy from one subcarrier spills into adjacent subcarriers. In OFDM, ICI is partially addressed by inserting a CP longer than the delay spread to maintain subcarrier orthogonality within a static frame. To further combat ICI, advanced equalization techniques (e.g., frequency domain equalization with Doppler compensation) are employed, although at increased complexity. 

\subsubsection{DAFT-domain ISI}
Unlike SCM and OFDM, AFDM carries information symbols into the DAFT domain for modulation. This means that the interference we need to consider is limited to that in the DAFT domain. This interference is essentially a form of ISI, which we refer to as DAFT-domain ISI. As shown in Fig. \ref{fig4-2}, the frequencies of the subcarriers in the digital TF domain appear as a series of straight lines with the same slopes in AFDM. These slopes are uniquely determined by the AFDM frequency modulation parameter $c_1$. In the DAFT domain, different copies of the same signal are also received, which is consistent with the principle of ISI generation. Adjusting the $c_1$ parameter enhances multipath discrimination capability, thereby mitigating DAFT-domain ISI.

\subsubsection{DD-domain ISI}
For the DD domain, the generation of interference is similar. Taking OTFS as an example, assuming that the channel model in \eqref{eqyhr0414.1} has 5 paths, with a maximum normalized delay of 4 and a maximum normalized Doppler shift of 2. Then, the components at the receiver end will be scattered simultaneously in both the delay dimension and the Doppler dimension, as shown in Fig. \ref{fig4-2}. At this point, if we separate the pilot components and data components and observe them separately on the time axis and frequency axis, we can again find that the interference caused by the superposition of different symbols is consistent with the cause of ISI. They appear as a series of pulse trains modulated by tones with different amplitudes in the TF domain, and if we observe them in the delay-Doppler plane, they appear as localized pulses on the lattice. Due to the influence of the DDC, the boundaries between the impacts become unclear, which can be called DD-domain ISI.

\begin{figure*}[htbp]\new
    \centering
    \subfloat[SCM-TDC]{
         \includegraphics[scale=0.35]{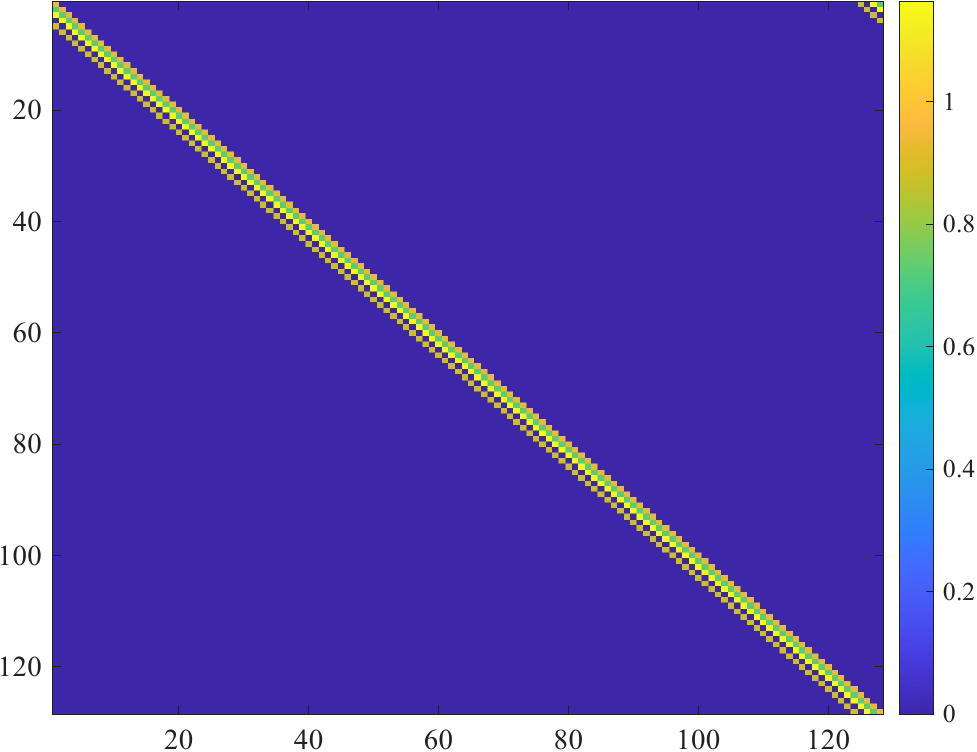}\label{ch_t_SCM}
    }
    \hfill
    \subfloat[SCM-FDC]{
         \includegraphics[scale=0.35]{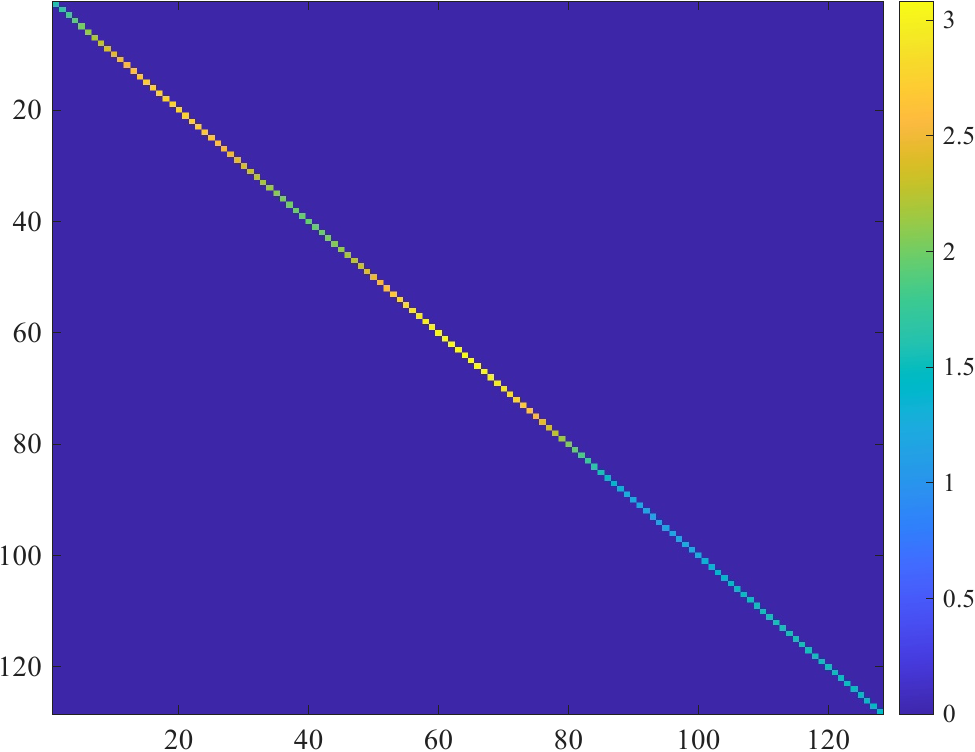}\label{ch_f_SCM}
    }
    \hfill
    \subfloat[SCM-DDC]{
         \includegraphics[scale=0.35]{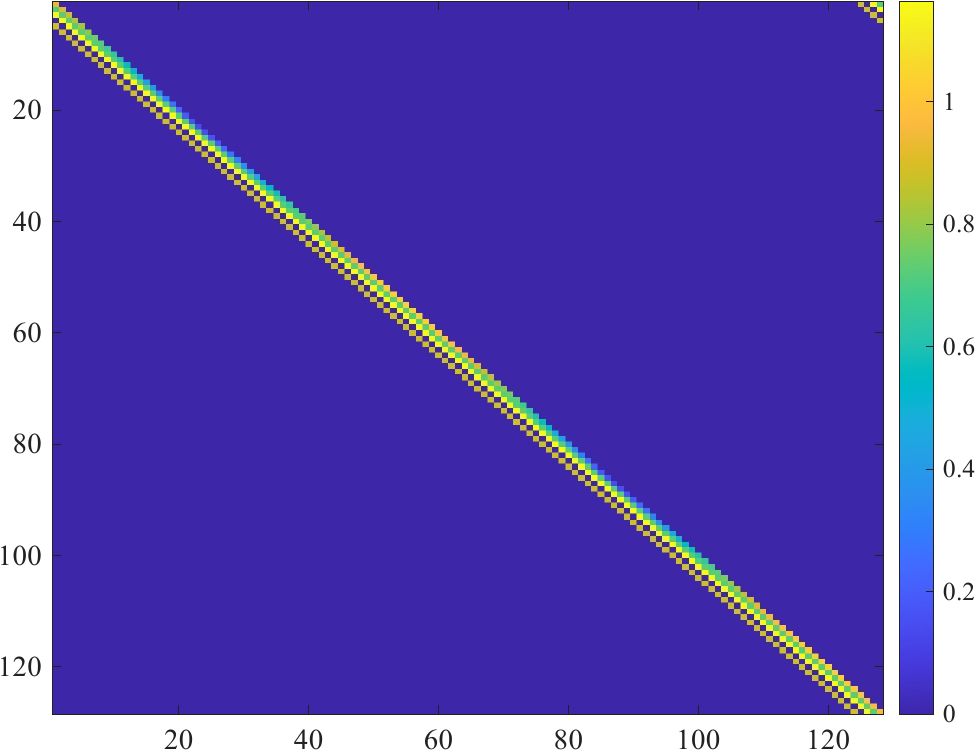}\label{ch_d_SCM}
    }
    \\

    \subfloat[OFDM-TDC]{
        \includegraphics[scale=0.35]{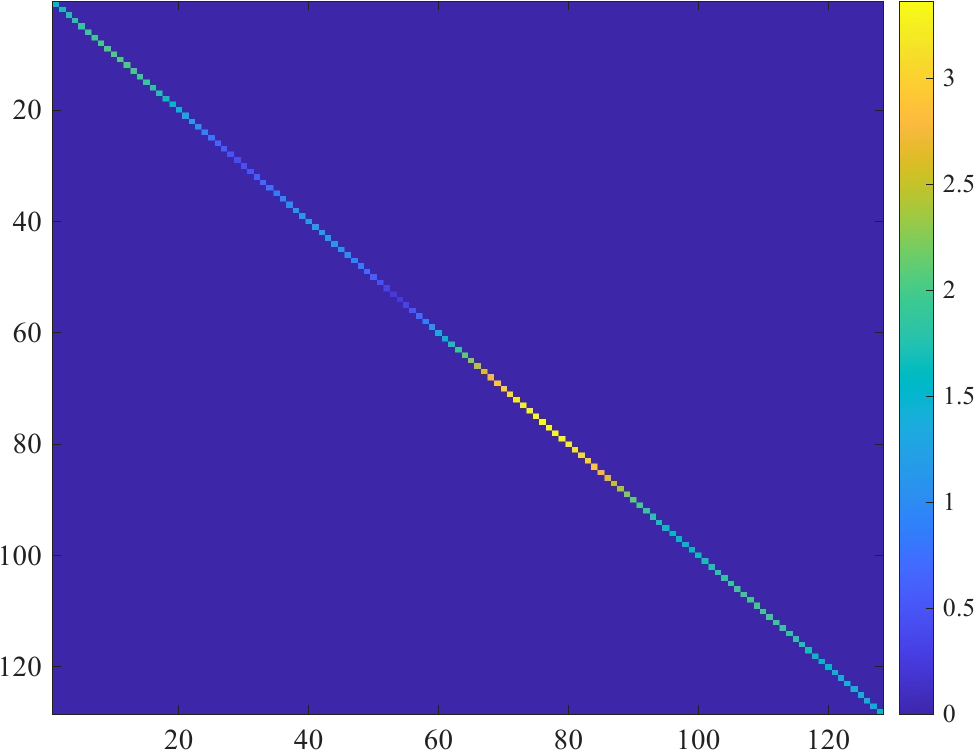}\label{ch_t_OFDM}
    }
   \hfill
    \subfloat[OFDM-FDC]{
        \includegraphics[scale=0.35]{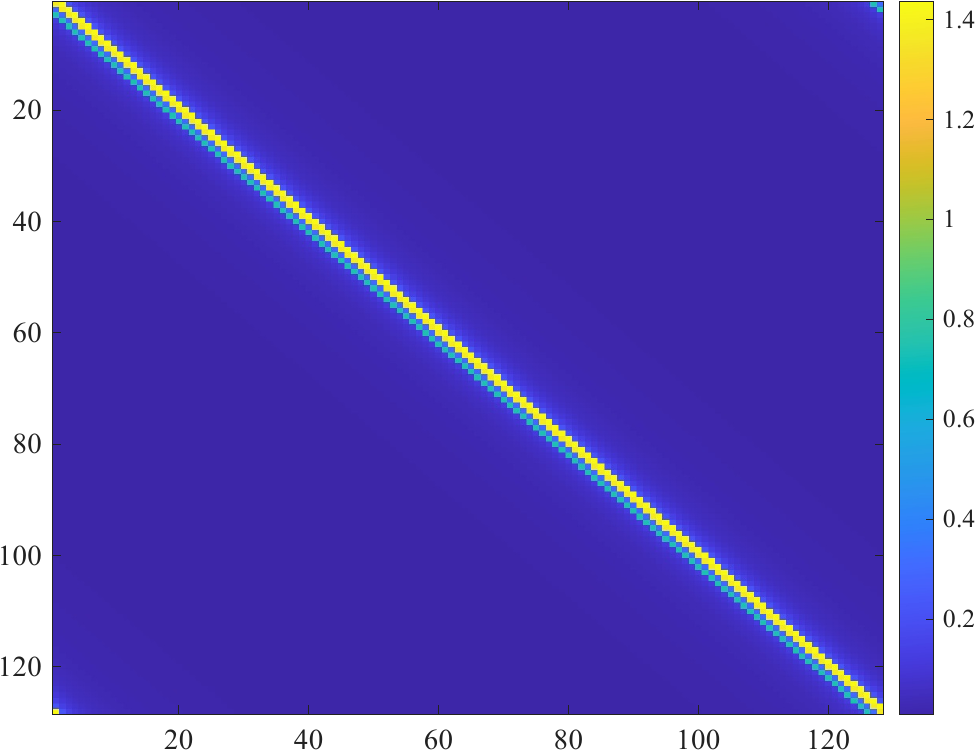}\label{ch_f_OFDM}
    }
   \hfill
   \subfloat[OFDM-DDC]{
        \includegraphics[scale=0.35]{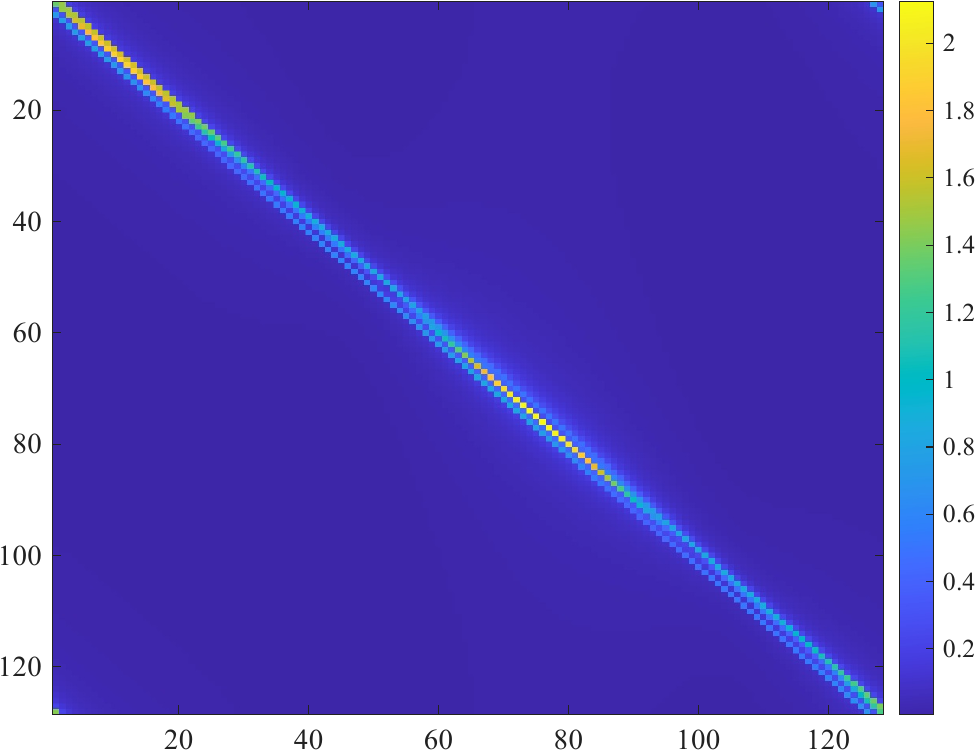}\label{ch_d_OFDM}
    }
    \\

    \subfloat[AFDM-TDC]{
         \includegraphics[scale=0.35]{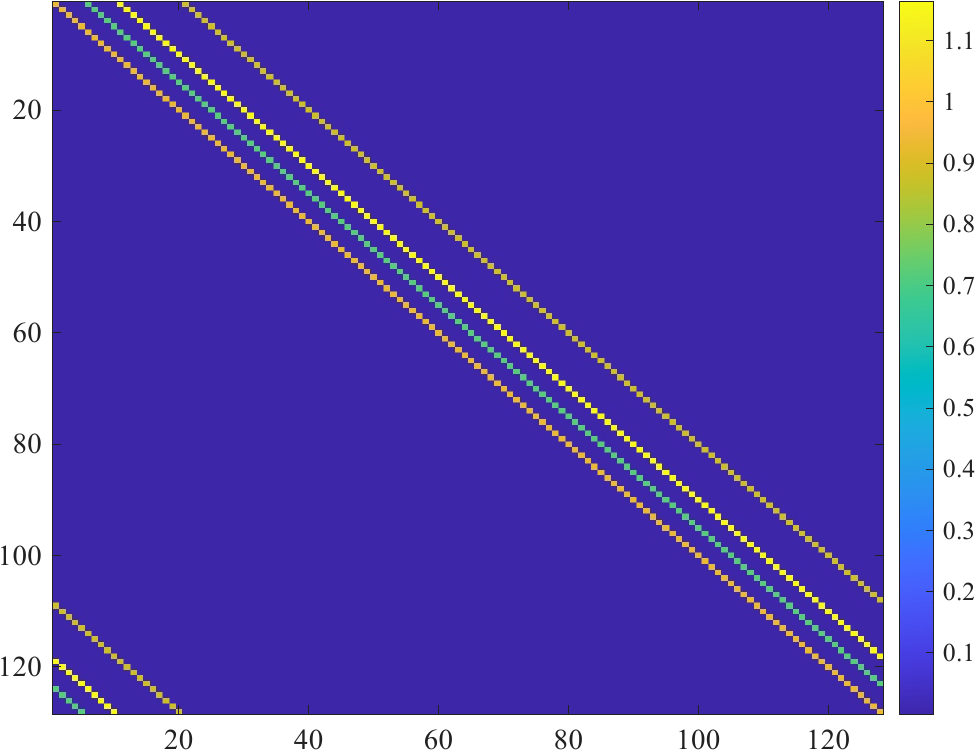}\label{ch_t_AFDM}
    }   
    \hfill
    \subfloat[AFDM-FDC]{
         \includegraphics[scale=0.35]{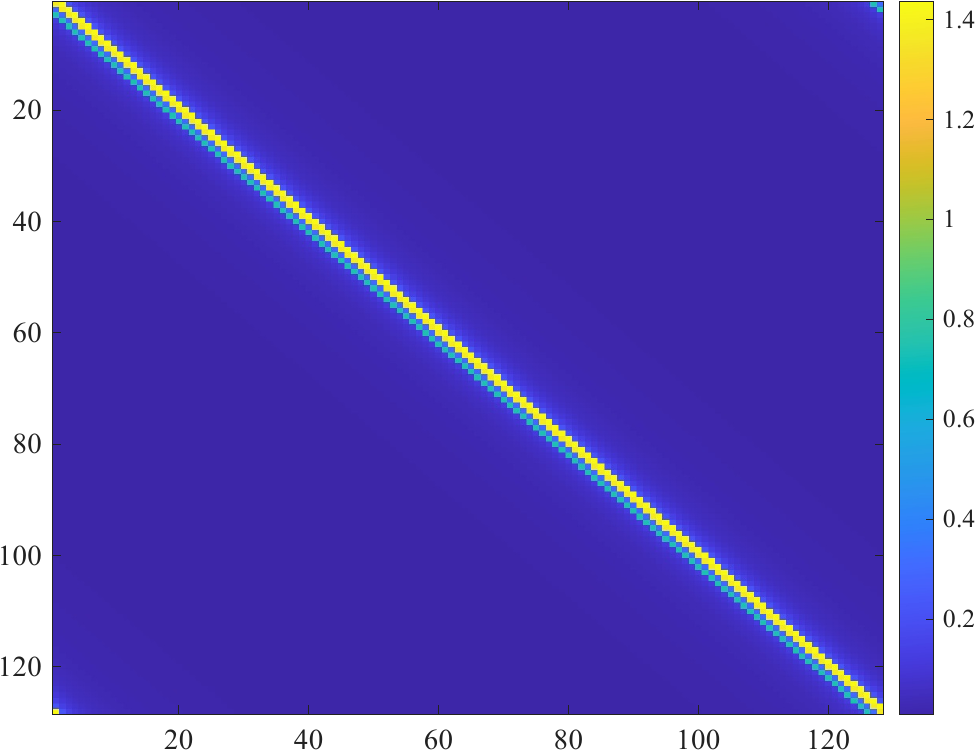}\label{ch_f_AFDM}
    }   
    \hfill
    \subfloat[AFDM-DDC]{
         \includegraphics[scale=0.35]{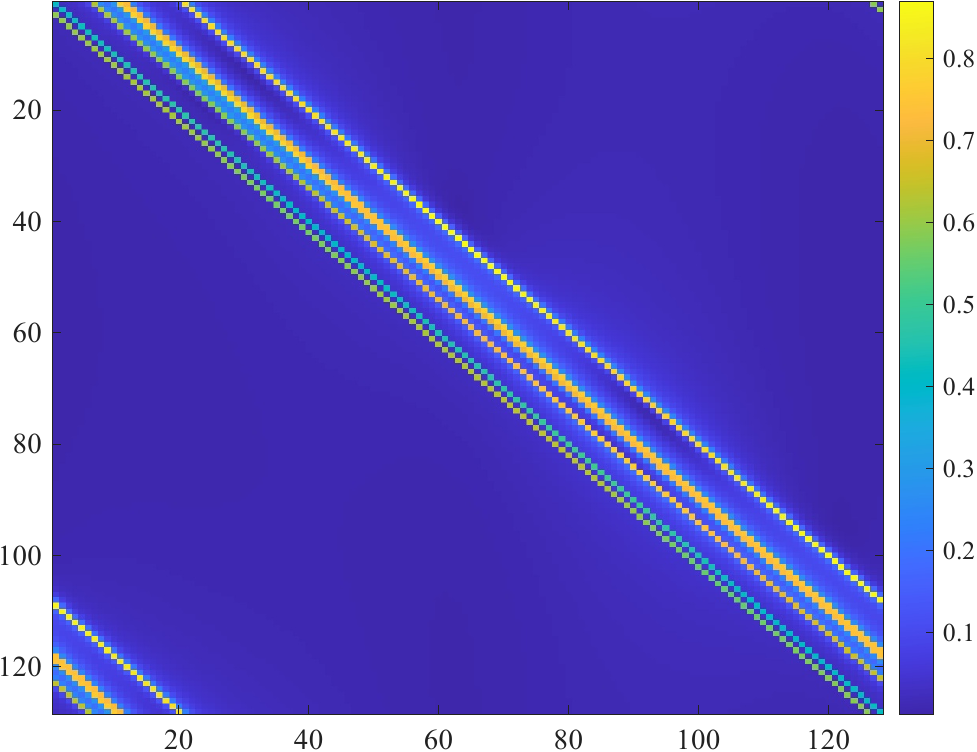}\label{ch_d_AFDM}
    }   
    \\

    \subfloat[OTFS-TDC]{
         \includegraphics[scale=0.35]{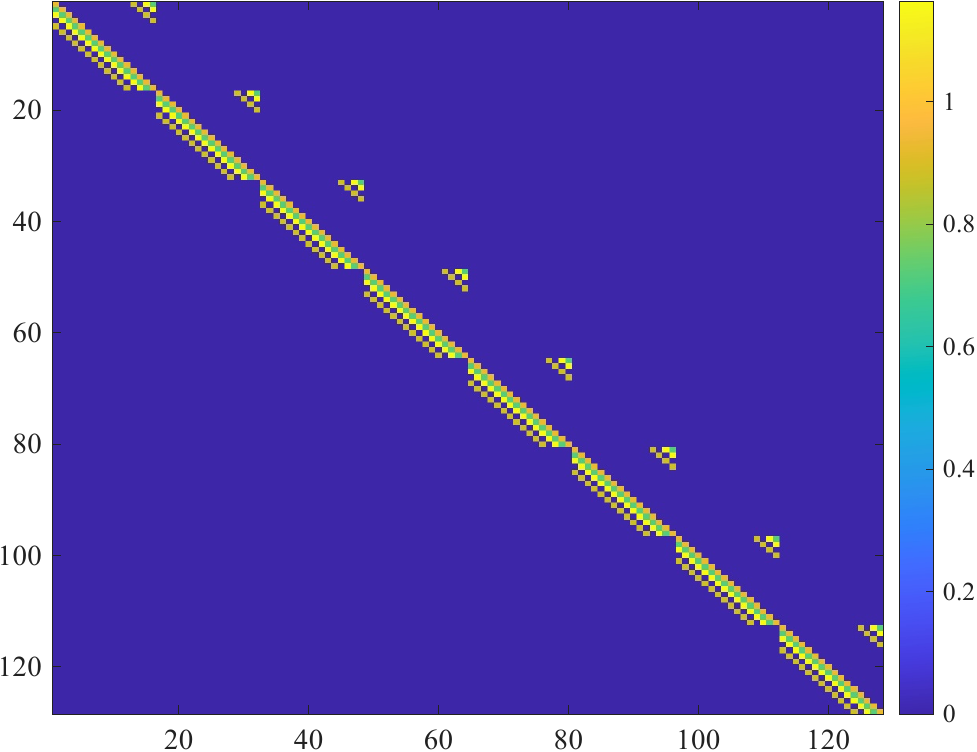}\label{ch_t_OTFS}
    } 
    \hfill
    \subfloat[OTFS-FDC]{
         \includegraphics[scale=0.35]{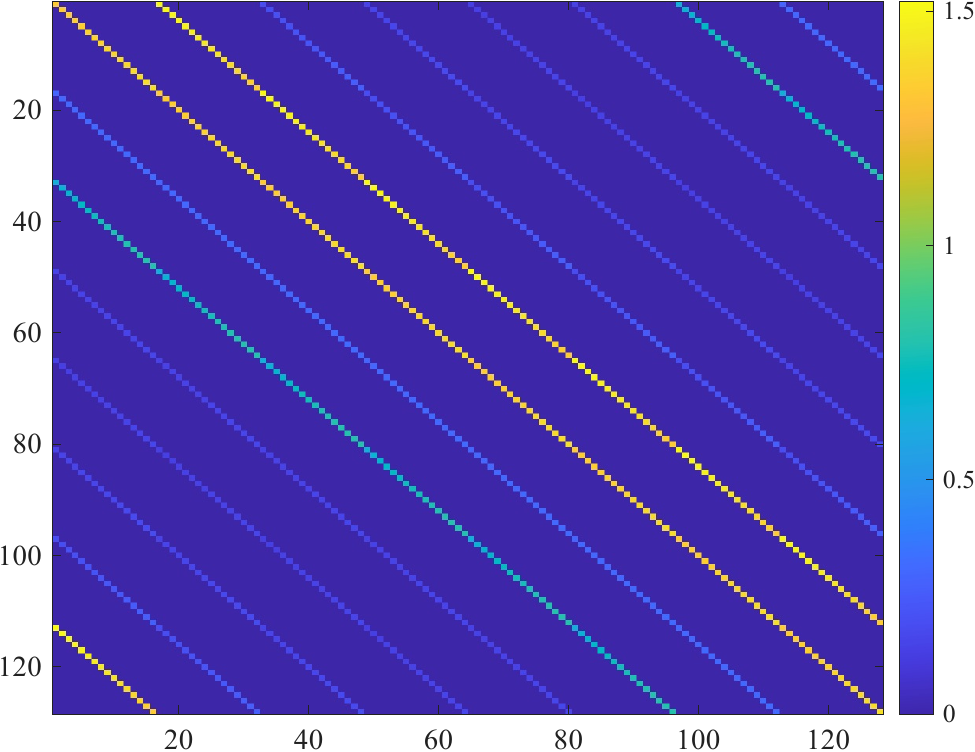}\label{ch_f_OTFS}
    } 
    \hfill
    \subfloat[OTFS-DDC]{
         \includegraphics[scale=0.35]{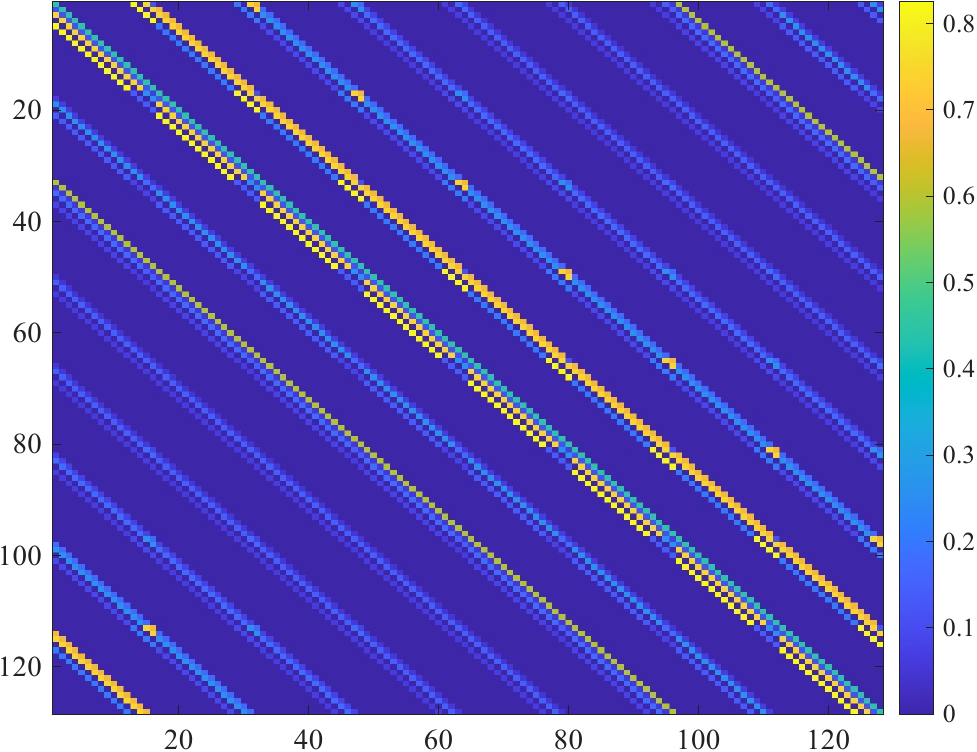}\label{ch_d_OTFS}
    } 
  
    \caption{Equivalent channel matrix structure of SCM, OFDM, AFDM, and OTFS under different channel conditions. For DDC, consider the channel with $P=5$ paths, with corresponding delays $\tau_i$, and path velocity $v_i$, which can be converted into Doppler shift $\nu_i$. Here, \{$\tau_0=0, \tau_1=0, \tau_2=0.39, \tau_3=1.17, \tau_4=2.34 $\} $\times 10^{-6}s$, and \{$v_0=0, v_1=-1080, v_2=648, v_3=270, v_4=108$\}$km/h$. For TDC and FDC, the delay and Doppler shift are consistent with the DDC settings, respectively. To ensure the same bandwidth during comparison, the carrier frequency is set to $24$ GHz, with a subcarrier spacing of $12$ kHz. The system size parameters are $M=128$ for SCM, OFDM, and AFDM, and $M=16, N=8$ for OTFS, respectively.}
    \label{fig4-1}
\end{figure*}

\begin{table}[tbhp]\new
\renewcommand\arraystretch{1.3}
\centering
\begin{threeparttable}
\caption{Parameter Assumptions for Performance Comparison}
\label{tab4-2}

\begin{tabular}{|m{0.4cm}|m{0.9cm}|m{1.05cm}|m{0.5cm}|m{0.6cm}|m{2.65cm}|}
\hline
Figs    & $B/f_c$ & $\tau_\text{max}$ (s)    & $\nu_\text{max}$ (kHz) & $f_s$ (MHz) & Normalization  \\ \hline

Fig. 18 & $6.4$e-5  & $2.34$e-6 & 24   & 1.536 
& $\Delta \hat{\tau}_\text{norm}=6.5104$e-7 s; $\Delta \hat{f}_\text{norm} = 12$kHz \\ \hline

Fig. 19 & $1.28$e-4 & $2.51$e-6 & 12   & 3.072
& $\Delta \hat{\tau}_\text{norm}=3.2552$e-7 s; $\Delta \hat{f}_\text{norm} = 3$kHz  \\ \hline

Fig. 20 & $4.57$e-3 & $312.5$e-9 & 0   & 128   
& $\Delta \hat{\tau}_\text{norm}=7.8125$e-9 s \\ \hline
\end{tabular}

\begin{tablenotes}
\footnotesize
\item \textit{Note:} $f_s$ denotes the equivalent discrete-time baseband sampling rate used in simulations, determined by $M\Delta f$. 
It is not the physical ADC sampling rate (which must satisfy $f_s \ge 2B$ in practical RF systems). 
All simulations in this paper are conducted at baseband using discrete-time multicarrier models.
\end{tablenotes}

\end{threeparttable}
\end{table}

Fig. \ref{fig4-1} illustrates the equivalent channel matrix structure of four typical waveforms under different channel conditions, showing that the equivalent channel matrices of AFDM and OTFS can distinguish the interference from different channel delay and Doppler components while exhibiting sparsity compared to SCM and OFDM. \new{To make the modeling assumptions explicit, Table \ref{tab4-2} summarizes the bandwidth-to-carrier ratio $B/f_c$, maximum delay $\tau_\text{max}$, maximum Doppler $\nu_\text{max}$, sampling rate $f_s$, and the normalization used for resolution metrics $\Delta \hat{\tau}_\text{norm}$ and $\Delta \hat{f}_\text{norm}$. Notably, the reported $B/f_c$ confirms operation in the narrowband regime for these cases, which means all simulated channels can be modeled using \eqref{eqyhr0414.1}.}

In general, the DD spread from the DDC introduces symbol spread in all waveforms in their underlying modulation domain, where the modulation domain refers to the domain where the symbols are multiplexed. Specifically, delay spread introduces time-domain symbol spread in SCM, Doppler spread introduces frequency-domain symbol spread in OFDM, and DD spreads induce DAFT-domain symbol spread in AFDM and DD-domain symbol spread in OTFS. These symbol spreads essentially cause ISI in the underlying modulation domains, as shown in Table \ref{tab4-1}, i.e., the time-domain ISI in SCM, the frequency-domain ICI in OFDM, the DAFT-domain ISI in AFDM, and the DD-domain ISI in OTFS. Insightfully, we refer to these interferences as MD-ISI. Consequently, each received modulation-domain symbol is determined by multiple transmitted modulation-domain symbols, which greatly complicates the symbol detection process at the receiver.

\begin{table}[ht]
\vspace{-12pt}
	\renewcommand\arraystretch{1}
	\centering
	\caption{Modulation-Domain ISI of Existing Waveforms}
	\label{tab4-1}
	\begin{tabular}{|c|c|c|}
\hline
\textbf{Dimension}  & \textbf{Waveforms} & \textbf{Modulation domain}     \\ \hline
\multirow{7}{*}{1D} & SCM               & Time                           \\ \cline{2-3} 
                    & OFDM              & \multirow{2}{*}{Frequency}     \\ \cline{2-2}
                    & DFT-s-OFDM        &                                \\ \cline{2-3} 
                    & FrFT-OFDM         & Fractional-frequency          \\ \cline{2-3} 
                    & OCDM              & Chirp                          \\ \cline{2-3} 
                    & IFDM              & Interleave-frequency           \\ \cline{2-3} 
                    & AFDM              & DAFT                           \\ \hline
\multirow{6}{*}{2D} & FBMC              & TF                 \\ \cline{2-3} 
                    & OTFS              & \multirow{3}{*}{Delay-Doppler} \\ \cline{2-2}
                    & ODDM              &                                \\ \cline{2-2}
                    & DDAM              &                                \\ \cline{2-3} 
                    & OTSM              & Delay-sequence                 \\ \cline{2-3} 
                    & ODSS              & Delay-scale                    \\ \hline
\end{tabular}
\vspace{-12pt}
\end{table}

\vspace{-5pt}
\subsection{Key Solutions against Modulation-domain ISI}
\label{sec4-2}
As demonstrated in the preceding subsection, the channel-induced MD-ISI causes modulation-domain symbols to be coupled at the receiver; thus, the single-tap equalization can not be applied directly. Instead, sequence-wise joint equalization should be performed, which requires the knowledge of the MD-ISI pattern, i.e., the modulation-domain input-output relationship (IOR) between the transmitted and received data symbols. To facilitate this process, three essential procedures, namely synchronization, channel estimation, and modulation-domain pulse shaping, should be conducted collaboratively.

\subsubsection{Synchronization}
Synchronization involves estimating and compensating for the timing offset (TO) and CFO. TO estimation in multicarrier waveforms refers to determining the starting point of each symbol, which enables the maximum energy collection of the desired signal with a limited sampling rate while avoiding ISI. In particular, the TO estimation process in multicarrier waveforms differs significantly from that in single-carrier systems, as there is no clear “eye-opening” to identify the optimal sampling point. Instead, each waveform symbol usually contains thousands of samples, as the required number of samples is proportional to the number of subcarriers. The use of a CP introduces tolerance to symbol timing errors and preserves orthogonality among subcarriers by extending the symbol duration. Moreover, accurate carrier frequency synchronization at the receiver is crucial. OFDM systems are very sensitive to CFO since they can only tolerate offsets that are a fraction of the spacing between the subcarriers without significant degradation in system performance. For Doppler-tolerant waveforms like AFDM and OTFS, inaccurate timing and carrier frequency synchronization may exaggerate the effective delay and Doppler spreads, respectively, which subsequently increase the channel estimation overhead and lower the spectral efficiency. Recent studies have proposed various approaches to compensate TO and CFO, including joint TO and CFO estimation by using delay-time periodic pilot structures without additional overhead, comb-type preamble waveforms designed to eliminate ambiguity in initial time synchronization, and pilot-aided methods leveraging the autocorrelation of maximum length sequences (MLS) for robust TO and channel estimation. These advancements enable OTFS systems to achieve near-ideal performance even under severe channel dynamics \cite{chung2024initial,sun2024pilot,bayat2022time}.

\subsubsection{Channel Estimation}
\new{Channel estimation is one of the most challenging tasks in practical communication systems, especially for high-mobility scenarios, as it is critical not only for accurate signal detection at the receiver but also for enabling the transmitter to make informed decisions. The transmitter typically acquires channel state information (CSI) through a pilot-based channel estimation process, where known pilots are sent along with the data, allowing the receiver to estimate channel parameters, i.e., the complex gains, delay shifts, and Doppler shifts of all propagation paths. Then, the acquisition of CSI at the transmitter is achieved either by exploiting channel reciprocity in time division duplex (TDD) systems or through explicit quantized feedback (e.g., using codebooks) over the uplink control channel in frequency division duplex (FDD) systems. Once acquired, this information enables the transmitter to make optimal decisions, such as optimizing transmission power, modulation schemes, and beamforming. Depending on the modulation domain and channel dynamics, different pilot structures can be employed to acquire and track the CSI efficiently.} For example, a time-domain pilot followed by a guard interval can be used in SCM to estimate the channel impulse response (CIR) of the channel. However, the CIR is time-dependent in the DDC channel, making the CIR estimated from the pilot symbol fail to correspond with that of the subsequent data signals. Therefore, it is necessary to send pilots to track the channel frequently. For OFDM, a frequency-domain pilot surrounded by guard bands can be adopted to estimate the frequency-domain impulse response (FDIR). However, in the case of DDC shown in Fig. \ref{fig4-2}, the FDIR is frequency-dependent, which means that the FDIR estimated by the pilot does not cohere with that of the data signals residing at a different frequency. Therefore, both SCM and OFDM are incapable of managing the unavoidable interference in DDC scenarios, due to their limitations in their modulation-domain IOR estimation.

In contrast, the modulation-domain pilot-aided channel estimation in AFDM and OTFS is more effective and efficient, as demonstrated by the embedded pilot-aided (EPA) channel estimation scheme proposed in \cite{bemani2023affinea, raviteja2019embedded,ranasinghe2025joint,li2022joint}. The rationale behind this is that the DD-domain and DAFT-domain channel representations of the DDC are DD-path-separable, quasi-static, and compact, which can be interpreted by the bijective relationship between these two domains \cite{9880774}. In particular, the DD-path-separable and quasi-static properties guarantee the efficient estimation of the channel parameters via a modulation-domain pilot, as demonstrated in Fig. \ref{fig4-2}. Moreover, the compact characteristic indicates a relatively small dispersion range in the underlying modulation domain, which enables a structure of embedding a guard-protected pilot within data symbols to enhance spectral efficiency. Furthermore, we can discard the adoption of guard symbols and gradually eliminate the mutual interference between the received pilot and data symbols by iteratively performing decision feedback channel estimation and signal detection to further reduce the channel estimation overhead \cite{mishra2021otfs,zhou2024gifree,yuan2021data,zheng2024channel}.

While the aforementioned methods focus on acquiring the channel parameters of all propagation paths and then calculating the modulation-domain effective channel matrix (ECM) to obtain the modulation-domain IOR, another effective way is to directly estimate the modulation-domain ECM. For example, the authors in \cite{yin2024diagonally} proposed an embedded pilot-aided diagonal reconstruction (EPA-DR) scheme that reconstructs the ECM directly from the received pilot symbols by exploring the diagonal reconstructability of AFDM ECM. By doing so, a more accurate ECM can be obtained with much lower computation complexity.

\subsubsection{Modulation-domain Pulse Shaping}
In general, the transmission process of all modulation waveforms can be interpreted as loading the symbols on a series of pulses that reside in the underlying modulation domain \cite{jung2007wssus}. Orthogonality and sidelobe decay speed are two key indicators that characterize modulation-domain pulses. \new{Specifically, assuming $g_{m k}(t)$ and $\gamma_{n l}(t)$ represent the transmit and receive pulses respectively, as given by \eqref{eq:g_lk(t)}, we consider the pulse design to be orthogonal or bi-orthogonal if $\left\langle g_{m k}(t), \gamma_{n l}(t)\right\rangle= \delta_{m n} \delta_{k l}$ is satisfied\cite{ bayat2024generalized,li2023pulse}; otherwise, it is non-orthogonal. In orthogonal designs, the same prototype filter is used at the transmitter and the receiver (\( p_{\mathrm{tx}}(t) = p_{\mathrm{rx}}(t) \)), while bi-orthogonal designs relax this constraint by allowing different prototype filters at the two ends (\( p_{\mathrm{tx}}(t) \neq p_{\mathrm{rx}}(t) \))\cite{sahin2014survey}. For example, SCM often employs Nyquist pulses (e.g., raised-cosine) to ensure strict time-domain orthogonality, while OFDM adopts rectangular pulses in the time domain for subcarrier orthogonality and achieves Nyquist pulse shaping in the frequency domain, where each subcarrier is shaped by a sinc-like spectrum to maintain orthogonality. In contrast, FBMC employs well-localized prototype filters to mitigate ICI and maintain real-field orthogonality; ODDM uses DDOP that are locally orthogonal to the delay and Doppler resolutions of the channel, which allows for achieving near bi-orthogonality while maintaining robustness against DDC\cite{lin2022orthogonal, lin2022delaydoppler}.}

Moreover, the sidelobe decay speed determines the degree of energy compaction of a pulse, where a quick sidelobe decay speed is preferred to alleviate the inter-pulse interference at the receiver after digital sampling. This has a significant impact on the degree of coupling among the data symbols and the pilot-data separability. Therefore, the MD-ISI at the receiver is the joint effect of channel-induced spreads and the pulse that we adopt at the transmitter. 

The channel condition significantly influences the choice of pulse shaping. Specifically, in the absence of modulation-domain pulse spread, for example, OFDM in LTI channels, orthogonality among the modulation-domain pulses is of the most importance. This is because it avoids the introduction of MD-ISI and hence enables single-tap equalization. While in the presence of modulation-domain pulse spread, the orthogonality among the transmitted modulation-domain pulses is not necessary. This can be attributed to the fact that the modulation-domain pulse spread will eventually destroy the orthogonality among the received modulation-domain pulses, regardless of whether the original transmitted pulses are orthogonal to each other or not. In this case, it is the sidelobe decay speed that deserves more attention to alleviate the MD-ISI, even at the cost of losing the orthogonality \cite{wei2021transmitter,yin2024evaluation}.

\section{Comparative KPI-based Analysis}
\label{sec5}
In the evolving paradigm of ISAC, where waveforms are designed to simultaneously fulfill communication and sensing objectives, quantifying performance through KPIs is a critical step for fair and systematic evaluation. By establishing a comprehensive set of KPIs, including CP overhead, BER, spectral efficiency, PAPR, and ambiguity functions, this section enables a sufficient comparison of diverse waveforms, such as 1D schemes like OFDM and 2D schemes like OTFS. Such comparisons not only reveal the trade-offs inherent in waveform design but also guide the development of a unified waveform framework capable of adapting to varying channel conditions and application requirements.

\vspace{-5pt}
\subsection{CP Overhead}
\label{sec5-1}
CP is a redundancy added to each OFDM symbol to combat ISI caused by multi-path propagation. It acts as a guard interval, ensuring that delayed versions of the transmitted block do not cause overlap with the subsequent blocks. To ensure the orthogonality of the received signal, the CP length, denoted by $T_\text{cp}$, should be no smaller than the delay spread, i.e., $T_\text{cp} \geq \tau_d$. However, CP introduces spectral inefficiency, as a portion of the transmitted power is wasted on redundant information. The CP overhead is typically expressed as
\begin{equation}
	\rho = \frac{T_\text{cp}}{T^\prime_\text{s}+T_\text{cp}},
	\label{CPoverhead}
\end{equation}
where $T^\prime_\text{s}=\frac{1}{\Delta f}$ denotes the OFDM symbol duration without CP and ${\Delta f}$ denotes the subcarrier spacing. To achieve high spectral efficiency, we need to make $\rho \to 0$, which implies minimizing $T_\text{cp}$. However, this is constrained by the requirement that $T_\text{cp} \geq \tau_d$, meaning that \(T_\text{cp}\) cannot be arbitrarily reduced if the channel exhibits significant delay spread. Therefore, an effective approach to increasing spectral efficiency is to manipulate the channel delay spread \(\tau_d\) to be as small as possible, which can be achieved by DDAM\cite{xiao2025rethinking}. 

\vspace{-5pt}
\subsection{BER}
\label{sec5-2}
BER is a key metric for assessing digital communication system reliability, measuring the ratio of erroneous bits to total transmitted bits. Comparing BER across modulation schemes is vital for optimizing system design and ensuring robust performance under varying channel conditions.
\begin{figure}[thbp]\new
	\centering
\includegraphics[width=0.45\textwidth]{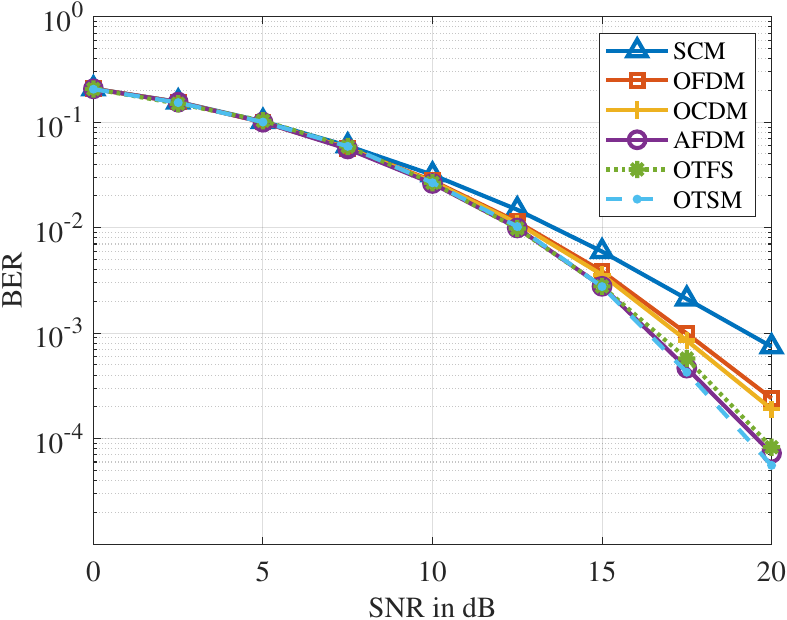}
	\caption{Comparison of BER performance of SCM, OFDM, OCDM, OTFS, AFDM, and OTSM modulated by 4-QAM with subcarrier number $M=1024$ for OFDM, OCDM, AFDM, and $M=N=32$ for OTFS and OTSM with EVA channel model using MMSE detector.}
	\label{fig-ber}
\end{figure}

Fig. \ref{fig-ber} illustrates the BER performance comparison for different waveforms, in which it can be seen that AFDM, OTFS, and OTSM all have better BER performance under DDC. \new{Here, all waveforms employ rectangular pulse shaping, and identical random seeds are used for bitstream generation and channel realizations so that all waveforms experience the same underlying fading process. In all simulations, the EVA channel model and Jakes Doppler spectrum are considered for each channel realization, i.e, the normalized Doppler shift of each path is generated using $\alpha_i=\alpha_{\max } \cos \left(\theta_i\right)$, where $\theta_i$ is uniformly distributed over $[-\pi, \pi]$ and the maximum normalized Doppler shift $\alpha_{\max }$ is corresponding to the maximum UE speed of 540 km/h. The parameters considered for the simulation are provided in Table \ref{tab:ber_params}.}

\begin{table}[tbhp]\new
\renewcommand\arraystretch{1.1}
\caption{Simulation Parameters for BER Evaluation}
\label{tab:ber_params}
\centering
\begin{tabular}{|m{5.6cm}|m{2cm}|}
\hline
\textbf{Parameter} & \textbf{Description} \\ \hline
Carrier frequency $f_c$ (GHz) & 24  \\ \hline
Bandwidth $B$ (MHz) & 3.072  \\ \hline
Subcarrier spacing $\Delta f_\text{1D}$ (kHz) of 1D waveforms& 3\\ \hline
Subcarrier spacing $\Delta f_\text{2D}$ (kHz) of 2D waveforms& 96\\ \hline
Frame size of 1D waveforms ($M\times 1$)  & $M$=1024  \\ \hline
Frame size of 2D waveforms ($M\times N$)  & $M$=32, $N$=32 \\ \hline
Modulation scheme & 4-QAM  \\ \hline
Channel model & EVA   \\ \hline
Number of multipaths $P$  & 9 \\ \hline
Maximum UE speed $v_\text{max}$ (km/h) & 540 \\ \hline
Maximum normalized delay $\tau_\text{max}/\Delta \hat{\tau}_\text{norm}$ & 8 \\ \hline
Maximum normalized Doppler $f_\text{max}/\Delta \hat{f}_\text{norm}$ & 4  \\ \hline
Channel realizations & 1000 \\ \hline
Channel estimation    & ideal CSI \\ \hline
Detector type & MMSE \\ \hline

\end{tabular}
\end{table}

\vspace{-12pt}
\subsection{PAPR}
\label{sec5-3}
PAPR is a key performance metric in multicarrier systems, defined as the ratio of the peak power to the average power of the transmitted signal, which can be expressed as
\begin{equation}
	\text{PAPR} = \frac{\max |s(t)|^2}{\mathbb{E}[|s(t)|^2]}.
\end{equation}
High PAPR affects the efficiency of power amplifiers, requiring higher linearity and leading to increased power consumption and cost. 
On the other hand, under fixed hardware constraints, the waveform with high PAPR requires lowering the average power to avoid nonlinear distortions and ensure reliable transmission, which degrades both communication and sensing performance.
This implies that reducing the PAPR of multicarrier waveforms is essential. \new{For OFDM, a larger number of subcarriers, which is denoted by $M$, generally results in higher PAPR. }

\begin{figure}[tbhp]\new
   \centering
   \includegraphics[width=0.45\textwidth]{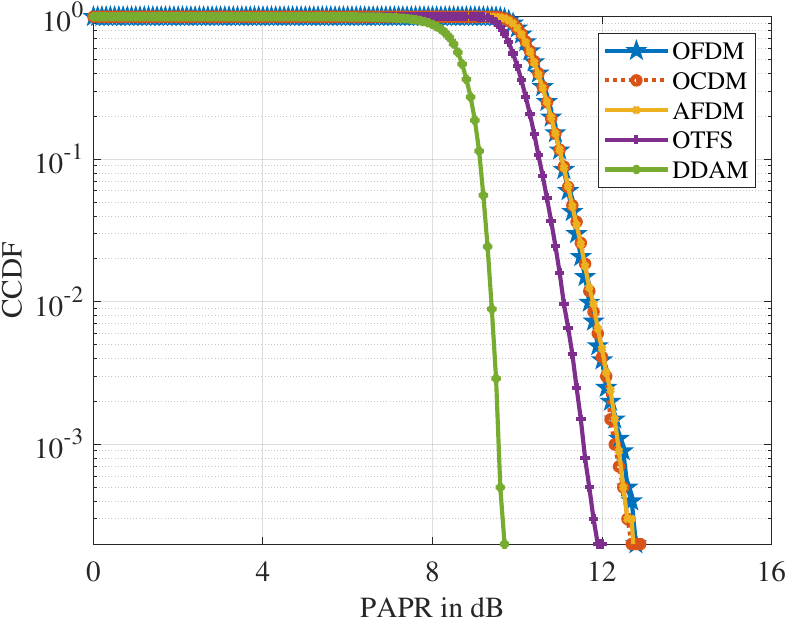}
       \vspace{-5pt}
   \caption{PAPR comparison for OFDM, OCDM, AFDM, OTFS, and DDAM.}
   \label{fig-papr}
    \vspace{-8pt}
\end{figure}

\new{Fig. \ref{fig-papr} plots the complementary cumulative distribution function (CCDF) of PAPR using $M=512$ sub-carriers with 128-QAM. The number of multi-paths is $P=5$, and the number of antennas is ${M_\text{t}} = 256$. All waveforms employ rectangular pulse shaping, and no additional windowing is applied, unless inherently required by the modulation. The detailed numerology and system parameters used for the evaluation are listed in Table~\ref{tab:papr_params}. This configuration represents a typical ISAC-oriented massive-MIMO downlink scenario. The PAPR of DDAM is considerably lower than that of OFDM, which is because the DDAM signal is constructed by superposing only $P$ path components, while the OFDM signal comprises $M \gg P$ sub-carriers. AFDM and OCDM exhibit PAPR levels similar to OFDM because they employ the same number of subcarriers as OFDM. In contrast, OTFS reduces PAPR due to its smaller number of subcarriers.}

\begin{table}[tbhp]\new
\renewcommand\arraystretch{1.1}
\caption{Simulation Parameters for PAPR Evaluation}
\vspace{5pt}
\label{tab:papr_params}
\centering
\begin{tabular}{|m{6.2cm}|m{1.8cm}|}
\hline
\textbf{Parameter} & \textbf{Description} \\ \hline
Carrier frequency $f_c$ (GHz) & 28  \\ \hline
Bandwidth $B$ (MHz) & 128  \\ \hline
Subcarrier spacing $\Delta f_\text{1D}$ (MHz) of 1D waveforms& 0.25\\ \hline
Subcarrier spacing $\Delta f_\text{2D}$ (MHz) of 2D waveforms& 4\\ \hline
Frame size of 1D waveforms ($M\times 1$)  & $M$=512  \\ \hline
Frame size of 2D waveforms ($M\times N$)  & $M$=32, $N$=16 \\ \hline
Number of time-domain symbols & 100 \\\hline
Modulation scheme & 128-QAM  \\ \hline
Number of transmit antennas $N_t$ & 256 \\\hline
Transmit power $P_t$ (dBm) & 30  \\\hline
Noise power spectral density $N_0$ (dBm/Hz) & $-174$  \\\hline
Number of multipaths $P$  & 5 \\ \hline
Maximum normalized delay $\tau_\text{max}/\Delta \hat{\tau}_\text{norm}$ & 40 \\ \hline
Channel realizations & 10000 \\ \hline
Channel estimation (for DDAM)    & ideal CSI \\ \hline
\end{tabular}
\end{table}

The significance of PAPR \cite{10812762} as a KPI lies in its role as a unifying metric for evaluating waveform trade-offs in ISAC systems. 1D waveforms like OFDM and DFT-s-OFDM can provide practical advantages due to their established implementations and compatibility with standards like 5G NR, but their PAPR characteristics necessitate mitigation techniques such as clipping, tone reservation, or precoding, which may compromise spectral efficiency or sensing resolution. 2D waveforms like OTFS and ODDM, which modulate in the delay-Doppler domain, exhibit moderate PAPR due to their TF spreads, but their complex implementation may limit practicality. Within a unified waveform framework, PAPR analysis guides the selection of modulation parameters (e.g., subcarrier number, precoding type) to balance communication efficiency and sensing performance.

\vspace{-5pt}
\subsection{Spectral Efficiency}
\label{sec5-4}
Spectral efficiency quantifies the data rate achievable per unit of bandwidth, typically measured in bits per second per Hertz (bps/Hz). For a multicarrier waveform, the spectral efficiency $\eta$ is defined as:
\begin{equation}\new
\eta = \frac{R}{B} = \frac{(1-k)\log_2(M_c) \cdot K}{T_s + T_{\text{cp}}} \cdot \frac{1}{B},
\label{eq:spectral_efficiency}
\end{equation}
where $R$ is the achievable data rate, $B$ is the occupied bandwidth, $M_c$ is the modulation order (e.g., quadrature phase shift keying (QPSK), 16-QAM, 64-QAM), $K$ is the number of subcarriers, $T_s = \frac{K}{B}$ is the OFDM symbol duration without CP, and $T_{\text{cp}}$ is the CP duration. The presence of CP, as discussed in \eqref{CPoverhead}, introduces overhead $\rho = \frac{T_{\text{cp}}}{T_s + T_{\text{cp}}}$, reducing the effective data rate and thus spectral efficiency. \new{Considering the pilot and guard overhead, the actual spectral efficiency should be adjusted by a factor of $(1-k)$, where $k$ represents the portion of pilots and guard intervals to all transmitted symbols.}

The primary role of spectral efficiency is to evaluate a waveform's ability to maximize data throughput within constrained spectrum resources. For instance, in OFDM, spectral efficiency is maximized by increasing $K$ or $M_c$, but this is limited by the CP overhead and channel impairments like delay spread $\tau_d$. As noted in the CP analysis, minimizing $T_{\text{cp}}$ (while ensuring $T_{\text{cp}} \geq \tau_d$) enhances $\eta$, but significant delay spread necessitates larger $T_{\text{cp}}$, reducing efficiency. 

In ISAC, spectral efficiency also impacts sensing performance, as bandwidth allocation affects both communication throughput and range resolution. A wider bandwidth $B$ enhances range resolution ($\Delta R = \frac{c}{2B}$, where $c$ is the speed of light), but dedicating spectrum to sensing (e.g., via pilots) reduces $\eta$ for communication. For example, OTFS operating in the delay-Doppler domain achieves robust sensing by spreading data across time and frequency, but its spectral efficiency may be lower than OFDM due to additional pilot overhead for channel estimation. Conversely, DFT-s-OFDM balances efficiency and robustness by precoding symbols, maintaining compatibility with 5G NR while supporting moderate sensing tasks. \new{For AFDM, the low pilot overhead is demonstrated by the EPA scheme\cite{bemani2023affinea, raviteja2019embedded}, which effectively reduces the pilot burden. Assuming the maximum normalized delay is $l_{\max }$, the maximum normalized Doppler shift is $\alpha_{\max }$, and the spacing factor used to combat the fractional Doppler is a non-negative integer $\xi_{\nu}$, then the pilot overhead of AFDM accounts for $2\left(l_{\max }+1\right)\left(2\left(\alpha_{\max }+\xi_\nu\right)+1\right)-1$ entries out of the $M$ entries of the AFDM symbols while OTFS requires $\left(4\left(\alpha_{\max }+\xi_\nu\right)+1\right)\left(2 l_{\max }+1\right)$ (for the integer Doppler shifts $\xi_\nu=0$ )\cite{bemani2023affinea,yin2024diagonally}. For example, considering the channel parameters with $ l_{\max }=8,  \alpha_{\max }=4$ in Table \ref{tab:ber_params}, the pilot overhead required for AFDM is 161, whereas OTFS requires 289, which is almost double that of AFDM. Overall, the trade-off between pilot overhead, guard intervals, and spectral efficiency is a critical consideration in ISAC systems. While a higher pilot density and larger guard intervals can improve channel estimation accuracy, they inevitably reduce the effective spectral efficiency.}

The significance of spectral efficiency also lies in its role as a benchmark for optimizing waveform performance in resource-constrained environments. 1D waveforms like OFDM and DFT-s-OFDM offer high spectral efficiency in TDCs, leveraging established standards, but their reliance on CP limits gains in dispersive environments. \new{Spectral efficiency can sometimes be enhanced by relaxing orthogonality constraints, such as the non-orthogonal AFDM scheme in \cite{yi2025nonorthogonal}.} 2D waveforms like OTFS and DDAM provide resilience to DDCs, potentially improving efficiency by reducing overhead, but their complex modulation schemes require careful optimization. Within a unified waveform framework, spectral efficiency analysis guides the selection of parameters (e.g., $K$, $T_{\text{cp}}$, modulation domain) to balance communication throughput and sensing resolution.

\begin{table*}[bhtp]\new
\renewcommand\arraystretch{1.2}
\caption{Modulation Complexity with Typical Equalizers}
\centering
\label{tab5-2}
\begin{tabular}{|cc|cc|cc|}
\hline
\multicolumn{2}{|c|}{\multirow{2}{*}{Waveform}}                           & \multicolumn{2}{c|}{TX}                                                                          & \multicolumn{2}{c|}{RX}                                                                           \\ \cline{3-6} 
\multicolumn{2}{|c|}{}                                                    & \multicolumn{1}{c|}{Operation}              & Complexity                                         & \multicolumn{1}{c|}{Operation}                                & Complexity                        \\ \hline
\multicolumn{2}{|c|}{\multirow{3}{*}{OFDM}}                               & \multicolumn{1}{c|}{\multirow{3}{*}{IFFT}}  & \multirow{3}{*}{$\mathcal{O}(M_tM\log_2 M)$}       & \multicolumn{1}{c|}{FFT}                                      & $\mathcal{O}(M\log_2 M)$          \\ \cline{5-6} 
\multicolumn{2}{|c|}{}                                                    & \multicolumn{1}{c|}{}                       &                                                    & \multicolumn{1}{c|}{Single-tap equalizer}                               & $\mathcal{O}(M)$                  \\ \cline{5-6} 
\multicolumn{2}{|c|}{}                                                    & \multicolumn{1}{c|}{}                       &                                                    & \multicolumn{1}{c|}{Constellation demapping}                  & $\mathcal{O}(1)$                  \\ \hline
\multicolumn{2}{|c|}{\multirow{5}{*}{AFDM}}                               & \multicolumn{1}{c|}{\multirow{5}{*}{IDAFT}} & \multirow{5}{*}{$\mathcal{O}(M_t(M\log_2 M+2M))$}  & \multicolumn{1}{c|}{DAFT}                                     & $\mathcal{O}(M\log_2 M+2M)$       \\ \cline{5-6} 
\multicolumn{2}{|c|}{}                                                    & \multicolumn{1}{c|}{}                       &                                                    & \multicolumn{1}{c|}{MMSE}                                     & $\mathcal{O}(M^3)$                \\ \cline{5-6} 
\multicolumn{2}{|c|}{}                                                    & \multicolumn{1}{c|}{}                       &                                                    & \multicolumn{1}{c|}{MP}                                       & $\mathcal{O}(MPT)$                \\ \cline{5-6} 
\multicolumn{2}{|c|}{}                                                    & \multicolumn{1}{c|}{}                       &                                                    & \multicolumn{1}{c|}{MAMP}                                     & $\mathcal{O}(MPT+PT^2+T^3)$       \\ \cline{5-6} 
\multicolumn{2}{|c|}{}                                                    & \multicolumn{1}{c|}{}                       &                                                    & \multicolumn{1}{c|}{Constellation demapping}                  & $\mathcal{O}(1)$                  \\ \hline
\multicolumn{1}{|c|}{\multirow{11}{*}{OTFS}} & \multirow{6}{*}{MC-OTFS}  & \multicolumn{1}{c|}{\multirow{3}{*}{ISFFT}} & \multirow{3}{*}{$\mathcal{O}(M_t(MN)\log_2 (MN))$} & \multicolumn{1}{c|}{SFFT}                                     & $\mathcal{O}(MN\log_2 (MN))$      \\ \cline{5-6} 
\multicolumn{1}{|c|}{}                       &                            & \multicolumn{1}{c|}{}                       &                                                    & \multicolumn{1}{c|}{FFT}                                      & $\mathcal{O}(\log_2N)$            \\ \cline{5-6} 
\multicolumn{1}{|c|}{}                       &                            & \multicolumn{1}{c|}{}                       &                                                    & \multicolumn{1}{c|}{MMSE}                                     & $\mathcal{O}((MN)^3)$             \\ \cline{3-6} 
\multicolumn{1}{|c|}{}                       &                            & \multicolumn{1}{c|}{\multirow{3}{*}{IFFT}}  & \multirow{3}{*}{$\mathcal{O}(\log_2N)$}            & \multicolumn{1}{c|}{MP}                                       & $\mathcal{O}(MNPT)$               \\ \cline{5-6} 
\multicolumn{1}{|c|}{}                       &                            & \multicolumn{1}{c|}{}                       &                                                    & \multicolumn{1}{c|}{MAMP}                                     & $\mathcal{O}(MNPT+PT^2+T^3)$      \\ \cline{5-6} 
\multicolumn{1}{|c|}{}                       &                            & \multicolumn{1}{c|}{}                       &                                                    & \multicolumn{1}{c|}{Constellation demapping}                  & $\mathcal{O}(1)$                  \\ \cline{2-6} 
\multicolumn{1}{|c|}{}                       & \multirow{5}{*}{Zak-OTFS} & \multicolumn{1}{c|}{\multirow{5}{*}{IDZT}}  & \multirow{5}{*}{$\mathcal{O}(M_t\log_2N)$}         & \multicolumn{1}{c|}{DZT}                                      & $\mathcal{O}(\log_2N)$            \\ \cline{5-6} 
\multicolumn{1}{|c|}{}                       &                            & \multicolumn{1}{c|}{}                       &                                                    & \multicolumn{1}{c|}{MMSE}                                     & $\mathcal{O}((MN)^3)$             \\ \cline{5-6} 
\multicolumn{1}{|c|}{}                       &                            & \multicolumn{1}{c|}{}                       &                                                    & \multicolumn{1}{c|}{MP}                                       & $\mathcal{O}(MNPT)$               \\ \cline{5-6} 
\multicolumn{1}{|c|}{}                       &                            & \multicolumn{1}{c|}{}                       &                                                    & \multicolumn{1}{c|}{MAMP}                                     & $\mathcal{O}(MNPT+PT^2+T^3)$      \\ \cline{5-6} 
\multicolumn{1}{|c|}{}                       &                            & \multicolumn{1}{c|}{}                       &                                                    & \multicolumn{1}{c|}{Constellation demapping}                  & $\mathcal{O}(1)$                  \\ \hline
\multicolumn{2}{|c|}{\multirow{3}{*}{DDAM}}                               & \multicolumn{1}{c|}{path-based MRT}         & $\mathcal{O}(M_tP)$                                & \multicolumn{1}{c|}{\multirow{3}{*}{Constellation demapping}} & \multirow{3}{*}{$\mathcal{O}(1)$} \\ \cline{3-4}
\multicolumn{2}{|c|}{}                                                    & \multicolumn{1}{c|}{path-based ZF}          & $\mathcal{O}(M_tP^2/N_s+M_tP)$                     & \multicolumn{1}{c|}{}                                         &                                   \\ \cline{3-4}
\multicolumn{2}{|c|}{}                                                    & \multicolumn{1}{c|}{path-based MMSE}        & $\mathcal{O}(M_t^3P^4/N_s+M_tP)$                   & \multicolumn{1}{c|}{}                                         &                                   \\ \hline
\end{tabular}

\vspace{1em}
\raggedright
\footnotesize
$M_t$: number of transmit antennas; 
$M$: number of subcarriers; 
$N$: number of symbols per OTFS frame; 
$P$: number of multi-paths; 
$T$: number of iterations for iterative algorithms (MP/MAMP);
$N_s$: number of total transmit information-bearing symbols per channel coherence block.
\vspace{-6pt}
\end{table*}

\vspace{-5pt}
\subsection{Modulation Complexity}
\label{sec5-5}
Modulation complexity quantifies the computational and implementation burden of generating and processing a modulated signal, typically measured in terms of the number of arithmetic operations required per symbol. For a multicarrier waveform with $M$ subcarriers and $N$ time slots, the modulation complexity is often expressed as the order of computational operations, as shown in Table \ref{tab5-2}. Here, the complexity arises \new{from transformations and receiver detectors}, with $M_t$ denoting the number of transmit antennas and $P$ denoting the number of multi-paths. \new{Both transmitter- and receiver-side complexities, including typical MMSE, MP\cite{raviteja2018interference,wu2024afdm}, and MAMP detectors\cite{liu2022memory,qi2025mamp}, are summarized for completeness.}

For OFDM, modulation involves an IFFT, making it computationally efficient and widely adopted in standards like 5G NR. In contrast, OTFS requires a two-stage transformation: mapping symbols to the DD domain and converting to the TF domain via ISFFT, resulting in a significantly higher complexity. This increased burden poses challenges for real-time processing, particularly in low-power ISAC devices like IoT nodes or vehicular transceivers. For DDAM, the DD compensation only involves a time shift and phase rotation of the transmit symbol sequence; thus, DDAM enjoys low complexity and small communication latency compared to OTFS\cite{xiao2025rethinking}.

The significance of modulation complexity as a KPI lies in its role in guiding waveform selection and optimization within a unified framework. 1D waveforms like OFDM and DFT-s-OFDM offer low complexity, enabling practical deployment in existing systems, but their performance in DDCs is limited. 2D waveforms like OTFS and ODDM provide superior robustness but at the cost of increased computational demands. \newer{A unified waveform framework effectively resolves this dilemma by enabling adaptive parameter configuration, such as reducing the time-slot dimension to construct a short-frame structure, which directly minimizes the inherent block-processing latency for URLLC applications. Furthermore, the integration of low-complexity approximate transforms or AI-driven optimization can significantly accelerate signal detection, ensuring that 2D waveforms remain computationally feasible even under strict latency constraints.} This KPI analysis underscores the need for adaptive designs to ensure the unified framework's practicality in diverse ISAC applications.

\vspace{-5pt}
\subsection{Ambiguity Functions}
\label{sec5-6}
The ambiguity function (AF) of a waveform fundamentally measures its range and velocity estimation ability with matched filtering (MF). In particular, the auto-ambiguity function (AAF) and cross-ambiguity function (CAF) correspond to mono-static and bi-static sensing configurations, respectively.

For a mono-static sensing receiver (MSR) that is co-located with the transmitter, the transmitted signal is fully known. Therefore, we perform MF on the echo signal, i.e., the received signal $r(t)$ with $s(t)$, which can be formulated as 
\begin{equation}
	R(\tau, \nu)= \int_{-\infty}^{\infty} r(t)s^{*}(t-\tau)e^{-j  2 \pi\nu t} \text{d}t.
	\label{eqyhr0414.7}
\end{equation}
Considering DDC by subsitituting  (\ref{eqyhr0414.2}) into (\ref{eqyhr0414.7}), we have
\begin{align}
	&R(\tau, \nu)  \notag\\
	=&\int_{-\infty}^{\infty} \left(\sum_{i=1}^{P} h_{i}  s(t-\tau_{i})e^{j2\pi\nu_{i}t} \right)s^{*}(t-\tau)e^{-j  2 \pi\nu t} \text{d}t  \notag \\
	=&\sum_{i=1}^{P} h_{i}e^{-j2\pi(\nu-\nu_{i})\tau_{i}}\int_{-\infty}^{\infty}s(t)s^{*}(t-(\tau-\tau_{i}))e^{-j2\pi(\nu-\nu_{i})t}\text{d}t \notag \\
	=&\sum_{i=1}^{P} h_{i}e^{-j2\pi(\nu-\nu_{i})\tau_{i}}A_{s,s}(\tau-\tau_{i}, \nu-\nu_{i})
	\label{eq24.01.09.5}
\end{align}
where $A_{s,s}(\tau, \nu)$ is the AF with a definition given by
\begin{equation}
	A_{a,b}(\tau, \nu)\triangleq \int_{-\infty}^{\infty}a(t)b^{*}(t-\tau)e^{-j2\pi\nu t}\text{d}t.
	\label{eq24.01.09.6}
\end{equation}
In particular, AF is known as AAF and CAF when $b(t)=a(t)$ and $b(t)\neq a(t)$, respectively. Moreover, as shown in (\ref{eq24.01.09.5}), the output of MF of $r(t)$ and $s(t)$ is the superposition of $P$ scaled and shifted AAF of $s(t)$, where the shifts along the delay domain and the Doppler domain are exactly the delay shifts and the Doppler shifts associated with the $P$ sensing targets, respectively. 
The above discussion highlights the close connection between the MF operation in sensing and the AF. 

\begin{figure*}[htbp]\new
    \centering
    \subfloat[AAF at zero-delay slice: Random data]{
        \includegraphics[width=0.48\textwidth]{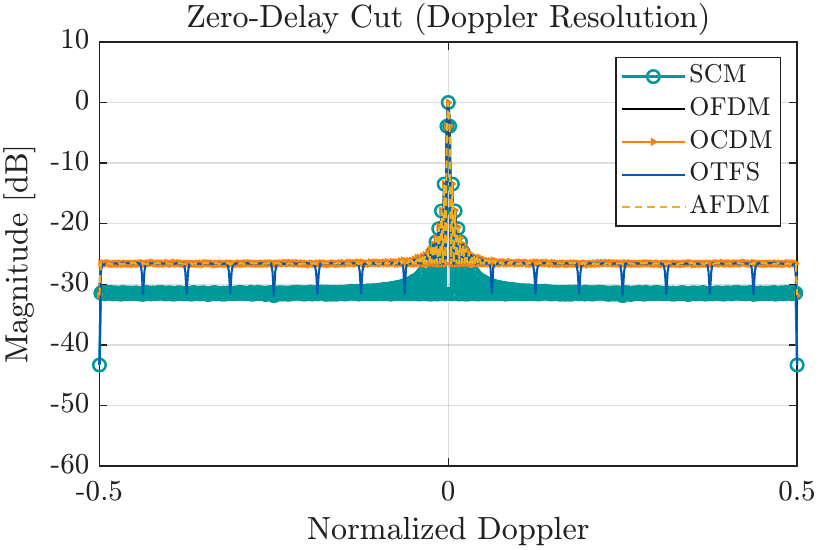}\label{fig-AF-tau=0-random}
    }
    \hfill
    \subfloat[AAF at zero-delay slice: All-one data]{
        \includegraphics[width=0.48\textwidth]{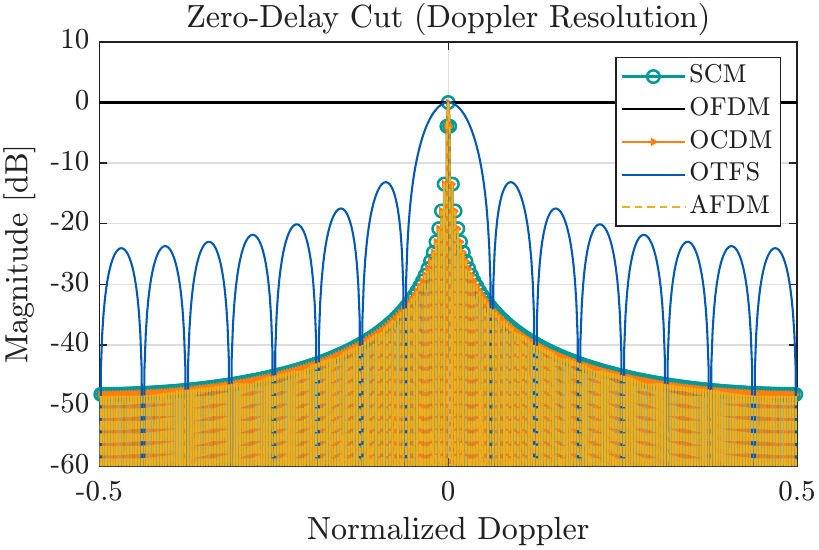}\label{fig-AF-tau=0-allone}
    }
    \\
    \caption{AAF comparison of SCM, OFDM, OCDM, AFDM, and OTFS at zero-delay slice with (a) random data and (b) all-one data.}
    \label{fig-AF-tau}
\end{figure*}

\begin{figure*}[hbpt]\new
    \centering
    \subfloat[AAF at zero-Doppler slice: Random data]{
        \includegraphics[width=0.48\textwidth]{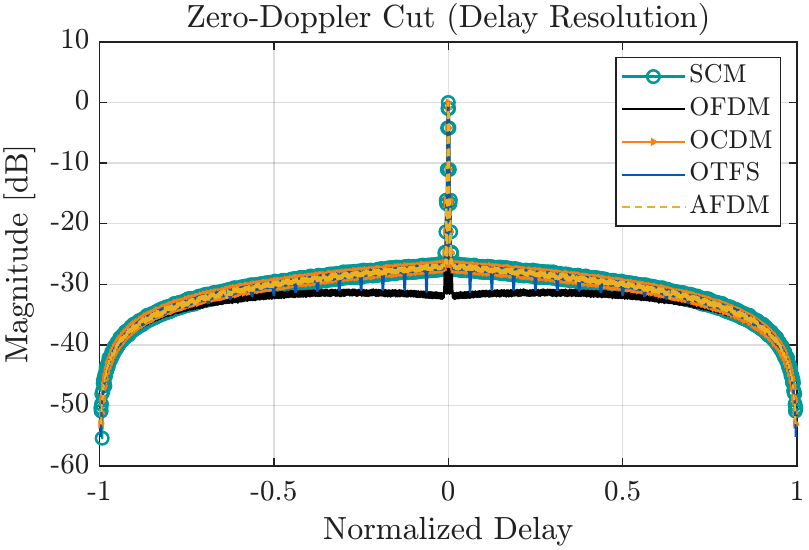}\label{fig-AF-nu=0-random}
    }
    \hfill
    \subfloat[AAF at zero-Doppler slice: All-one data]{
        \includegraphics[width=0.48\textwidth]{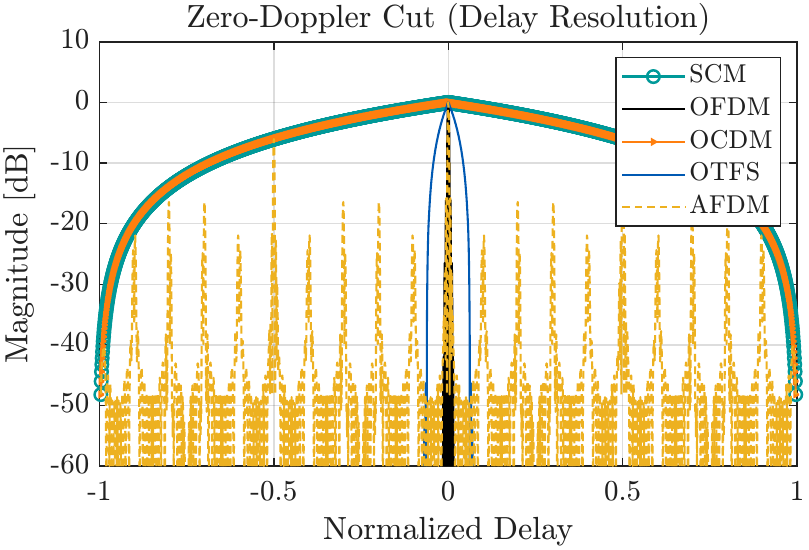}\label{fig-AF-nu=0-allone}
    }
    \caption{AAF comparison of SCM, OFDM, OCDM, AFDM, and OTFS at zero-Doppler slice with (a) random data and (b) all-one data.}
    \label{fig-AF-nu}
\end{figure*}

\new{The key sidelobe metrics, namely the peak-to-sidelobe ratio (PSLR) and integrated sidelobe ratio (ISLR), are used to quantify the relative amplitude and total energy of sidelobes with respect to the mainlobe, respectively. The PSLR measures the relative amplitude of the highest sidelobe to the mainlobe peak, indicating the strength of sidelobe suppression and false-target probability. It is defined as\cite{bedeer2025ambiguity,rou2025normalized}}
\begin{equation}\new
\mathrm{PSLR}_{\xi} = 20\log_{10}\left(\frac{\max_{\xi \in \text{sidelobes}} |A(\xi)|}{\max_{\xi} |A(\xi)|}\right),
\label{eq:PSLR}
\end{equation}
\new{where $\xi \in \{\tau, \nu\}$ denotes the delay or Doppler domain. A lower (more negative) PSLR value indicates better sidelobe suppression.}

\new{On the other hand, the ISLR characterizes the ratio between the total sidelobe energy and the mainlobe energy, reflecting the degree of energy concentration within the mainlobe. It is given by\cite{bedeer2025ambiguity,rou2025normalized}}
\begin{equation}\new
\mathrm{ISLR}_{\xi} = 10\log_{10}\left(\frac{\sum_{\xi \in \text{sidelobes}} |A(\xi)|^{2}}{\sum_{\xi \in \text{mainlobe}} |A(\xi)|^{2}}\right),
\label{eq:ISLR}
\end{equation}
\new{where a smaller ISLR value implies stronger energy focusing and lower leakage beyond the mainlobe.
}

As shown in Fig. \ref{fig-AF-tau} and Fig. \ref{fig-AF-nu}, the zero-delay and zero-Doppler slices AAFs of SCM \cite{10838324}, OFDM \cite{11037613}, OCDM, AFDM \cite{yin2026ambiguity}, and OTFS \cite{li2023pulse} with \new{both random data and all-one data} are compared. \new{The corresponding mainlobe widths, PSLR, and ISLR metrics are summarized in Table~\ref{tab:AF metrics}. We can observe that OFDM exhibits the narrowest delay-domain mainlobe and lowest sidelobe levels, confirming its excellent range resolution. In contrast, OCDM, AFDM, and OTFS achieve narrower Doppler-domain mainlobes, indicating superior velocity resolution.} 

\begin{table*}[hbtp]\new
\caption{Normalized AF metrics extracted from 1D cuts for the unit symbol case.}
\label{tab:AF metrics}
\renewcommand\arraystretch{1.2}
\centering
\begin{tabular}{@{\hspace{2em}}l@{\hspace{2em}}|@{\hspace{2em}}r@{\hspace{4em}} r@{\hspace{4em}} r@{\hspace{4em}} r@{\hspace{4em}} r@{\hspace{4em}} r@{\hspace{2em}}}
\hline Waveform & $\Delta \tau_{3 \mathrm{~dB}}$ & $\Delta \nu_{3 \mathrm{~dB}}$ & { PSLR $_\tau$} & { ISLR $_\tau$} & { PSLR $_\nu$} & { ISLR $_\nu$} \\
\hline 
SCM  & 0.5842 & 0.0031 & 0.0000 & -13.4643 & 21.3393 & -9.6311 \\
OFDM & 0.0033 & 1.7988 & -16.0975 & -313.0712 & -14.8007 & -186.1829 \\
OCDM & 0.5842 & 0.0031 & 0.0000 & -13.4643 & 21.3393 & -9.6311 \\
OTFS & 0.0369 & 0.0554 & -46.3360 & -13.1480 & -55.2364 & -7.6628 \\
AFDM & 0.0033 & 0.0031 & -6.0206 & -13.4643 & -0.7303 & -9.6311 \\
\hline
\end{tabular}
\vspace{-6pt}
\end{table*}

\subsection{Hardware and RF Impairments}
\label{sec5-7}
\new{Practical transceiver implementations are inevitably affected by various hardware and RF imperfections, which can significantly distort the transmitted and received signals, especially in wideband and high-frequency scenarios. The most critical impairments include PN, CFO, power amplifier (PA) nonlinearity, and timing jitter. These effects jointly deteriorate the orthogonality among subcarriers or symbols and thus increase MD-ISI.}

\new{Among these, PN originates from oscillator instability and can be decomposed into a slow common phase error (CPE) and fast ICI components. CFO results from frequency mismatches between the transmitter and receiver oscillators, producing a deterministic phase slope across symbols and severe ICI when the offset is large. PA introduces amplitude and phase distortions, typically modeled by amplitude-to-amplitude (AM/AM) and amplitude-to-phase (AM/PM) curves, which result in severe signal saturation and spectral spreading, posing a significant challenge for high-PAPR multicarrier schemes. Finally, timing jitter and sampling offset destroy symbol alignment, leading to loss of orthogonality and synchronization mismatch.}

\new{Recent studies have investigated the robustness of different waveforms under these impairments. For OFDM-based systems, it has been demonstrated that PN and PA nonlinearity are the most detrimental factors in mmWave and sub-THz bands, which can be effectively counteracted through a combination of digital predistortion, adaptive phase tracking, and spectral shaping techniques such as windowing\cite{chen2024modeling,wang2019waveform}. In contrast, OTFS exhibits strong resilience against PN and CFO due to its delay-Doppler domain representation. Experimental emulation confirms that OTFS maintains up to one order of magnitude lower BER than OFDM under combined PN, CFO, and PA distortions\cite{abushattal2023comprehensive}.} 

\new{It is worth noting that OTSM also shows promising robustness. Analytical and simulation results indicate that OTSM can jointly compensate IQ imbalance, CFO, and PN through low-complexity estimation in the delay-subcarrier domain, maintaining near-ideal BER even under imperfect CSI conditions\cite{neelam2023jointa,doosti-aref2024performance}. Regarding AFDM, while systematic studies are still limited, its DAFT-domain chirp basis offers potential robustness to frequency offsets and time variations owing to its inherent full-diversity property\cite{li2026affinea}. Further investigation of phase noise and PA distortion effects on chirp parameter selection is an open research topic.}

\subsection{Summary and Practical Insights}
\label{sec5-8}
The overall performance of waveforms can be effectively evaluated using the typical KPIs mentioned above. As summarized in Table \ref{tab5-1}, we can see that there are many trade-offs in performance between waveforms. \new{For 1D waveforms, OFDM and DFT-s-OFDM remain suitable for multipath or uplink scenarios due to their low complexity and maturity, whereas FrFT-OFDM and IFDM provide robustness against non-stationary and high-mobility environments through flexible domain transformations. AFDM demonstrates strong adaptability in DDC owing to its flexible structure and full-diversity gain.
For 2D waveforms, OTFS and ODDM offer improved performance under high Doppler and TF selectivity, achieving better BER and adaptability at the cost of increased modulation complexity. Additionally, Zak-OTFS and ODDM achieve enhanced practicality through the Zak-transform and pulse design utilizing DDOP, respectively. Specifically, DDAM is suited for large-scale MIMO applications by leveraging spatial dimensions with lower PAPR. OTSM achieves advantages in computational complexity and BER performance through Walsh-Hadamard sequences, while FBMC and ODSS are more suitable for asynchronous multi-user and wideband channels, respectively, due to their spectral localization. By examining various practical channel models, we observe that eMBB scenarios benefit from waveforms emphasizing spectral efficiency (e.g., FBMC, DDAM), while URLLC and high-mobility applications favor low-latency and Doppler-resilient schemes such as AFDM, OTFS, or ODDM.} \newer{This is primarily due to their superior Doppler resilience, ensuring ultra-reliability, while the low-latency requirement can be addressed through short-frame structures and low-complexity detection designs.}

Therefore, when considering advanced application scenarios, it is necessary to select the appropriate waveform based on KPIs to achieve the best performance. Here, by constructing the unified waveform framework, we can better perform performance comparisons and analyses, thereby guiding waveform selection and switching in the future.

\begin{table*}[thbp]
\renewcommand\arraystretch{1.2}
\centering
\caption{KPI evaluation of existing waveforms}
\label{tab5-1}
\begin{threeparttable}
\begin{tabular}{|m{0.6cm}|m{1.62cm}|m{1.2cm}|m{1.2cm}|m{1.2cm}|m{1.2cm}|m{1.4cm}|m{1.5cm}|m{4.5cm}|}
\hline
\textbf{1D/2D}  & \textbf{Waveforms} & \textbf{CP Overhead}             & \textbf{BER in DDC}              & \textbf{PAPR}                    & \textbf{Spectral Efficiency}     & \textbf{Modulation Complexity} & \new{\textbf{RF Impairment Robustness}}  & \textbf{Adaptability}                     \\ \hline
\multirow{6}{*}{1D} & OFDM               & $\bigstar$                       & $\bigstar$                       & $\bigstar$                       & $\bigstar$ $\bigstar$            & $\bigstar$ $\bigstar$ $\bigstar$& \new{$\bigstar$} & Multipath time-dispersive channels        \\ \cline{2-9} 
                    & DFT-s-OFDM         & $\bigstar$                       & $\bigstar$                       & $\bigstar$ $\bigstar$            & $\bigstar$                       & $\bigstar$ $\bigstar$            &  \new{$\bigstar$ $\bigstar$} & Uplink transmission                       \\ \cline{2-9} 
                    & FrFT-OFDM          & $\bigstar$ $\bigstar$            & $\bigstar$ $\bigstar$            & $\bigstar$                       & $\bigstar$ $\bigstar$            & $\bigstar$ $\bigstar$       & \new{$\bigstar$ $\bigstar$}       & Transmission over non-stationary channels \\ \cline{2-9} 
                    & OCDM               & $\bigstar$ $\bigstar$            & $\bigstar$ $\bigstar$            & $\bigstar$                       & $\bigstar$ $\bigstar$            & $\bigstar$ $\bigstar$     & \new{$\bigstar$ $\bigstar$}         & Underwater communication                  \\ \cline{2-9} 
                    & IFDM               & $\bigstar$ $\bigstar$            & $\bigstar$ $\bigstar$ $\bigstar$ & $\bigstar$ $\bigstar$            & $\bigstar$ $\bigstar$            & $\bigstar$ $\bigstar$      & \new{$\bigstar$ $\bigstar$}        & Relying on high-performance detectors     \\ \cline{2-9} 
                    & AFDM               & $\bigstar$ $\bigstar$            & $\bigstar$ $\bigstar$ $\bigstar$ & $\bigstar$                       & $\bigstar$ $\bigstar$ $\bigstar$ & $\bigstar$ $\bigstar$       & \new{$\bigstar$ $\bigstar$ $\bigstar$}    & High mobility scenarios                   \\ \hline
\multirow{6}{*}{2D} & FBMC               & $\bigstar$ $\bigstar$ $\bigstar$ & $\bigstar$ $\bigstar$            & $\bigstar$ $\bigstar$            & $\bigstar$ $\bigstar$ $\bigstar$ & $\bigstar$           & \new{ $\bigstar$ $\bigstar$ }         & Multi-user asynchronous scenarios         \\ \cline{2-9} 
                    & OTFS               & $\bigstar$ $\bigstar$            & $\bigstar$ $\bigstar$ $\bigstar$ & $\bigstar$ $\bigstar$            & $\bigstar$ $\bigstar$            & $\bigstar$ $\bigstar$   & \new{$\bigstar$ $\bigstar$ $\bigstar$}          & High mobility scenarios                   \\ \cline{2-9} 
                    & ODDM               & $\bigstar$ $\bigstar$            & $\bigstar$ $\bigstar$ $\bigstar$ & $\bigstar$ $\bigstar$            & $\bigstar$ $\bigstar$            & $\bigstar$ $\bigstar$     & \new{$\bigstar$ $\bigstar$}       & High mobility scenarios                   \\ \cline{2-9} 
                    & DDAM               & $\bigstar$ $\bigstar$ $\bigstar$ & $\bigstar$ $\bigstar$ $\bigstar$ & $\bigstar$ $\bigstar$ $\bigstar$ & $\bigstar$ $\bigstar$ $\bigstar$ & $\bigstar$ $\bigstar$     & \new{$\bigstar$ $\bigstar$ $\bigstar$}       & \new{High mobility scenarios, especially in mmWave/THz}                               \\ \cline{2-9} 
                    & OTSM               & $\bigstar$ $\bigstar$            & $\bigstar$ $\bigstar$ $\bigstar$ & $\bigstar$ $\bigstar$            & $\bigstar$ $\bigstar$            & $\bigstar$ $\bigstar$     & \new{$\bigstar$ $\bigstar$}         & High mobility scenarios                 \\ \cline{2-9} 
                    & ODSS               & $\bigstar$ $\bigstar$            & $\bigstar$ $\bigstar$ $\bigstar$ & $\bigstar$ $\bigstar$            & $\bigstar$ $\bigstar$            & $\bigstar$ $\bigstar$      & \new{ $\bigstar$ $\bigstar$}    & Wideband doubly-dispersive channels       \\ \hline
\end{tabular}
\begin{tablenotes}
\footnotesize
\item $\bigstar$: Three stars indicate attractive performance, two stars indicate medium performance, and one star indicates less attractive performance. For example, we want the BER to be as low as possible, so waveforms with low BER in the BER column are awarded three stars (most attractive).
\end{tablenotes}
\end{threeparttable}
\vspace{-10pt}
\end{table*}

\vspace{-5pt}
\section{Advanced Applications of Multicarrier Waveforms}
\label{sec6}
In this section, a comprehensive analysis of advanced applications is provided based on the KPIs analysis mentioned in Section \ref{sec5}. For each application, the performance requirements are examined, and the suitability of specific multicarrier waveforms is evaluated based on their resilience to channel-induced interference, flexibility for multi-user and multi-antenna systems, PHY security, compatibility with OFDM-based systems, and robustness to integrated communication and multi-application. By leveraging the KPIs of 1D waveforms and 2D waveforms, researchers can offer tailored solutions to meet the needs of diverse scenarios.

\vspace{-8pt}
\subsection{Multiple Access}
\label{sec6-1}
\new{Multiple access (MA) is of critical importance for modern communication networks, particularly in enabling concurrent uplink and downlink transmissions under mobility, power, and synchronization constraints\cite{clerckx2024multiple, jorswieck2024nextgeneration}. The waveform chosen fundamentally shapes how users share resources and determines how effectively interference can be suppressed or exploited\cite{wunder20145gnow}.}

Orthogonal multiple access (OMA) schemes include frequency division multiple access (FDMA), time division multiple access (TDMA), code division multiple access (CDMA), and orthogonal frequency division multiple access (OFDMA)\cite{yin2006ofdma}. CDMA and OFDMA are well-developed and comprehensive OMA schemes. In CDMA\cite{adachi2005broadband, prasad1996overview,lie-liangyang2003multicarrier}, each user is assigned a unique orthogonal code (such as Walsh codes), allowing users to transmit signals simultaneously at the same time and frequency, with different users distinguished by the orthogonality of the code domain. In OFDM, subcarriers provide orthogonal resource units, thereby inherently offering the advantages of MA. However, OFDMA is sensitive to frequency synchronization, limiting its efficiency in high mobility scenarios\cite{morelli2007synchronization}. \new{In the uplink, where user power budgets and synchronization accuracy are often limited, waveforms with low PAPR and strong spectral localization offer distinct advantages. In this case, DFT-s-OFDM can support power-efficient grant-free access and has been adopted in the 5G NR uplink. At the same time, FBMC with low sidelobes and strong TF localization may be employed for asynchronous uplink access.}

Non-orthogonal multiple access (NOMA) schemes include \newer{code-domain NOMA (CD-NOMA)\cite{liu2021sparse} with sparse code multiple access (SCMA)\cite{hoshyar2008novel,nikopour2013sparse} as a promising scheme,} power domain NOMA (PD-NOMA)\cite{ge2021otfs}, space division multiple access (SDMA)\cite{chen2006mber,vandenameele2006space}, and \new{rate-splitting multiple access} (RSMA)\cite{mao2022ratesplitting,10804646,huai2024crossdomain}. \newer{Specifically, CD-NOMA may be regarded as a generic CDMA with the major aim of supporting massive machine-type communications. In SCMA, the idea is to map the user data to a number of subcarriers (underpinned by a multicarrier waveform such as OFDM), whereby the sparse codebook property and message passing algorithm (MPA) are exploited to attain low-complexity near-optimal decoding performance.} In \cite{luo2024building,luo2024afdmscma}, Luo {\em{et al.}} systematically studied AFDM-empowered SCMA and proposed a codebook design based on the input-output relationship; in \cite{wen2023otfsscma,ge2023otfs}, the integration of OTFS and SCMA was studied to provide large-scale connectivity and support the coordinated multi-point coverage scenario. 

\newer{On the other hand, RSMA generalizes and unifies PD-NOMA, SDMA, and physical multicasting in an elegant way\cite{clerckx2024multiple}. By splitting the user data into common messages and private messages, RSMA is able to attain certain operating points that cannot be achieved by PD-NOMA and SDMA. Through proper subcarrier assignment and power allocation, multicarrier RSMA was shown to outperform OFDMA and OFDM-PD-NOMA in \cite{sahin2023multicarrier,sahin2025ofdmrsma}.} 

\new{Therefore, future systems may benefit from waveform-MA co-design, where user multiplexing, interference control, and modulation domain (e.g., DD, chirp, or beam domain) are jointly optimized for specific uplink/downlink applications.} \newer{As an example, it is interesting to design novel waveforms and their signal processing algorithms to enable low-complexity downlink multiuser detection whereby every user can efficiently recover its own messages with minimum decoding of other users' data symbols.}

\vspace{-8pt}
\subsection{MIMO}
\label{sec6-2}

\new{The integration of MIMO with different waveforms fundamentally determines how spatial diversity, multiplexing, and channel estimation are realized in practice. The underlying waveform not only shapes how MIMO channels are resolved in the time-frequency-space domain but also directly affects receiver detection complexity.}

\new{Depending on whether spatial resources are allocated to a single or multiple users, MIMO can operate in single-user MIMO (SU-MIMO) or multi-user MIMO (MU-MIMO) mode, whose effectiveness strongly depends on the underlying waveform structure. In OFDM-based systems\cite{studer2016quantized}, orthogonal subcarriers simplify MIMO equalization using ZF or MMSE detectors but become sensitive to CFO and ICI. MIMO-FBMC achieves higher spectral efficiency without CP but suffers from imaginary-part interference, which motivates complex pilot designs such as complex training sequence decomposition (CTSD) for accurate channel reconstruction\cite{hu2017training}.}  

\new{Beyond conventional multicarrier designs, modulation-domain waveforms enable more flexible and scalable MIMO implementations. 
When integrated with MIMO, these waveforms inherently increase receiver detection complexity due to the need for joint processing across antennas and modulation domains. 
However, their structured sparsity and domain separability make it possible to design efficient low-complexity detection algorithms, which alleviate this computational burden. In \cite{luo2026joint}, the joint sparse graph was applied in MIMO-AFDM to achieve a low complexity receiver design. In \cite{cheng2024mimooddma}, a novel signal detection approach by leveraging the spatial-based generative adversarial network was used for MIMO-ODDM.
In addition, accurate CSI acquisition and adaptive precoding are crucial for fully exploiting spatial multiplexing and diversity gains.  
In \cite{yinCDDS}, a scalable precoding scheme, called cyclic delay-Doppler shift (CDDS), was developed to achieve optimal transmit diversity gain in MIMO-AFDM and MIMO-OTFS systems. \newer{Distinct from traditional space-time block coding (e.g., Alamouti) that typically requires joint decoding across multiple time slots, CDDS functions as an intra-frame precoder to extract full diversity within a single frame. This capability allows for low-latency one-shot transmission without temporal extension, making it suitable for URLLC scenarios.} 
In \cite{savaux2024spatial}, the application potential of AFDM in MU-MIMO was studied through frequency-domain precoding. 
Notably, DDAM naturally supports large-scale MIMO systems by separating aligned delays and Doppler components in the spatial domain, thereby reducing matrix inversion complexity and simplifying detection\cite{xiao2025rethinking}.} 
Additionally, since the integration of MIMO and MA is essential, waveform design must consider the prerequisites for MA\cite{shen2022random}.

\new{Overall, while MIMO integration inevitably increases the complexity of signal detection and channel estimation, advanced modulation waveforms provide the structural sparsity and domain separability required to design lightweight algorithms that maintain high performance in dynamic environments.}

\vspace{-8pt}
\subsection{Full Duplex}
\label{sec6-3}
Full-duplex (FD) communication enables simultaneous transmission and reception in the same frequency band, significantly enhancing spectral efficiency. This is particularly important in the context of ISAC, where a single base station must concurrently receive radar and communication signals, necessitating the use of FD operation. However, the coupling between the transmitter and receiver, known as self-interference (SI), represents the greatest challenge in FD systems, as it severely impacts the reception of the desired signal. 

Typically, SI cancellation methods can be categorized into three domains: the antenna domain, the analog domain, and the digital domain\cite{smida2024inband}. By combining these three domains, SI can be reduced to the noise level. In terms of waveform design, an FD ISAC waveform design is proposed in\cite{xiao2022waveform} that transmits communication signals during the idle time of conventional pulsed radar while effectively canceling SI.

As for the waveforms currently researched, full-duplex communication is already being considered as a trend, with various waveforms specifically designed to exploit their unique characteristics:

\begin{itemize}
    \item{FD-OFDM exploits frequency-domain characteristics for differential interference cancellation, enabling simultaneous static SI cancellation and dynamic target detection\cite{duan2024frequencydomain}.}
    \item{FD-OTFS utilizes delay-time domain properties to jointly achieve sensing and communication functions, incorporating an OTFS-specific SI cancellation scheme\cite{luo2023inband}.}
    \item{FD-AFDM takes advantage of the chirp characteristic for low-cost SI cancellation\cite{bemani2024integrated}, and employs symbol design in the DAFT domain to realize FD ISAC functionality\cite{luxianjie}.}
\end{itemize}

Thus, waveform design in the full-duplex domain is an area of significant research value, with each waveform requiring the leveraging of its specific characteristics to maximize full-duplex capabilities and meet the demands of ISAC.

\vspace{-8pt}
\subsection{Security and Privacy}
\label{sec6-4}
Due to the broadcast nature of wireless communication or ISAC signals, legitimate nodes may face potential security risks and privacy concerns\cite{lu2024integrated}. On the one hand, attackers can eavesdrop on the signals between legitimate communication parties to obtain legitimate information or analyze the received signals to obtain the location and mobility trajectories of legitimate nodes. On the other hand, attackers can also actively attack the legitimate receiver via signal spoofing or impersonation attacks. 

To address these challenges, many waveforms exploit their inherent characteristics for specific security designs. In \cite{naderi2022channel}, an adaptive subcarrier selection method based on the frequency response of OFDM subcarriers is proposed, achieving encryption by adjusting the length of the secure cyclic prefix of each OFDM symbol. Moreover, in \cite{ma2023physical}, an index modulation-assisted orthogonal time-frequency space (IM-OTFS) modulation system working in an FDD mode is proposed, where the angle reciprocity of the legitimate link is used to construct secure mapping rules between information bits, modulation symbols, and symbol indices, thereby generating chaotic sequences for encryption. A simple yet effective approach to mitigate jamming interference is to introduce resource hopping \cite{deng2023jamming}, wherein the system pseudorandomly varies the resource bins assigned to different users from one OTFS block to the next. Furthermore, since the AFDM waveform has two adjustable chirp parameters, eavesdroppers are unable to decrypt the information without knowing the chirp parameters; thus, AFDM has a natural advantage in waveform security design\cite{chenhaibo}. In \cite{tie2024securityenhanced}, a secure transmission scheme for AFDM systems is proposed, where a base station sends a set of AFDM NOMA signals to users to sense channel characteristics, allowing the base station and users to obtain channel parameters and calculate chirp parameters, thus achieving secure negotiation of chirp parameters. In \cite{rou2025chirppermuteda}, an AFDM encryption scheme based on index modulation is proposed, where each index is pre-assigned to a specific set of chirp parameters, and the encryption is achieved by activating the chirp parameters corresponding to a particular index. The main challenges that need to be addressed for security and privacy protection in the future include the following aspects:

\begin{itemize}
\item{Enhancement of Confidentiality: In eavesdropping scenarios, combining artificial noise and other techniques can increase the disparity between legitimate channels and eavesdropping channels, making it difficult for eavesdroppers to intercept the signal. Additionally, high-dynamic channel key extraction algorithms need to be designed to encrypt waveform information.}

\item{Low Probability of Detection: When the locations of eavesdroppers are known, the transmitted signal can be designed to make it difficult for eavesdroppers to intercept. Moreover, various covert communication techniques can be used to conceal the locations of legitimate communication parties, making it difficult for eavesdroppers to track them\cite{wei2022multifunctional}.}

\end{itemize}

\vspace{-8pt}                      
\subsection{Index Modulation}
\label{sec6-5}
Index modulation (IM) is a novel technique that embeds information into the activation pattern of transmission resources~\cite{Wen2017IM,bacsar2013orthogonal,Wen2021IM}. It has been widely integrated into various waveform frameworks, which effectively enhances spectral efficiency, diversity gain, and interference resilience.
This section provides a summary of IM-enhanced schemes designed for such DDCs, with a focus on their design principles and key techniques.

Currently, based on the type of modulation domain, existing IM techniques for DDCs can be classified into three categories: DD domain-based IM~\cite{zhang2024otfs}, DAFT domain-based IM~\cite{tao2024affine,zhu2024design,liu2025pre}, and delay-sequency domain-based IM~\cite{doosti2023sequency}.
Among them, the representative technique of IM in the DD domain is OTFS-IM. Initial research on the OTFS-IM system mainly concentrated on expanding the IM dimensions and designing the overall framework~\cite{feng2022phase,zhang2024enhanced,zhang2025orthogonal}. In~\cite{zhao2021orthogonal}, Zhao {\em{et al.}} introduced the concept of dual-mode IM into OTFS, where bit mapping diversity is achieved by switching between activation patterns and constellation modulation. In~\cite{li2023spatial} and~\cite{yang2023spatial}, the authors investigated the design of IM in the spatial domain.
In~\cite{qian2023block}, a block-wise IM scheme was proposed by dividing the DD grid into independently activated sub-blocks, enabling effective exploitation of block sparsity under high mobility.
On this basis, a joint delay-Doppler domain IM scheme was presented, which simultaneously utilizes both delay and Doppler dimensions for index mapping, thereby improving throughput without increasing constellation size~\cite{tek2024joint}.

Apart from OTFS-IM, alternative modulation-domain waveforms such as AFDM have also been explored due to their unique potential in enabling IM over DDCs.
In~\cite{tao2025affine}, Tao {\em{et al.}} proposed a complete AFDM-IM system model, analyzed the conditions for achieving full diversity gain, and developed a low-complexity MPA for joint detection.
Building upon this, Liu {\em{et al.}} developed a novel AFDM scheme with the pre-chirp index modulation (PIM) philosophy, which embeds additional information by dynamically assigning pre-chirp parameters for index modulation~\cite{liu2025pre}.
In addition, Anoop {\em{et al.}} proposed a dual-mode IM scheme that combines modulation-domain and activation-pattern-based mapping~\cite{anoop2025dual}. Sui {\em{et al.}} incorporated generalized spatial modulation (GSM) into AFDM, enabling joint spatial-frequency domain index modulation and offering new design insights for multi-antenna systems~\cite{sui2026generalized}.

In summary, OTFS-IM and AFDM-IM each exhibit distinct advantages in terms of structure, spectral efficiency, and complexity management. Although notable progress has been made, challenges remain in complexity reduction, dynamic index design, and joint resource scheduling. Future work may focus on multi-domain joint indexing, AI-assisted IM pattern optimization, communication-sensing integration, and hardware prototyping, paving the way for IM-enhanced waveform deployment in next-generation wireless systems.

\begin{figure}[thbp]\new
	\centering
    \includegraphics[width=0.45\textwidth]{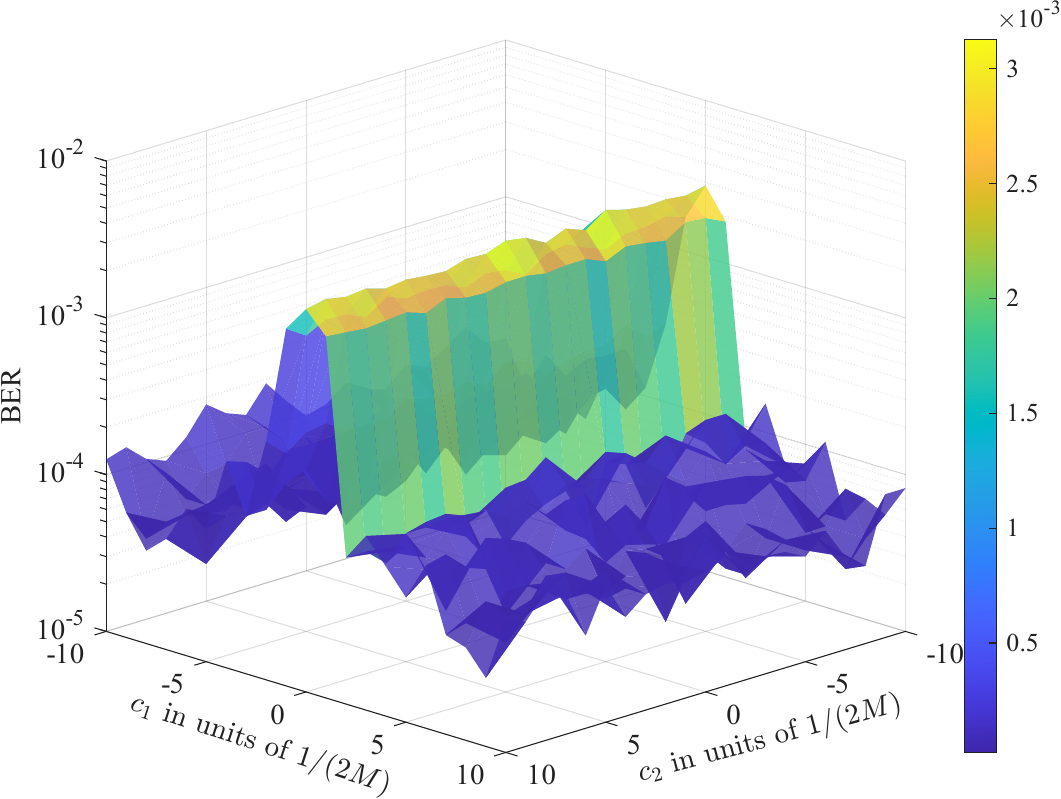}
	\caption{BER performance of AFDM under different $c_1$ and $c_2$ modulated by QPSK with subcarrier number $M=128$ with fixed channel parameters consistent with Fig.~\ref{fig4-1} using MMSE detector.}
	\label{fig-BER-c1c2}
    \vspace{-8pt}
\end{figure}

\vspace{-8pt}
\subsection{Integrated Sensing and Communications}
\label{sec6-6}

\begin{figure*}[hbpt]\new
    \centering
    \subfloat[]{
        \includegraphics[width=0.48\textwidth]{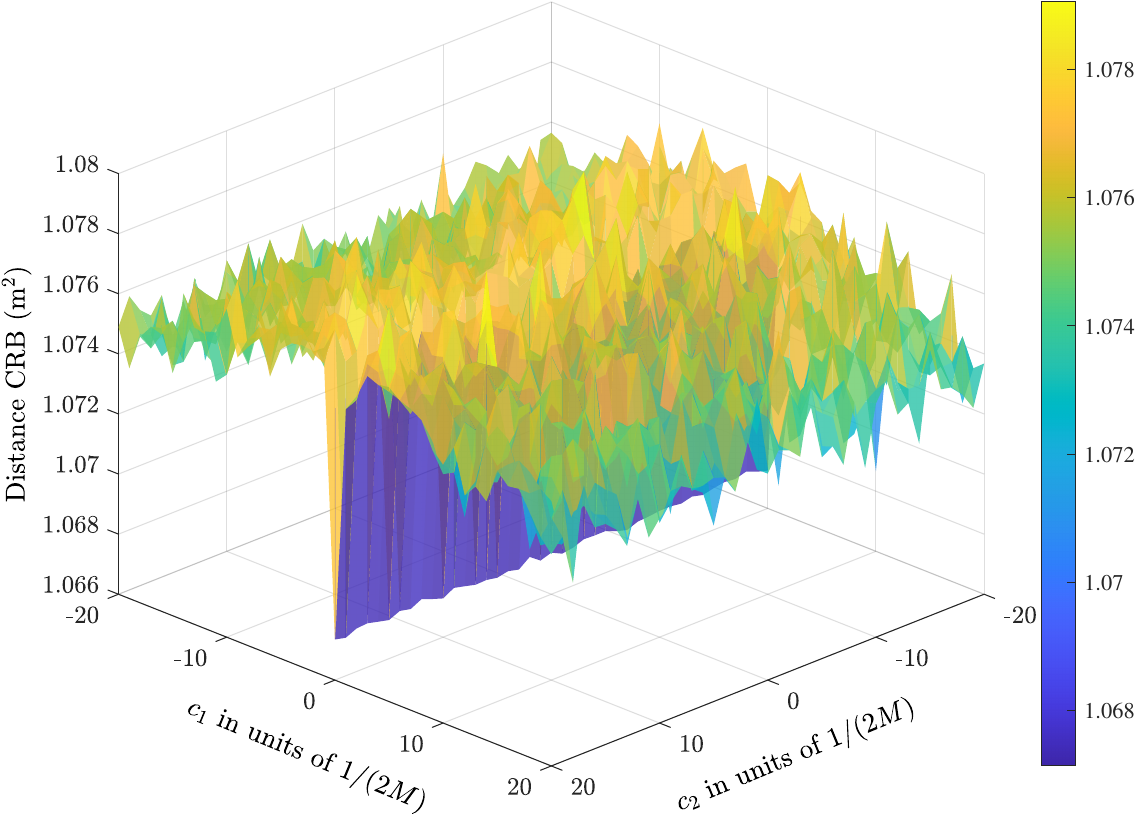}\label{fig-CRB-D}
    }
    \hfill
    \subfloat[]{
        \includegraphics[width=0.48\textwidth]{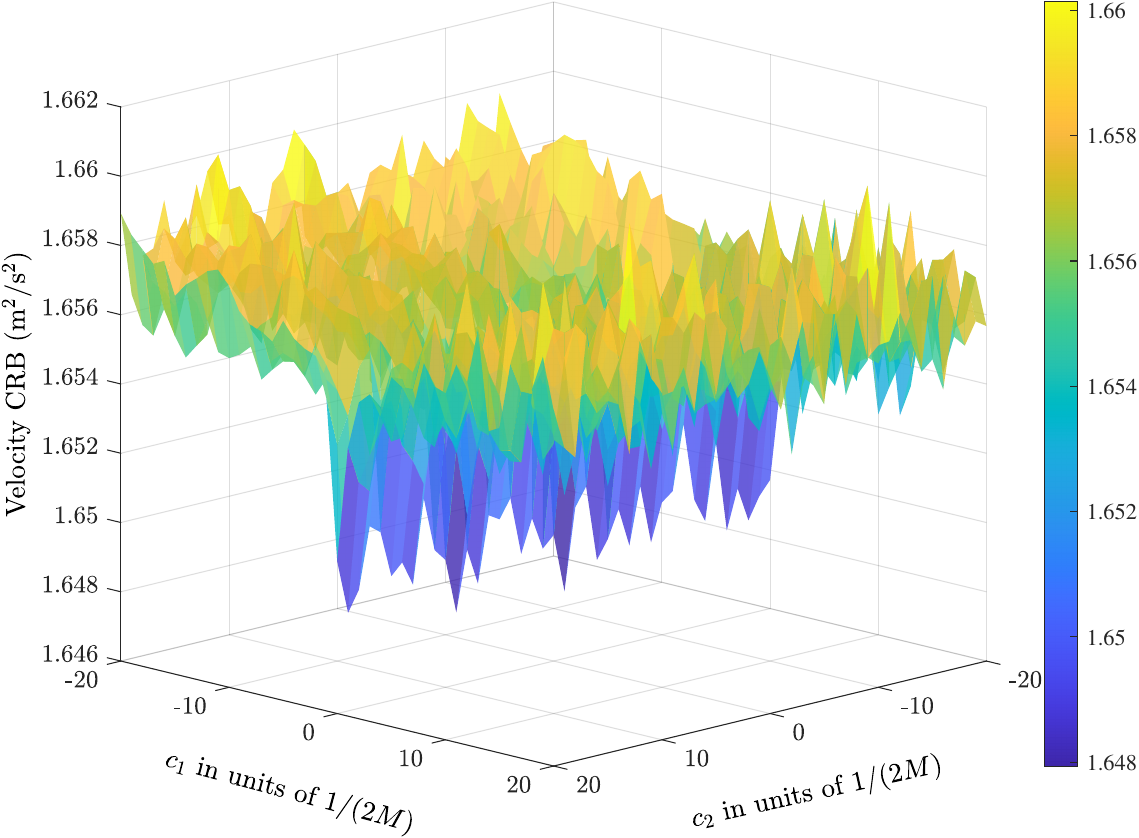}\label{fig-CRB-V}
    }
    \caption{Averaged CRBs of AFDM-ISAC symbols with random data under different $c_1$ and $c_2$. For the hypothetical single target with a detection range of $350\,\text{m}$ and a velocity of $1080\,\text{km/h}$, all other conditions remain consistent with those in Fig~\ref{fig4-1}. This includes consideration of AFDM symbols with $M=128$ using QPSK modulation and the SNR=10\,dB. (a) averaged distance CRB; (b) averaged velocity CRB.}
    \label{fig-CRB}
    \vspace{-12pt}
\end{figure*}

The design of ISAC at the PHY can be primarily categorized into three predominant methods: sensing-centric design, communication-centric design, and joint design. It is asserted that communication-centric design is considered the most promising ISAC paradigm for practical implementation in next-generation networks\cite{liu2022integrated}. This can be attributed to its full compatibility with existing cellular infrastructure, which enables low-cost and large-scale deployment. 
Currently, communication-centric design mainly focuses on the following types of waveforms: 1D waveforms represented by OFDM and AFDM, and 2D waveforms represented by OTFS, DDAM, and ODDM in the DD domain. 

The key to the design of the AFDM-ISAC system lies in balancing the performance of communication and sensing by utilizing the degrees of freedom of AFDM from the adjustable parameters\cite{zhang2026afdmenabled,luo2025novel}. \new{When using different chirp parameters to evaluate communication performance on a fixed channel, it can be observed that AFDM demonstrates significant performance improvements over both OFDM ($c_1=c_2=0$) and OCDM ($c_1=c_2=1/(2M)$), as illustrated in Fig.~\ref{fig-BER-c1c2}. As shown in Fig.~\ref{fig-CRB}, the theoretical lower bounds for the distance and velocity estimates of the AFDM-ISAC symbols also exhibit variations when different $c_1$ and $c_2$ parameters are selected. Here, we employ the average $\text{Cram\'er-Rao bound}$ (CRB) simulated from 5,000 Monte Carlo simulations to represent the theoretical values for single-target ISAC estimation. The results demonstrate that the parameters of AFDM can deliver robust sensing accuracy\cite{yin2025ofdm}.} Ni {\em{et al.}} proposed an effective AFDM-based ISAC scheme and designed a specialized metric set to extract delay and Doppler shift, thereby achieving synergy between communication signals and sensing functions\cite{ni2025integrated}. Zhu {\em{et al.}} proposed a low-complexity bistatic ISAC system based on AFDM, achieving high-resolution ranging of moving targets while achieving optimal sensing resolution\cite{zhu2024afdmbased}. Bemani {\em{et al.}} utilized a single-pilot scheme for AFDM-ISAC, achieving efficient SI cancellation\cite{bemani2024integrated}.
On the other hand, OTFS-ISAC\cite{zhang2023radar} and ODDM-ISAC\cite{lin2025joint, wang2024exploring} systems leverage the advantages of signals in the DD domain lattice to support high-resolution sensing and reliable communication. \new{Among these, velocity detection plays a crucial role in ISAC, particularly in high-speed mobility scenarios. In practice, combined detection and interference cancellation methods are often employed to enhance the accuracy of velocity detection\cite{zegrar2024otfsbased,li2024otfsbased,ranasinghe2025joint2}.} In \cite{yuan2024otfs}, the paradigm shift from OTFS to DD-ISAC is analyzed, which is considered a promising future direction for ISAC development. 

Nevertheless, the research of Liu {\em{et al.}} indicates that OFDM still has irreplaceable advantages in terms of ranging sidelobe levels, which is crucial for sensing performance\cite{11037613}. On the other hand, DDAM theoretically shares the same maximum unambiguous distance, distance resolution, and velocity resolution as OFDM. However, DDAM is more robust in Doppler frequency estimation, with its maximum unambiguous Doppler frequency being on the same order of magnitude as the system bandwidth $B$, rather than $B/K$ in OFDM systems~\cite{xiao2023integrated}. Therefore, in high-mobility scenarios, DDAM's sensing performance significantly outperforms OFDM\cite{xiao2023exploiting}.

To enhance the performance of communication-centric ISAC designs, synergistic optimization of communication and sensing can be achieved by optimizing parameters such as subcarrier spacing, modulation domain selection, and CP length. In addition, the integrated channel sounding and communication (ICSC), which is designed to accurately measure channels, will also better empower ISAC\cite{zhouyu,zhounanhao}. When evaluating the suitability of waveforms in ISAC, careful consideration should be given to aspects such as the AF, SI cancellation techniques, and PAPR\cite{wang2025optimal}. In the future, the unified waveform framework should support dynamic waveform switching to enable scene adaptation in ISAC systems.

\vspace{-8pt}
\subsection{Integrated Localization and Communications}
\label{sec6-7}
Integrated localization and communications (ILAC) is expected to play a key role in a multitude of applications for the B5G and 6G wireless networks\cite{morselli2023soft,xiao2021overview}, such as the IoT\cite{abuyaghi2024positioning}, smart environments\cite{weng2024intelligent}, and autonomous driving\cite{kuutti2018survey}. In particular, mmWave networks can enable new applications via large-bandwidth exploitation and multiantenna processing\cite{kwon2023integrated}. However, the mmwave networks also face a number of challenges. Among them, the path loss is extremely severe at high carrier frequencies, which must be compensated through beamforming, while the key to beamforming lies in the knowledge of the location of the target. 

The propagation channel is an important bridge between localization and communication. Typical location-related channel parameters include round-trip delay, angle of arrival (AoA), angle of departure (AoD), and so on. The key to obtaining these parameters lies in the design and processing of the signal in the PHY of the ILAC system\cite{gong2023simultaneous}.

Among the existing multicarrier waveforms, research on ILAC based on OFDM waveforms has been extensively explored\cite{keskin2021mimo}. Studies have demonstrated that OFDM can achieve radar performance with accuracy comparable to that of frequency modulated continuous wave (FMCW)\cite{gaudio2020effectiveness}. Waveforms such as OTFS and AFDM\cite{yi2024highprecision} are designed based on a complete delay-Doppler representation of time-varying multipath channels, inherently capable of estimating delay and Doppler shifts, and exhibiting commendable communication and sensing capabilities\cite{rou2024orthogonal}. In summary, waveform design is a pivotal component in ILAC systems.

\vspace{-5pt}
\section{Open Challenges and Future Directions}
\label{sec7}
In this section, we discuss the challenges and future directions of multicarrier waveforms from three perspectives: theoretical, application, and implementation, aiming to promote the practical application of the unified framework for PHY waveforms.
\vspace{-10pt}
\subsection{Theoretical Limits}
\label{sec7-1}
The theoretical limits of waveform design determine the performance bounds of future 6G systems, particularly in spectral efficiency, sensing resolution, and interference immunity. Here, we outline the primary theoretical challenges in developing a unified waveform framework:

\begin{itemize}
    \item \textbf{Information Capacity Trade-off in ISAC}: The informatics capacity of ISAC systems remains unresolved due to the competition between communication throughput and perception accuracy for spectrum resources \cite{liu2026sensing}. For instance, increasing pilot frequency density enhances ranging accuracy but reduces the effective data rate, necessitating a joint capacity framework to quantify this trade-off.
    \item \textbf{Complexity of Waveform Optimization}: Deriving closed-form solutions for generalized parametric waveform models is theoretically complex. The optimization of \(\mathbf{U}\) regarding waveform set and inborn waveform parameters for multi-scenario adaptation, balancing spectral efficiency and interference suppression, poses significant mathematical challenges \cite{10909646}.
\end{itemize}

Future research should focus on developing a unified information theory model for ISAC to delineate performance boundaries between communication and perception. Exploring nonlinear signal processing techniques may overcome traditional orthogonality constraints, while deriving closed-form solutions for parametric waveform design will guide framework optimization. These efforts will enhance the theoretical foundation of waveform design and support 6G multi-scenario convergence.

\vspace{-8pt}
\subsection{Application Challenges}
\label{sec7-2}
Next-generation wireless networks, spanning high-mobility vehicular networking, low-latency industrial IoT, and space-air-ground integrated networks (SAGIN), demand a unified waveform framework to enable efficient communication and accurate sensing in heterogeneous environments. However, diverse KPIs (e.g., spectral efficiency, ranging accuracy, and latency) and compatibility with existing standards present significant application challenges. Below, we outline the primary challenges in deploying the unified waveform framework:

\begin{itemize}
    \item \textbf{Scenario Diversity and Waveform Flexibility}: Diverse scenarios impose stringent requirements on waveform adaptability. The unified framework must dynamically adjust coarse-grained (e.g., waveform types) and fine-grained (e.g., waveform parameters\cite{zhang2024daft}) to meet these needs, but real-time switching under varying channel conditions remains challenging.
    \item \new{\textbf{AI-assisted Waveform Decision}: Although AI-assisted waveform decision-making demonstrates significant potential for achieving adaptive waveform switching, its practical application still faces numerous challenges. These include the need for large-scale training data, high computational overhead from real-time inference, and difficulties in generalizing trained models.}
    \item \textbf{Resource Competition}: Maximizing performance for competing demands (e.g., conflicts in ISAC) within limited resources (e.g., power, spectrum, time, and space) is a key issue that waveform design needs to address. Optimizing resource allocation to balance the competing demands lacks mature solutions \cite{3gppisac}.
    \item \textbf{Compatibility with Existing Standards}: Standard compatibility is imperative for ensuring strong viability. The framework must support emerging 6G scenarios while ensuring seamless integration with existing infrastructures \new{such as 5G NR~\cite{3gpp38211}}. This requires new protocol designs or spectrum-sharing mechanisms, posing significant compatibility challenges.
\end{itemize}

Future research should develop scenario-adaptive waveform switching algorithms leveraging real-time CSI to adjust parameters dynamically, e.g., transitioning from OFDM to Doppler-resistant waveforms for high-mobility scenarios. Additionally, joint resource allocation strategies for multi-user ISAC, alongside cross-industry standardization efforts to define unified ISAC waveform specifications, will facilitate practical deployment and ensure the framework's centrality in 6G applications.

\vspace{-8pt}
\subsection{Implementation Challenges}
\label{sec7-3}
The unified waveform framework faces significant implementation challenges, including computational complexity, hardware constraints, system integration, and interference control \cite{10736552}. While 1D waveforms (e.g., OFDM, DFT-s-OFDM) benefit from low complexity and mature 5G deployments, 2D waveforms and ISAC co-design increase deployment difficulties. Below, we outline the primary implementation challenges:

\begin{itemize}
    \item \textbf{Multifunctional Integration}: Multifunctional systems (e.g., ISAC, ILAC) necessitate integrated communication and other applications on a single RF front-end, sharing antennas and bandwidth. This integration demands complex signal processing and precise synchronization mechanisms\cite{10188491, 9585321}.
    \item \textbf{Computational Complexity}: The unified framework requires balancing performance and complexity through parametric design (e.g., reduced grid resolution). However, achieving efficient real-time computation,  especially for some 2D waveforms, remains a significant bottleneck. 
    \item \textbf{Multi-granularity Resource Allocation}: Coarse-grained allocation uses larger blocks for high-throughput tasks, while fine-grained allocation enables precise partitioning for latency-sensitive or high-resolution needs. Multi-granularity resource allocation requires the development of adaptive algorithms to balance granularity while ensuring quality of service dynamically\cite{du2023overview}.
\end{itemize}

Future research should focus on low-complexity algorithms, such as approximate transformations or dimensionality reduction for 2D waveforms. AI-based waveform optimization can dynamically adjust parameters (e.g., subcarrier count, modulation domain) to mitigate hardware limitations. Developing modular ISAC hardware platforms supporting flexible waveform switching and function expansion will enhance deployment, enabling the unified framework to meet 6G communication and sensing requirements.

\section{Conclusion}
\label{sec8}
This paper undertakes a systematic investigation into advanced PHY waveform design, with a particular focus on the development of a unified waveform framework for next-generation wireless networks. A classification strategy of 1D waveforms (i.e., OFDM, DFT-s-OFDM, FrFT-OFDM, OCDM, IFDM, AFDM) and 2D waveforms (i.e., FBMC, OTFS, ODDM, DDAM, OTSM, ODSS) is proposed. This strategy facilitates a more profound comprehension of their design principles. A detailed investigation is then undertaken to ascertain the underlying causes of waveform interference suppression. This analysis is approached from two distinct vantage points: firstly, the interference caused by both the transmitter and receiver is investigated, and secondly, the channel matrix structure is examined. The analysis of communication and sensing KPIs, including CP overhead, BER, PAPR, spectral efficiency, modulation complexity, and AF, demonstrates the compatibility of the considered waveforms with different channels. Furthermore, the development and robustness of existing waveforms in advanced applications such as MA, MIMO, IM, and ISAC are investigated, with the challenges being analyzed on an individual basis. Moreover, by examining theoretical limits, application challenges, and implementation challenges, the study identifies several critical future directions, with the aim of contributing to the advancement of next-generation wireless communication technologies.

\bibliographystyle{IEEEtran}
\bibliography{refs}

\end{document}